\documentclass[journal]{new-aiaa}
\usepackage[utf8]{inputenc}
\usepackage{textcomp}

\usepackage[toc,page]{appendix}
\usepackage{graphicx}
\usepackage{amsmath}
\usepackage[version=4]{mhchem}
\usepackage{siunitx}
\usepackage{longtable,tabularx}
\title{Optimization-Based Formation Flight on Libration Point Orbits
\footnote{\small{A preliminary version of this work was presented as manuscript 25-717 at the AAS/AIAA Astrodynamics Specialist Conference, in Boston, MA, on Aug. 2025.}}}

\author{
    Yuri Shimane\footnote{Assistant Professor, Department of Mechanical and Aerospace Engineering, Member AIAA; yshimane@uci.edu (Corresponding Author).}
}
\affil{University of California, Irvine, CA 92617}
\author{
    Purnanand Elango\footnote{Research Scientist, Member AIAA.}, and 
    Avishai Weiss\footnote{Senior Principal Research Scientist, Member AIAA.}
}
\affil{Mitsubishi Electric Research Laboratories, Cambridge, MA 02139}

\usepackage{amsmath}
\usepackage{amsfonts}   % \usepackage{amssymb}  % redundant with amsfonts?
\usepackage{siunitx}
\usepackage{booktabs}
\usepackage{mathtools}% http://ctan.org/pkg/mathtools
\usepackage{caption}
\usepackage{subcaption}
\usepackage{multirow}
\usepackage{lscape}
\usepackage{upgreek}
\usepackage{algorithm}
\usepackage{algpseudocode}
\usepackage{todonotes}
\usepackage{multirow}
\usepackage{tikz}
\usetikzlibrary{shapes.geometric, arrows, calc}
\newcommand{\Bbold}{\boldsymbol{B}}

\newcommand{\Tbold}{\boldsymbol{T}}

\newcommand{\abold}{\boldsymbol{a}}
\newcommand{\dbold}{\boldsymbol{d}}
\newcommand{\rbold}{\boldsymbol{r}}
\newcommand{\vbold}{\boldsymbol{v}}
\newcommand{\ubold}{\boldsymbol{u}}
\newcommand{\Ubold}{\boldsymbol{U}}

\newcommand{\ellbold}{\boldsymbol{\ell}}
\newcommand{\xbold}{\boldsymbol{x}}
\newcommand{\Xbold}{\boldsymbol{X}}
\newcommand{\ybold}{\boldsymbol{y}}
\newcommand{\Zbold}{\boldsymbol{Z}}

\newcommand{\fbold}{\boldsymbol{f}}
\newcommand{\Ibold}{\boldsymbol{I}}

\newcommand{\xibold}{\boldsymbol{\xi}}

\newcommand{\st}{\text{ subject to }}
\DeclareMathOperator*{\esssup}{ess\,sup}

\newcommand{\red}[1]{\textcolor{black}{#1}}
\newcommand{\minimize}[1]{\underset{#1}{\operatorname{min}}}

\newcommand*{\permcomb}[4][0mu]{{{}^{#3}\mkern#1#2_{#4}}}

\newcommand*{\comb}[1][-1mu]{\permcomb[#1]{C}}
\newcommand{\rev}[1]{\textcolor{black}{#1}}

\tikzstyle{startstop} = [rectangle, rounded corners, 
minimum width=3cm, 
minimum height=1cm,
text centered, 
draw=black,
fill=gray!20]

\tikzstyle{check} = [diamond, 
aspect=2.5,
text centered, 
draw=black, 
fill=gray!5]

\tikzstyle{io} = [trapezium, 
trapezium stretches=true, % A later addition
trapezium left angle=70, 
trapezium right angle=110, 
minimum width=3cm, 
minimum height=1cm, text centered, 
draw=black]
\tikzstyle{process} = [rectangle, 
minimum width=3.6cm, 
minimum height=1cm, 
text centered, 
text width=4.6cm, 
draw=black]

\tikzstyle{processerror} = [rectangle, rounded corners,
minimum width=3.6cm, 
minimum height=1cm, 
text centered, 
text width=4.6cm, 
draw=red,
fill=red!10]

\tikzstyle{arrow} = [thick,->,>=stealth]
\begin{document}

\maketitle

\begin{abstract}
A model predictive control (MPC) framework is developed for station-keeping in spacecraft formation flight along libration point orbits.
At each control period, the MPC policy solves a multi-vehicle optimal control problem (MVOCP) that tracks a reference trajectory, while enforcing path constraints on the relative motion of the formation.
The control policy makes use of a limited set of control nodes consistent with operational constraints that allow only a small number of maneuver opportunities per revolution.
To promote recursive feasibility, path constraints are progressively tightened across the prediction horizon. An isoperimetric reformulation of the constraints is used to prevent inter-sample violations.
The resulting MVOCP is a nonconvex program, which is solved via sequential convex programming.
The proposed approach is evaluated in a high-fidelity ephemeris model under uncertainties for a formation along the near-rectilinear halo orbit (NRHO), and subject to path constraints on inter-spacecraft separation and relative Sun phase angle.
The results demonstrate maintenance of a spacecraft formation that satisfies the path constraints with realistic cumulative propellant consumption.
\end{abstract}

\section*{Nomenclature}
{\renewcommand\arraystretch{1.0}
\noindent\begin{longtable*}{@{}l @{\quad=\quad} l@{}}
% letters
$\abold$    & acceleration vector, \SI{}{m/s^2} \\
$\mathcal{F}$ & reference frame \\
$\fbold$    & dynamics function \\
$g$         & path constraint function \\
$\Tilde{g}$ & tightened path constraint function \\
$h$         & specific angular momentum, \SI{}{m^2/s} \\
$\ellbold$  & line of sight vector \\
$M$         & number of spacecraft \\
$N$         & number of control nodes \\
$N_{\rm rev}$ & number of revolutions in prediction horizon \\
$\rbold$    & position vector, \SI{}{m}  \\
$\Tbold^A_B$ & transformation matrix from frame $A$ to frame $B$ \\
$t$         & time, \SI{}{s}  \\
$\Ubold$    & concatenated control vector, \SI{}{m/s} \\
$\ubold$    & impulsive control vector, \SI{}{m/s} \\
$\vbold$    & velocity vector, \SI{}{m/s}  \\
$\Xbold$    & concatenated state vector \\
$\xbold$    & spacecraft state vector \\
$\ybold$    & isoperimetric path constraint slack variable vector \\
% greek
$\zeta$     & constraint tightening function \\ 
$\eta$      & constraint tightening margin \\ 
$\theta$    & osculating true anomaly, \SI{}{rad} \\
$\kappa$    & constraint tightening parameter \\
$\mu$       & gravitational parameter, \SI{}{m^3/s^2} \\
$\phi$      & relative Sun phase angle, \SI{}{rad} \\
% symbols
$\mathbb{R}^n$ & length-$n$ column vector of real numbers \\
$\mathbb{R}^{n \times m}$ & $n$-by-$m$ matrix of real numbers \\
$\mathbb{R}_+$ & positive real number 
% $\mathbb{R}^{n \times m}_+$ & $n$-by-$m$ matrix of positive real numbers  \\
% % acronyms
% CT          & Continuous-time \\
% CTCS        & Continuous-time constraint satisfaction \\
% HFEM        & High-Fidelity Ephemeris Model \\
% MPC	        & Model Predictive Control \\
% NRHO        & Near Rectilinear Halo Orbit \\
% OCP         & Optimal Control Problem
\end{longtable*}}

% ===================================================================== %
\section{Introduction}
% \lettrine{T}{his} document is a \LaTeX{} template for preparation of papers for AIAA Technical Journals. If you are reading a hard-copy or .pdf version of this document, download the electronic file, new-aiaa.cls, and use it to prepare your manuscript.
Safe guidance schemes for formation flight around libration point orbits (LPOs) are of growing interest due to their relevance to future missions~\cite{Levinson-Muth2022-ep,Sherman2023,Capannolo2023} as well as for proximity operations and rendezvous with assets such as the Gateway along its near-rectilinear halo orbit (NRHO)~\cite{Bucci2018,Bucchioni2020,Blazquez2020,Sandel2024-sp}.

Motivated by the success of the \rev{coordinates describing relative motion in Earth orbit such as relative orbital elements}, several works have developed \rev{similar mechanisms} for the LPO context~\cite{Calico1984,Hsiao2002,Scheeres2003,Franzini2019-if,Elliott2021,Elliott2022,Burnett2022-jr,Takubo2026-bo}.
Specifically, the center manifold of the LPO, which corresponds to a non-expanding, oscillatory eigenvector of a periodic trajectory, gives rise to quasi-periodic motions that remain in the vicinity of the LPO, known as quasi-periodic tori (QPT).
The natural tendency of QPTs to remain in the vicinity of an LPO has been exploited to provide dynamical insight into relative motion.
For example, Calico and Wiesel~\cite{Calico1984} characterize Floquet modes to provide insight into relative motion in the vicinity of LPOs. 
Hsiao and Scheeres~\cite{Hsiao2002} introduced linearized \rev{relative motion descriptors} to design control laws that produce oscillatory motions similar to the motion in the center eigenspace in Scheeres et al.~\cite{Scheeres2003}. 
Recently, local toroidal coordinates (LTCs)~\cite{Elliott2022} have emerged as a geometric framework for relative motion around LPOs in simplified dynamics models such as the circular restricted three-body problem (CR3BP).
LTCs are constructed from the oscillatory Floquet mode associated with the center manifold and are well-suited for studying bounded relative motion and guidance problems.
%An LTC is defined using the non-expanding, oscillatory mode of the LPO's monodromy matrix and is well-suited to study bounded motion and guidance problems in the vicinity of the LPO. 
Elliott and Bosanac~\cite{Elliott2021} developed a targeting-based guidance scheme in the LTC defined in the CR3BP and demonstrated its performance in a high-fidelity ephemeris model (HFEM).
Takubo et al.~\cite{Takubo2026-bo} also adopted the LTC system and developed optimization-based guidance problems in the CR3BP that enforce passive safety through geometric constraints.

% A challenge remains in formation flight for HFEM, where only quasi-periodic trajectories can be constructed, and therefore, no exact center manifold exists.
% Despite these advances, significant challenges remain for formation flight in high-fidelity ephemeris models. In the HFEM, only quasi-periodic trajectories can be constructed, and therefore, no exact center manifold exists.
% To construct a QPT in the HFEM, a QPT from simplified models such as the CR3BP may be transitioned through differential correction~\cite{Howell2005}.
% To ensure the transitioned QPT exhibits the desirable properties exhibited in the simplified model, the transition process must be carefully monitored and tuned by the trajectory designer.
% Once a QPT is obtained, LTC-based guidance developed in simplified models has been reported to perform well in HFEM under uncertainties as well, due to the similarity of the relative motion's eigenstructure between simplified models and the HFEM~\cite{Elliott2021,Takubo2026-bo}.
% Still, the inherent need for transitioning from simpler dynamics models renders any formal guarantees difficult to establish in the HFEM.

However, \rev{using} the favorable properties of QPT for formation flight in the HFEM requires additional care.
Because of the time-varying, aperiodic nature of the dynamics in the HFEM, LPO reference trajectories are no longer strictly periodic\footnote{We maintain the term ``libration point \textit{orbit}'' to refer to quasi-periodic orbits around libration points in the HFEM.}. Consequently, the center manifold cannot be defined in the strict sense because a monodromy matrix is not well defined.
Nevertheless, the local eigenstructure of LPO trajectories in simplified models such as the CR3BP has been shown to closely resemble that observed in HFEM dynamics~\cite{Elliott2021,Elango2022-jv,Takubo2026-bo}, and consequently, quasi-periodic motions analogous to QPTs can still be constructed in the HFEM, although typically only over finite durations~\cite{Howell2005}. In practice, these finite-duration QPTs can be generated from almost-purely oscillatory modes and used as reference trajectories for spacecraft formations. For example, Foss et al.~\cite{Foss2025-sm} construct such finite-duration QPTs in the HFEM to define the desired relative motion within a spacecraft formation.

Because of the similarity in eigenstructure, guidance strategies based on LTCs constructed in the CR3BP have been shown empirically to perform adequately for relative motion control in HFEM simulations~\cite{Elliott2021,Takubo2026-bo}.
The eigenmotion controller in~\cite{Elango2022-jv} or the Cauchy-Green Tensor-based controller in~\cite{Williams2023-dr} and references therein are examples in which non-expanding eigenmodes constructed directly in the HFEM over finite time intervals are utilized for station-keeping of a single spacecraft.
While finite-duration QPTs provide a useful structure for designing the relative motion of a spacecraft formation, because these center modes are only approximate in the HFEM, establishing formal guarantees of passive safety remains challenging. 
To overcome this limitation, the control scheme in~\cite{Foss2025-sm} consists of decoupled control blocks that achieve, respectively, \rev{(i)} \textit{absolute} station-keeping, i.e., ensuring each spacecraft \rev{tracks the absolute motion of the reference LPO, and (ii) \textit{relative} station-keeping, i.e., maintaining the desired formation geometry, while ensuring passive safety between all pairs of spacecraft in the formation.
Safety is enforced through} a control barrier function (CBF), which becomes active in case maneuvers generated by either the absolute or relative station-keeping blocks are predicted to violate passive safety constraints.
% remains near the reference LPO, 
% \textit{relative} station-keeping, i.e., maintaining the desired formation geometry, and passive safety using a control barrier function (CBF), in case maneuvers generated by either the absolute or relative station-keeping blocks are predicted to violate passive safety constraints.
Although this modular architecture facilitates integration with existing absolute station-keeping controllers, the decoupling of these objectives introduces an inherent loss of propellant optimality.
The sub-optimality of such a modular framework, along with its need for additional a posteriori checks on safety through a CBF, motivates a guidance framework that simultaneously achieves absolute and relative station-keeping while enforcing safety and operational constraints directly within the control formulation, without relying on approximate dynamical structures.

% % Literature table
% \begin{table}[]
% \centering
% \caption{TODO}
% \label{tab:literature_comparison}
% \begin{tabular}{@{}lll@{}}
% \toprule
% Reference & Approach & Application \\ \midrule
% Foss et al.~\cite{Foss2025-sm}
%     & LTC      & Formation          \\
% A
%     &          &             \\ 
% Elango et al.~\cite{Elango2022-jv}
%     & Eigenmotion MPC &             \\ 
% Shimane et al.~\cite{Shimane2025}
%     & MPC      & Single vehicle     \\
% (This work)
%     & MPC & Formation \\ \bottomrule
% \end{tabular}
% \end{table}

In this work, we develop a model predictive control (MPC) framework that simultaneously regulates the absolute and relative motion of all spacecraft in the formation.
In the proposed approach, passive safety requirements, together with other mission constraints on the formation geometry, are enforced explicitly as constraints within the optimization problem, thereby removing the need to explicitly rely on dynamic structures such as QPTs.
By enforcing mission requirements on the formation directly through the optimization problem, dynamics-agnostic path constraints, such as bounded relative Sun phase angle, can be incorporated in a straightforward and systematic manner.
Our formulation provides a unified mechanism to simultaneously enforce formation safety, operational constraints, and propellant efficiency within a single control framework.
\rev{The centralized structure of the proposed MPC, as well as its optimization-based approach eliminating the need for QPTs within the guidance scheme, is in contrast to the decoupled approach in~\cite{Foss2025-sm}.}

The main contributions of this work are threefold and are summarized as follows.
\rev{First,} we formulate a multi-vehicle optimal control problem (MVOCP) that simultaneously achieves absolute and relative station-keeping, while enforcing mission requirements as continuous-time (CT) path constraints; the problem is posed with an operationally realistic control cadence of two maneuvers per LPO revolution.
Second, the MVOCP is incorporated into an MPC scheme, where the CT path constraints are handled through (a) progressive constraint tightening along the prediction horizon to empirically promote recursive feasibility, and (b) an isoperimetric reformulation to enforce CT path constraints~\cite{Teo1987,Elango2024}.
Finally, the proposed MPC framework is evaluated via Monte Carlo simulations in the HFEM, incorporating initial orbit insertion, dynamics modeling, navigation, and control execution errors. CT path constraints are imposed on inter-spacecraft separation and relative optical visibility.
The results demonstrate reliable station-keeping with projected annual propellant costs on the order of a few hundred \SI{}{cm/s} per spacecraft.

The remainder of this paper is organized as follows: in Section~\ref{sec:background}, we provide \rev{a review of relevant literature,} an overview of the HFEM dynamics, the NRHO reference trajectory, and the environment uncertainties considered in this study.
Section~\ref{sec:mvocp} introduces the CT path constraints and formulates the MVOCP.
 Section~\ref{sec:solution_approach} presents the constraint tightening and CT \rev{path constraints enforcement} strategies, along with details of the SCP solution method, as applied to the MVOCP for station-keeping in formation flight. 
 %We also provide a discussion on the SCP scheme adopted to solve the problem.
The proposed method is demonstrated through Monte Carlo experiments presented in Section~\ref{sec:numerical_experiments}.
Finally, conclusions are provided in Section~\ref{sec:conclusions}.

% ===================================================================== %
\section{Background}
\label{sec:background}
We present the background necessary to develop the proposed MVOCP and the corresponding MPC scheme.
We begin by introducing the literature related to this work in~\ref{sec:background_literature}, followed by the reference frames in~\ref{sec:background_frames}, and the HFEM dynamics in~\ref{sec:background_HFEM}.
Then, we provide a short discussion on the NRHO with respect to which the spacecraft formation is studied in~\ref{sec:background_NRHO}.
We then introduce the uncertainties that are considered within subsequent numerical experiments in~\ref{sec:background_uncertainties}.

\subsection{Related Literature}
\label{sec:background_literature}
\rev{A summary of related station-keeping works is presented in Table~\ref{tab:literature_comparison}.
In the remaining subsections, we provide an overview on sequential convex programming (SCP) in the context of MPC, constraint tightening, and CT path constraints.
}

% \rev{The related literature is grouped in terms of }
\subsubsection{Recursive Sequential Convex Programming}
The reformulated MVOCP solved within the MPC framework results in a nonconvex nonlinear program (NLP), which we solve using SCP.
In effect, this work extends the single-vehicle station-keeping MPC along LPOs~\cite{Shimane2025} to the multi-vehicle formation setting, \rev{with the introduction of CT path constraints to adhere to formation flight-specific mission requirements}.
SCP algorithms such as the Augmented Lagrangian-based method~\cite{Oguri2023} and the prox-linear method~\cite{Lewis2008,Drusvyatskiy2018} provide theoretical convergence guarantees to a local feasible minimizer of the original nonconvex problem. These methods have been successfully applied to cislunar trajectory optimization~\cite{Kumagai2024,Kumagai2025-vu,Spada2025-eh}, making them well-suited for solving the MVOCP.
In this work, we adopt the Augmented Lagrangian-based SCP \rev{algorithm}~\cite{Oguri2023}.
\rev{As far as the MVOCP formulation is concerned, any other NLP algorithms such as sequential quadratic programming or interior point methods may be employed interchangeably; for a taxonomy of NLP algorithms, see~\cite{Vanaret2026-zf}.}

\subsubsection{Constraint Tightening}
The constraint tightening scheme employed here is closely related to techniques used in robust MPC.
Constraint tightening, together with a terminal constraint defined by a positively invariant set resulting from a linear feedback policy, can ensure recursive feasibility and constraint satisfaction in the presence of persistent, bounded disturbances~\cite{Richards2005-ud}.
In the context of robust MPC, constraint tightening has been employed in order to retain a margin that can be utilized in the future by a selected feedback policy, originally for linear MPC~\cite{Gossner1997-hb,Chisci2001-or,Richards2007-ud,Kuwata2007-dv} and recently for nonlinear MPC~\cite{Kohler2021-dn}.
Tuning the tightening based on a feedback policy is challenging in the formation flight context: station-keeping along LPOs typically involves sparse maneuver opportunities, often limited to one or two control actions per orbit, with free-drift intervals spanning several days. Under such long control cadences, disturbances may perturb the system faster than the feedback compensation can regulate the system back to its desired response.
\rev{Chance constraints may be introduced as an alternative mechanism for providing probabilistic guarantees of feasibility; while their tractable reformulations are known to result in conservatism~\cite{Mesbah2016-cn,Heirung2018-ub}, calibrating the constraint tightening directly can improve the performance of the MPC~\cite{Lorenzen2017-se}.
To prioritize the incurred control cost over providing formal guarantees, we adopt a constraint tightening function with tunable parameters to promote recursive feasibility empirically.
Such a direct calibration of effectively a safety margin allows for a mission designer to determine a desirable trade-off between propellant consumption and risk of constraint violation that aligns with a given mission.}
% \rev{While providing a theoretical guarantee for recursive feasibility is beyond the scope of the proposed approach,} we empirically show that an appropriately tuned tightening schedule promotes recursive feasibility of the proposed MPC framework.

\subsubsection{Continuous-Time Constraint Satisfaction}
CT path constraints are inherently functional inequality constraints that must hold over an infinite set of time instants. 
To enforce such requirements within a finite-dimensional optimization problem, we adopt an isoperimetric reformulation~\cite{Teo1987} into finitely many constraints.
The theoretical properties of incorporating isoperimetric constraints into nonlinear programming formulations are discussed in detail in Elango et al.~\cite{Elango2024}. 
The incorporation of isoperimetric constraints has been applied to the powered descent and landing problem~\cite{Elango2025}, cislunar spacecraft rendezvous~\cite{Elango2025RDV}, and for geostationary satellite station-keeping~\cite{Pavlasek2025}.
Pavlasek et al.~\cite{Pavlasek2025} consider an MPC scheme with CT path constraints for station-keeping of geostationary satellites and use SCP to solve the receding-horizon optimization problems, similarly to the approach adopted in this work.

% Literature table
\begin{table}[]
\centering
\footnotesize
\caption{Summary of related station-keeping works}
\label{tab:literature_comparison}
\begin{tabular}{@{}lllll@{}}
\toprule
Reference & Approach & Orbit & Application & Note
\\ \midrule
Takubo et al.~\cite{Takubo2026-bo}
    & Shrinking horizon MPC & LPO & Formation & LTC-based guidance
\\
Elango et al.~\cite{Elango2022-jv}
    & MPC & LPO & Single vehicle  & Targeting stable mode of reference LPO
\\ 
Foss et al.~\cite{Foss2025-sm}
    & MPC      & LPO & Formation              & Targeting state along QPT, safety with control barrier function
\\ 
Shimane et al.~\cite{Shimane2025}
    & MPC      & LPO & Single vehicle         & Operations-consistent maneuver cadence
\\
Elango et al.~\cite{Elango2025RDV}
    & Stochastic OCP & LPO & Rendezvous     & CT constraints
\\
Pavlasek et al.~\cite{Pavlasek2025}
    & MPC      & GEO & Single vehicle & CT constraints
\\
(This work)
    & MPC & LPO & Formation                   & CT constraints with constraint tightening
\\ \bottomrule
\end{tabular}
\end{table}

\subsection{Reference Frames}
\label{sec:background_frames}
Several reference frames are used throughout this work to describe spacecraft dynamics, environmental effects, and relative motion within the formation. The dynamics of each spacecraft are propagated in an \textit{inertial frame} $\mathcal{F}_{\rm Inr}$ in which Newton's laws of motion apply, defined by the J2000 convention~\cite{Acton2018} and centered at the Moon.
The dynamics also include terms related to the local gravitational potential irregularities, which are resolved in the \textit{principal axes frame} of the Moon $\mathcal{F}_{\rm PA}$~\cite{Song2011}.
The motion of the reference NRHO is often analyzed in the Moon-centered \textit{Earth-Moon rotating frame} $\mathcal{F}_{\rm EM}$, defined with basis vectors
\begin{equation}
    \boldsymbol{e}_1(t) = \dfrac{-\dbold_{\oplus}(t)}{\| \dbold_{\oplus}(t) \|_2}
    ,\quad
    \boldsymbol{e}_2(t) = \boldsymbol{e}_3(t) \times \boldsymbol{e}_1(t)
    ,\quad
    \boldsymbol{e}_3(t) = \dfrac{\dbold_{\oplus}(t) \times \vbold_{\oplus}(t)}{\| \dbold_{\oplus}(t) \times \vbold_{\oplus}(t) \|_2}
    ,
\end{equation}
where $\dbold_{\oplus}$ and $\vbold_{\oplus}$ denote the position and velocity vectors of the Earth relative to the Moon with respect to~$\mathcal{F}_{\rm Inr}$.
To analyze the formation trajectory together with illumination conditions, we also define the Moon-centered Sun–Moon rotating frame $\mathcal{F}_{\rm SM}$; the basis vectors for this frame are given by
\begin{equation}
    \boldsymbol{e}_1(t) = \dfrac{-\dbold_{\odot}(t)}{\| \dbold_{\odot}(t) \|_2}
    ,\quad
    \boldsymbol{e}_2(t) = \boldsymbol{e}_3(t) \times \boldsymbol{e}_1(t)
    ,\quad
    \boldsymbol{e}_3(t) = \dfrac{\dbold_{\odot}(t) \times \vbold_{\odot}(t)}{\| \dbold_{\odot}(t) \times \vbold_{\odot}(t) \|_2}
    ,
\end{equation}
where $\dbold_{\odot}$ and $\vbold_{\odot}$ denote the position and velocity vectors of the Sun relative to the Moon with respect to~$\mathcal{F}_{\rm Inr}$.
Frames relative to a reference trajectory provide key insight into formation flight~\cite{Takubo2026-bo}.
Let $\rbold_{\rm ref}$ and $\vbold_{\rm ref}$ denote the position and velocity vectors of a reference trajectory in~$\mathcal{F}_{\rm Inr}$.
The \textit{radial, along-track, cross-track} (\rev{RSW}) frame~$\mathcal{F}_{\rm RSW}$, following the convention in~\cite[pg.~155]{Vallado2001}, is defined by 
\begin{equation}
    \boldsymbol{e}_1(t) = \dfrac{\rbold_{\rm ref}(t)}{\| \rbold_{\rm ref}(t) \|_2}
    ,\quad
    \boldsymbol{e}_2(t) = \boldsymbol{e}_3(t) \times \boldsymbol{e}_1(t)
    ,\quad
    \boldsymbol{e}_3(t) = \dfrac{
        \rbold_{\rm ref}(t) \times \vbold_{\rm ref}(t)
    }{
        \| \rbold_{\rm ref}(t) \times \vbold_{\rm ref}(t) \|_2
    }
    ,
\end{equation}
and centered at the reference trajectory.
Finally, we denote by $\Tbold^{\rm A}_{\rm B} \in \mathbb{R}^{3 \times 3}$ the transformation matrix from $\mathcal{F}_{\rm A}$ to $\mathcal{F}_{\rm B}$.

\subsection{High-Fidelity Ephemeris Model Dynamics}
\label{sec:background_HFEM}
We consider a formation involving $i\in \mathcal{I} = \{ 0,\ldots,M-1 \}$ spacecraft.
Let $\rbold_i \in \mathbb{R}^3$ denote the position of the $i^{\rm th}$ spacecraft in $\mathcal{F}_{\rm Inr}$, and $\dot{\rbold} \triangleq \vbold$ denote the time derivative of $\rbold$ in $\mathcal{F}_{\rm Inr}$.
The dynamics of the $i^{\rm th}$ spacecraft state $\xbold_i \triangleq [\rbold_i^T, \vbold_i^T]^T$ is given by 
\begin{equation}    \label{eq:HFEM_dynamics}
    \dot{\xbold}_i = \fbold(\xbold_i,t) 
    = -\dfrac{\mu}{r_i^3} \rbold_i + 
    \abold_{\rm SH}(\xbold_i,t) +
    \abold_{\rm SRP}(\xbold_i,t) + 
    \sum_{b \in \{\oplus,\odot\}} \abold_{\mathrm{3bd},b}(\xbold_i,t)
    ,
\end{equation}
where $r_i = \|\rbold_i\|_2$, $\mu$ is the gravitational parameter of the Moon, $\abold_{\rm SH}$ is the acceleration due to spherical harmonics terms of the Moon, $\abold_{\rm SRP}$ is the acceleration due to solar radiation pressure (SRP), and $\abold_{\mathrm{3bd},b}$ is the third-body acceleration due to body $b$.
In this work, we consider spherical harmonics terms up to $n_{\max} = 4^{\rm th}$ degree and order, the cannonball model for SRP, and third-body effects due to the Earth, denoted by $(\cdot)_{\oplus}$, and the Sun, denoted by $(\cdot)_{\odot}$.
Expressions for acceleration terms $\abold_{\rm SH}$, $\abold_{\rm SRP}$, and $\abold_{\mathrm{3bd},b}$ 
\rev{can be found in classical texts such as~\cite[chap.~3]{montenbruck2000satellite}.}
% We reproduce here the Canonball model expression for $\abold_{\rm SRP}$, 
% We
% The perturbation acceleration due to SRP is
% \begin{align}
%     \abold_{\rm SRP}(t) &= 
%     P_{\rm \odot} \left( \dfrac{\mathrm{AU}}{\|\rbold(t) - \rbold_{\odot}(t)\|_2} \right)^2 C_r \dfrac{A}{m} \dfrac{\rbold(t) - \rbold_{\odot}(t)}{\|\rbold(t) - \rbold_{\odot}(t)\|_2}
%     ,
% \end{align}
% where $P_{\rm \odot}$ is the Sun's pressure at $1$ astronomical unit $\mathrm{AU} = 149.6\times10^6$ \SI{}{km}, $C_r$ is the radiation pressure coefficient, and $A/m$ is the pressure area-to-mass ratio of the spacecraft.
% }
%are provided in Appendix~\ref{appendix:dynamics_expr}.

We define the concatenated state $\Xbold \in \mathbb{R}^{6M}$ by collecting the states of all $M$ spacecraft,
\begin{equation}
    \Xbold(t) = \begin{bmatrix}
        \xbold_0^T(t) & \cdots & \xbold_{M-1}^T(t)
    \end{bmatrix}^T
    ,
\end{equation}
with a corresponding concatenated dynamics,
\begin{equation}    \label{eq:f_concat}
    \dot{\Xbold}(t) = \fbold_{\rm concat} (\Xbold(t), t)
    = \begin{bmatrix}
        \fbold(\xbold_0(t),t) \\ \vdots \\ \fbold(\xbold_{M-1}(t),t)
    \end{bmatrix}
    .
\end{equation}
Note that while the dynamics are assumed to be identical for all spacecraft, such an assumption may be removed by defining independent dynamics $\fbold_i$ for each spacecraft.
% We also introduce a linear approximation model for the propagation of state perturbation, to be used within convex subproblems of the SCP.
% Under first-order approximation, the additive perturbation on the state at some initial time $t_k$, denoted by $\delta \xbold_i(t_k)$, is linearly mapped to some future time $t \geq t_k$ via the state-transition matrix (STM) $\Phibold_i \in \mathbb{R}^{6 \times 6}$,
% \begin{equation}
%     \delta \xbold_i(t) = \Phibold_i(t,t_k) \delta \xbold_i(t_k),
% \end{equation}
% where $\Phibold(t,t_k)$ is obtained by integrating the matrix-valued initial value problem (IVP),
% \begin{equation}    \label{eq:STM_propagation}
%     \begin{aligned}
%         \dot{\boldsymbol{\Phi}}_i (t,t_k)
%         &= \dfrac{\partial \fbold(\xbold_i(t), t)}{\partial \xbold_i} \boldsymbol{\Phi}(t,t_k)
%         ,
%         \\
%         \boldsymbol{\Phi}_i(t_k,t_k) &= \Ibold_{6}
%         .
%     \end{aligned}
% \end{equation}

Each spacecraft is assumed to have impulsive control capability, resulting in instantaneous velocity changes.
This approximation is justified by the typically small magnitude of station-keeping maneuvers required on NRHO, which are typically on the order of a few \SI{}{cm/s} per maneuver~\cite{Shimane2025}.
Let $\ubold_{i,k} \in \mathbb{R}^3$ denote an impulsive control on the $i^{\rm th}$ spacecraft at time $t_k$. 
The state of the spacecraft at time $t > t_k$ is obtained by
\begin{equation}    \label{eq:integration_single}
    \xbold_i(t) = \xbold_i(t_k) + \Bbold_i \ubold_i
    + \int_{t_k}^t \fbold(\xbold_i(\tau), \tau) \mathrm{d}\tau
    ,
\end{equation}
where $\Bbold_i = \begin{bmatrix} \boldsymbol{0}_{3\times3} & \Ibold_3 \end{bmatrix}^T$.
For conciseness, we also define the concatenated control $\Ubold_k \in \mathbb{R}^{3M}$, such that~\eqref{eq:integration_single} for all $M$ spacecraft is expressed as
\begin{equation}
    \Xbold(t) = \Xbold(t_k) + \Bbold_{\rm concat} \Ubold_k + \int_{t_k}^t \fbold_{\rm concat} (\Xbold(\tau), \tau) \mathrm{d}\tau
    ,
\end{equation}
where $\Ubold_k = \begin{bmatrix}\ubold_{0,k}^T & \cdots & \ubold_{M-1,k}^T \end{bmatrix}^T$ and $\Bbold_{\rm concat} = \operatorname{blkdiag}{(\Bbold_0,\ldots,\Bbold_{M-1})}$, where $\operatorname{blkdiag}(\cdot)$ returns the block-wise diagonal concatenation of the input matrices.

% -------------------------------------------------------------------- %
\subsection{Characteristics along Near-Rectilinear Halo Orbit}
\label{sec:background_NRHO}
\begin{figure}[t]
     \centering
     \begin{subfigure}[b]{0.33\textwidth}
         \centering
         \includegraphics[width=\textwidth]{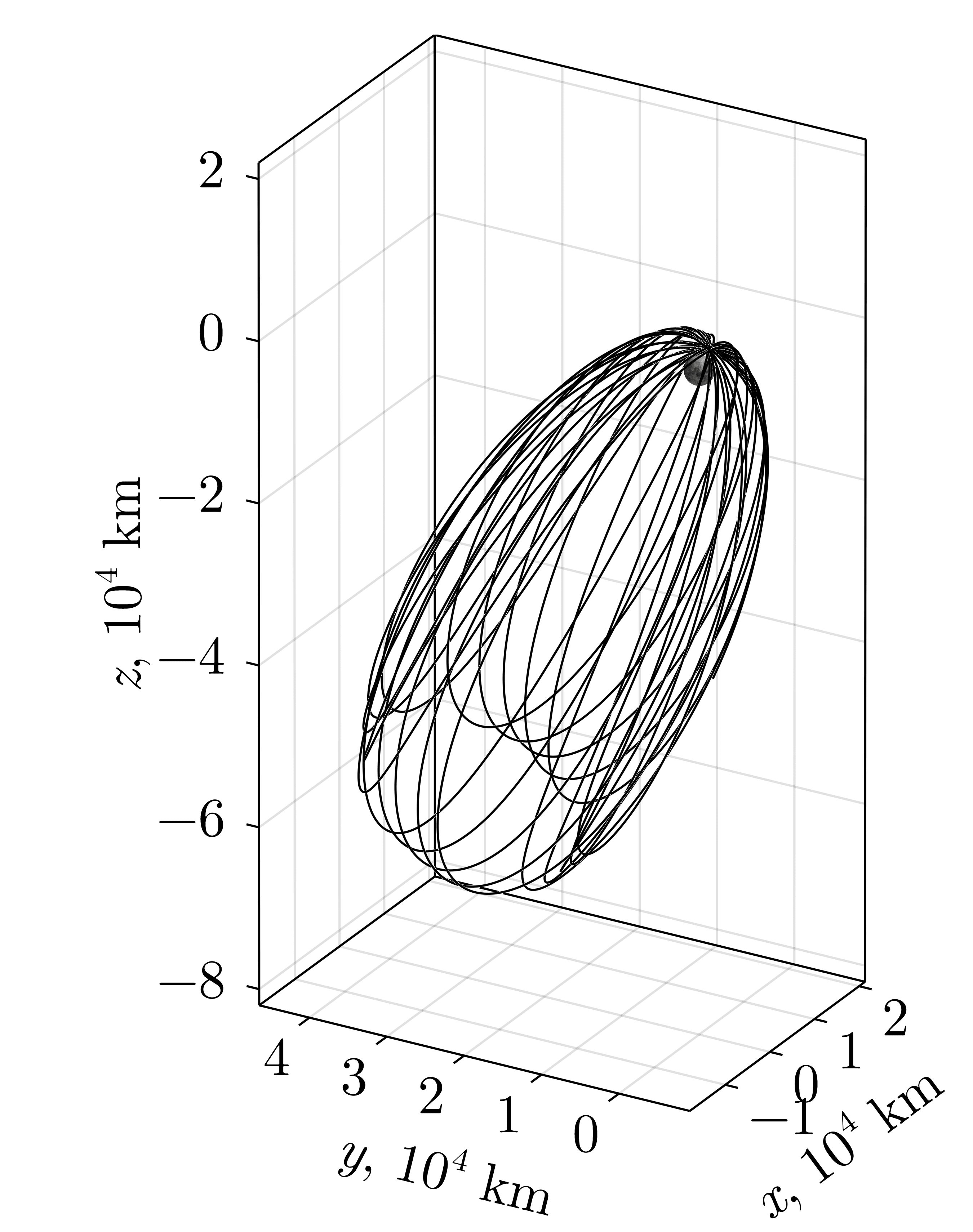}
         \caption{J2000 inertial frame}
         \label{fig:NRHO_traj_k2000}
     \end{subfigure}
     \hfill %\\
     \begin{subfigure}[b]{0.33\textwidth}
         \centering
         \includegraphics[width=\textwidth]{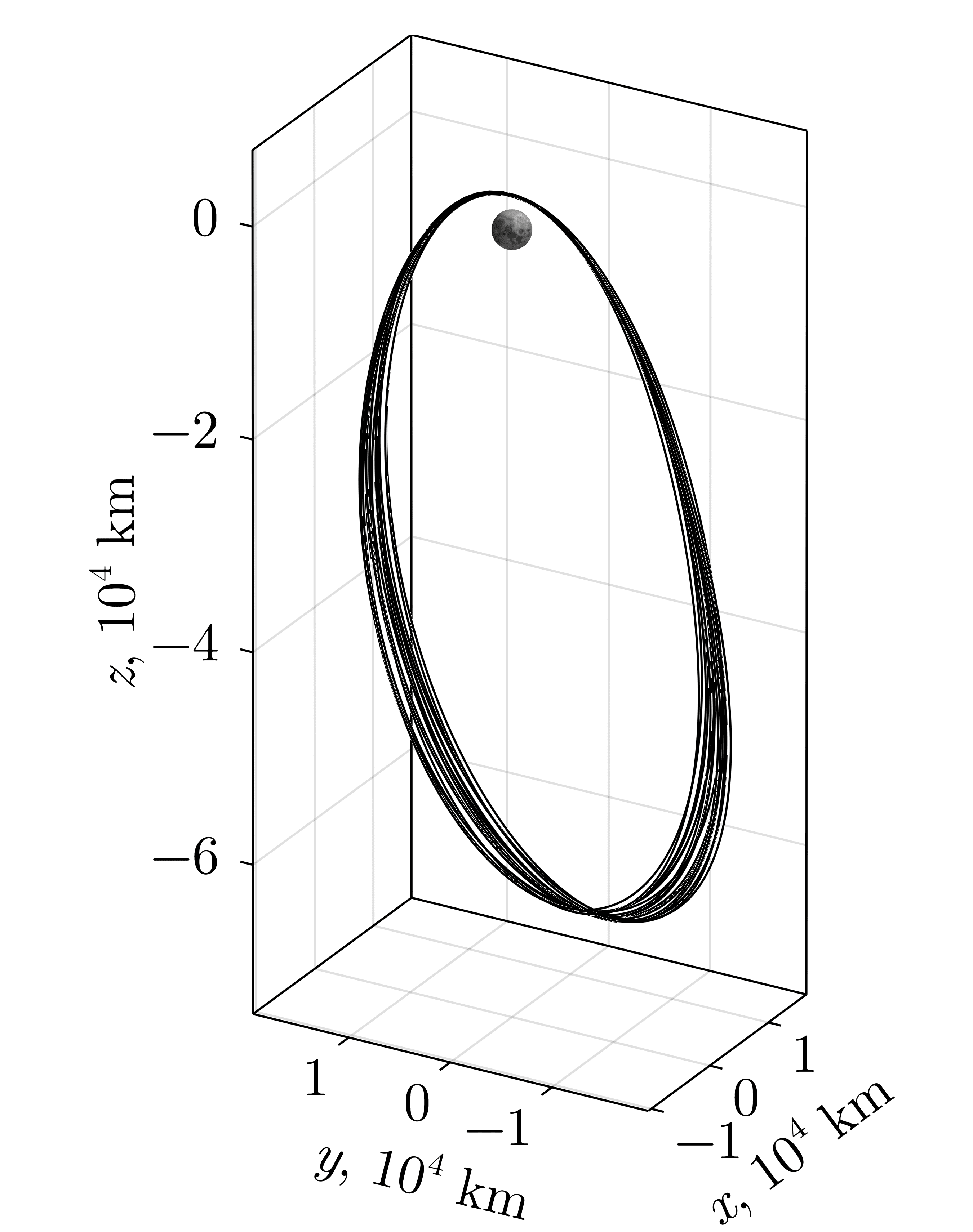}
         \caption{Earth-Moon rotating frame}
         \label{fig:NRHO_traj_EMrot}
     \end{subfigure}
     \hfill %\\
     \begin{subfigure}[b]{0.33\textwidth}
         \centering
         \includegraphics[width=\textwidth]{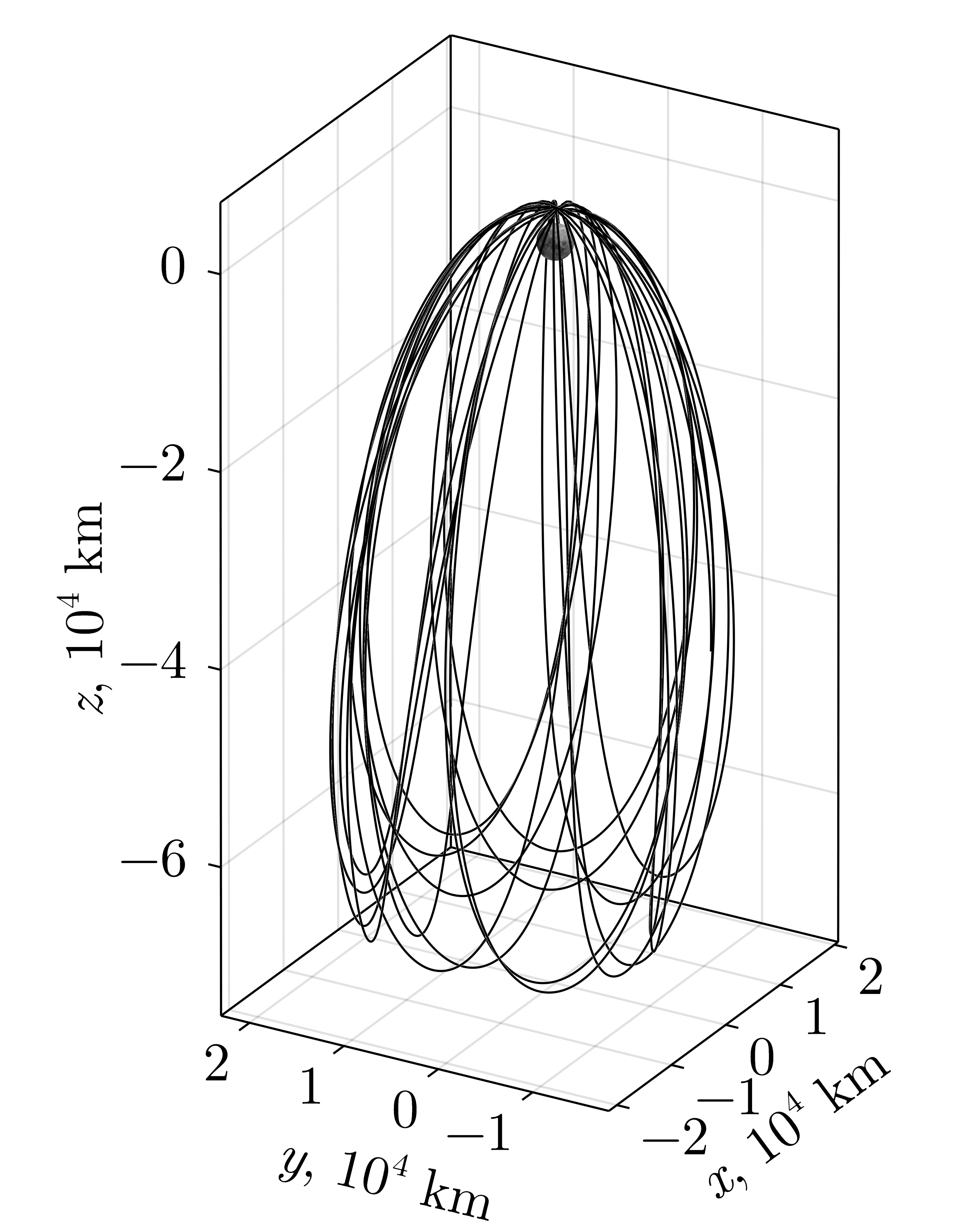}
         \caption{Sun-Moon rotating frame}
         \label{fig:NRHO_traj_SMrot}
     \end{subfigure}
    \caption{Reference Near-Rectilinear Halo Orbit over 20 revolutions in Moon-centered frames}
    \label{fig:NRHO_reference_trajectory}
\end{figure}

We briefly introduce the NRHO and the osculating true anomaly, which is used to parameterize the station-keeping control scheme subsequently.
In this work, we adopt the 15-year NRHO baseline originally generated by NASA~\cite{Lee2019} for the Gateway as the \textit{reference} orbit for all spacecraft within the formation.
%Quantities associated with the reference NRHO are hereafter denoted by $(\cdot)_{\rm ref}$. \todo[]{Need to fix?}
Figure~\ref{fig:NRHO_reference_trajectory} shows the reference NRHO in $\mathcal{F}_{\rm Inr}$ and the Earth-Moon rotating frame, $\mathcal{F}_{\rm EM}$, over the course of 20 revolutions.
In $\mathcal{F}_{\rm EM}$, the NRHO exhibits its well-reported symmetry about the $xz$-plane.
A thorough treatment of the dynamical properties of the NRHO is provided in~\cite{ZimovanSpreen2020} and references therein.

% NRHO range & anomalies

\subsubsection{Osculating True Anomaly}
Although motion along the NRHO is strongly influenced by non-Keplerian effects, the osculating true anomaly with respect to the Moon~\cite[pg.~28]{Vallado2001},
\begin{equation}
    \theta = \operatorname{atan2}{
    \left( h v_r, h^2 / \| \rbold\|_2 - \mu \right)
    }
    , \,
    h = \| \rbold \times \vbold \|_2
    , \,
    v_r = \dfrac{\rbold \cdot \vbold}{r}
    ,
\end{equation}
provides a convenient angular parametrization of spacecraft position along the NRHO, which is useful for planning operations, such as specifying locations where the spacecraft maneuvers.
For example, both CAPSTONE and the planned Gateway operations schedule maneuvers at $\theta = 200^{\circ}$.

\begin{figure}[t]
     \centering
     \begin{subfigure}[b]{0.49\textwidth}
         \centering
         \includegraphics[width=\textwidth]{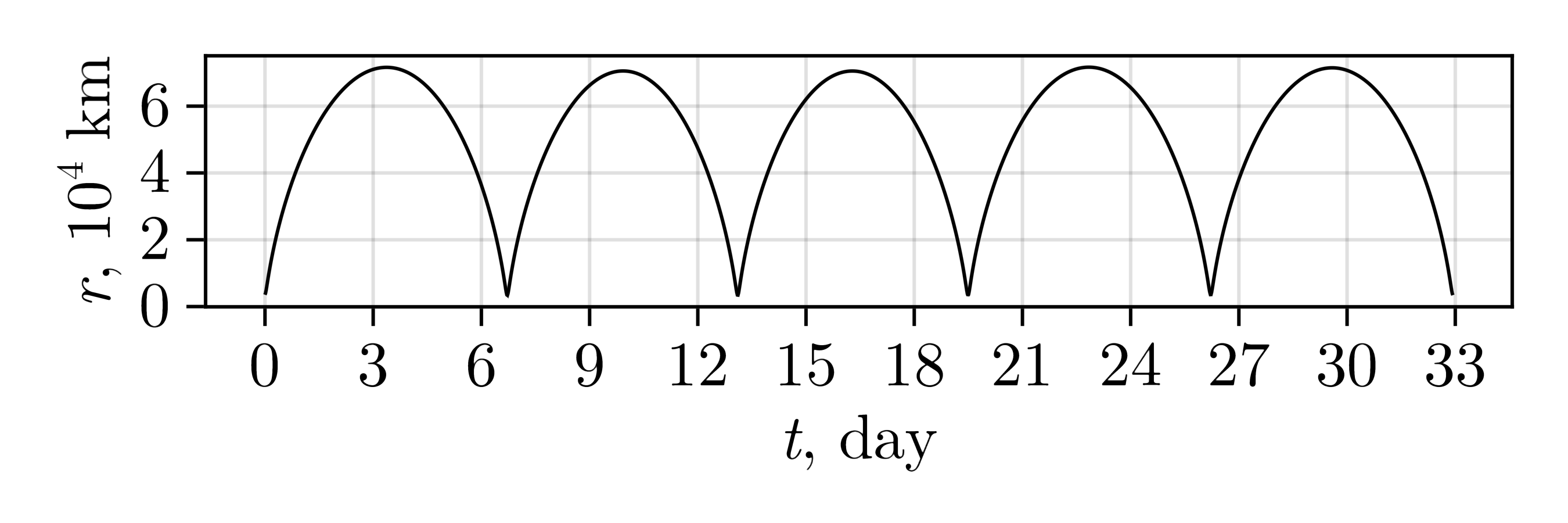}
         \caption{Range with respect to the Moon}
         \label{fig:NRHO_range}
     \end{subfigure}
     \hfill %\\
     \begin{subfigure}[b]{0.49\textwidth}
         \centering
         \includegraphics[width=\textwidth]{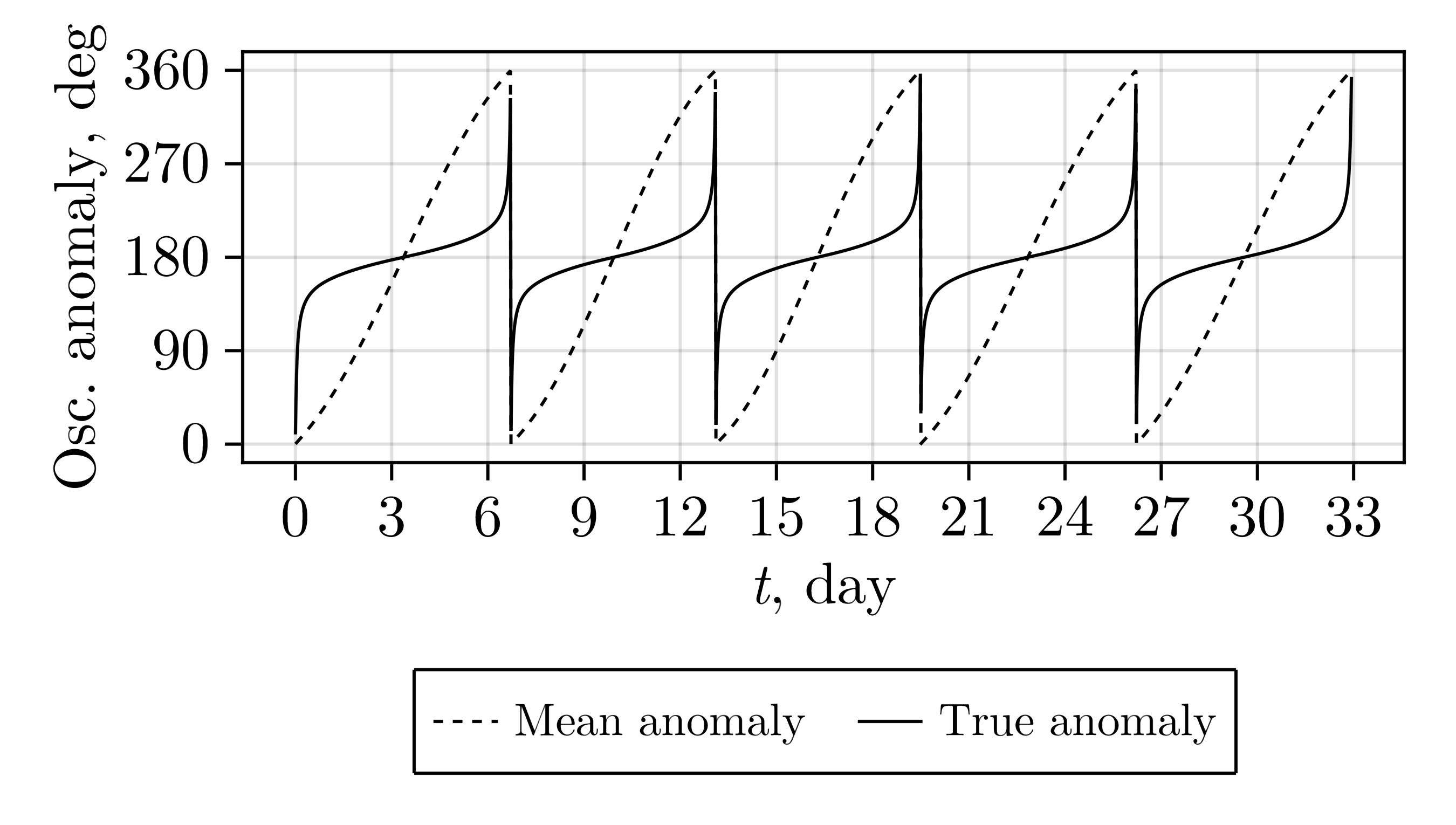}
         \caption{Mean and true anomaly with respect to the Moon}
         \label{fig:NRHO_anomalies}
     \end{subfigure}
    \caption{Range and osculating anomaly history along reference Near-Rectilinear Halo Orbit}
    \label{fig:NRHO_anatomy}
\end{figure}

Figure~\ref{fig:NRHO_anatomy} shows the range, mean anomaly, and true anomaly histories over 5 revolutions of the reference NRHO.
The true anomaly history illustrates the quick perilune transit at $\theta \sim 0^{\circ}$, and the much longer dwell near apolune for $160^{\circ} \leq \theta \leq 200^{\circ}$.
Because of the significant non-Keplerian effects along the NRHO trajectory, the mean anomaly does not evolve linearly with time.

% NRHO range & anomalies

% -------------------------------------------------------------------- %
\subsubsection{Relative Motions}
\begin{figure}
    \centering 
    \includegraphics[width=0.38\linewidth]{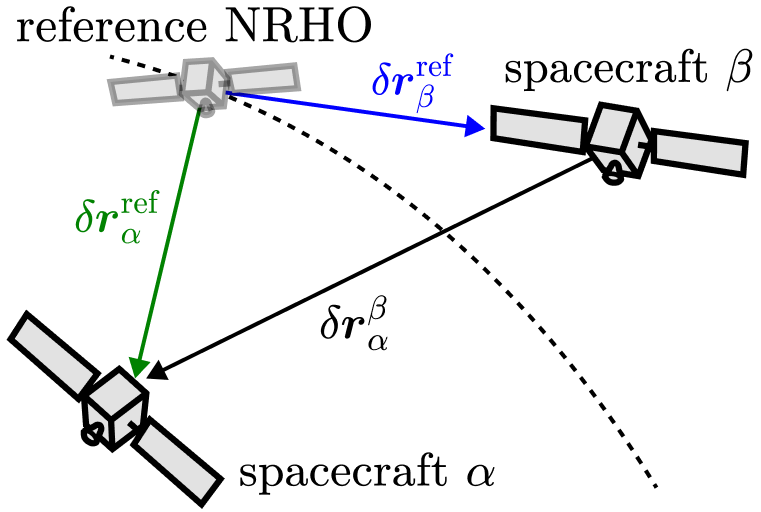}
    \caption{Illustration of reference-relative and formation-relative position vectors}
    \label{fig:RelativeMotionDef}
\end{figure}

Within the context of this work, \rev{we introduce two definitions of relative state, $\delta \xbold$, namely 1) the \textit{reference-relative} state, $\delta \xbold_{i}^{\mathrm{ref}}$, i.e., $\delta \xbold$ of $i^{\rm th}$ spacecraft with respect to the reference NRHO,
\begin{equation}
    \delta \xbold_{i}^{\mathrm{ref}}(t) 
    = \begin{bmatrix}
        \delta \rbold_{i}^{\mathrm{ref}}(t) \\ \delta \vbold_{i}^{\mathrm{ref}}(t)
    \end{bmatrix}
    = \xbold_i(t) - \xbold_{\rm ref}(t)
    ,
\end{equation}
and 2) the \textit{formation-relative} state $\delta \xbold_{\alpha}^{\beta}$, i.e., $\delta \xbold$ of the $\alpha^{\rm th}$ spacecraft relative to the $\beta^{\rm th}$ spacecraft,
\begin{equation}
    \delta \xbold_{\alpha}^{\beta}(t) 
    = \begin{bmatrix}
        \delta \rbold_{\alpha}^{\beta}(t) \\ \delta \vbold_{\alpha}^{\beta}(t)
    \end{bmatrix}
    = \xbold_{\alpha} (t) - \xbold_{\beta}(t)
    .
\end{equation}
}
Figure~\ref{fig:RelativeMotionDef} illustrates $\delta \xbold_{i}^{\mathrm{ref}}$ and $\delta \xbold_{\alpha}^{\beta}$.
% we denote the \textit{reference-relative} state, $\delta \xbold_{i}^{\mathrm{ref}}$, as the state of the $i^{\rm th}$ spacecraft, $\delta \xbold_{i}^{\mathrm{ref}}$, relative to the reference NRHO,
% \begin{equation}
%     \delta \xbold_{i}^{\mathrm{ref}}(t) 
%     = \begin{bmatrix}
%         \delta \rbold_{i}^{\mathrm{ref}}(t) \\ \delta \vbold_{i}^{\mathrm{ref}}(t)
%     \end{bmatrix}
%     = \xbold_i(t) - \xbold_{\rm ref}(t)
%     .
% \end{equation}
The reference-relative state~$\delta \xbold_{i}^{\mathrm{ref}}$ is critical for defining ``absolute'' station-keeping~\cite{Shimane2025,Foss2025-sm} with respect to the reference NRHO, and characterizing the approximate\footnote{The LTC is only approximate due to the quasi-periodic nature of NRHO in the HFEM.} LTC~\cite{Takubo2026-bo}, \rev{allowing for assessment of} the dynamical behavior of the formation.
A detailed discussion of the dynamics of $\delta \xbold_{i}^{\mathrm{ref}}$ can be found in~\cite{Takubo2026-bo}.

% We also define the \textit{formation-relative} state $\delta \xbold_{\alpha}^{\beta}$ between any two spacecraft $\alpha$ and $\beta$ as the state of the $\alpha^{\rm th}$ spacecraft relative to the $\beta^{\rm th}$ spacecraft,
% \begin{equation}
%     \delta \xbold_{\alpha}^{\beta}(t) 
%     = \begin{bmatrix}
%         \delta \rbold_{\alpha}^{\beta}(t) \\ \delta \vbold_{\alpha}^{\beta}(t)
%     \end{bmatrix}
%     = \xbold_{\alpha} (t) - \xbold_{\beta}(t)
%     .
% \end{equation}
The formation-relative state~$\delta \xbold_{\alpha}^{\beta}$ is used to enforce desired properties of the inter-spacecraft configuration within the formation, such as bounded separations or relative Sun phase angle.
These requirements fall within the scope of ``relative'' station-keeping.

% In this work, we devise a single control scheme that simultaneously achieves absolute and relative station-keeping of all spacecraft in the formation; this is achieved by defining the problem in absolute dynamics and incorporating path constraints pertaining to both reference-relative motion and formation-relative motion for all combinations of $i$, $\beta$.
% Our approach is contrasted to the LTC-based guidance~\cite{Takubo2026-bo} or the modular guidance with CBF-based safety~\cite{Foss2025-sm}, where absolute and relative station-keeping are independently conducted.
% While we enforce desired properties of the formation directly in the form of constraints, the LTC-based guidance enables passive safety guarantees, i.e., a formation-relative property, by defining the relative station-keeping problem in reference-relative dynamics and leveraging the QPT structure~\cite{Takubo2026-bo}.
% In~\cite{Foss2025-sm}, passive safety is enforced through a CBF that defines a stochastic keep-out zone, while capitalizing on the rarity of collision-risk events.

There are advantages and disadvantages to using either separate absolute and relative station-keeping schemes, as in~\cite{Foss2025-sm}, or a unified station-keeping formulation, as adopted in this work. Decoupling the problem into absolute and relative components enables a modular guidance architecture. For example, a flight-proven absolute station-keeping strategy, such as the $x$-axis crossing control scheme deployed on CAPSTONE~\cite{Cheetham2022}, may be combined with a separate relative station-keeping controller. Furthermore, developing the relative station-keeping problem in reference-relative dynamics provides analytical insight into bounded motion, improving understanding of the formation’s dynamical behavior.
However, solving absolute and relative guidance problems separately is inherently suboptimal and typically leads to higher propellant usage. In contrast, a unified station-keeping formulation can simultaneously regulate absolute and relative motion, generally resulting in lower propellant consumption. 
Finally, in either approach, formation-relative path constraints can be introduced to enforce additional behaviors that may be difficult to capture through the QPT structure alone.

% There are advantages and disadvantages to either having a single station-keeping scheme, as is done in this work, or separate absolute and relative control schemes, as is done in~\cite{Foss2025-sm}. 
% In terms of performance, defining a single scheme will generally result in lower cumulative and recursive propellant cost compared to defining separate guidance problems, which is inherently suboptimal.
% %
% In return, by decoupling into absolute and relative station-keeping problems, a modularized guidance design can be achieved.
% In effect, a flight-proven absolute station-keeping scheme, such as the $x$-axis crossing control scheme deployed on CAPSTONE~\cite{Cheetham2022}, may be complemented with the relative station-keeping scheme.
% Furthermore, developing the relative station-keeping problem in reference-relative dynamics comes with analytical insights into the bounded motion, which enhances our understanding of the formation's dynamical behavior.
% %
% Finally, with both approaches, formation-relative path constraints can be introduced to enforce additional behaviors that may be hard to define through the QPT structure alone.

% -------------------------------------------------------------------- %
\subsection{Uncertainties in Station-Keeping Problem}
\label{sec:background_uncertainties}
Several sources of error \rev{affect} both the absolute and relative motion of the spacecraft within the formation.
Following previous works on station-keeping along the NRHO~\cite{Davis2022,Shimane2025-by}, we incorporate initial orbit insertion error, dynamics modeling error in the SRP term, navigation error, and control execution error.
Here, we omit errors due to momentum wheel desaturation maneuvers occurring around perilune considered for the Gateway~\cite{Davis2022}.
This choice, also used in~\cite{Foss2025-sm}, is based on the assumption that the spacecraft in the context of this work are smaller uncrewed assets, for which desaturation maneuvers need not happen around perilune, where the dynamics are most sensitive to errors.
\rev{
For example, placing desaturation maneuvers once a revolution immediately before a station-keeping maneuver would result in minimal effect on its position, and the induced disturbance on the velocity could be absorbed by the station-keeping maneuver.
If, on the other hand, desaturation maneuvers around perilune are operationally unavoidable, the resulting impulse can significantly disturb the trajectory and violate any desirable path constraints; while an immediate station-keeping maneuver is also possible here, a correction around perilune is risky for the same dynamics sensitivity reasons, and is preferably avoided.}

\subsubsection{Initial Orbit Insertion Error}
We consider variation on the initial ``starting state'' of a given spacecraft to be perturbed by errors due to an inevitably imperfect execution of the orbit insertion maneuver.
At the beginning of the mission $t_0$, the initial state for the $i^{\rm th}$ spacecraft, $\xbold_{i}(t_0)$, is defined by a desired reference-relative state $\delta \xbold_{i,\rm des}^{\rm ref}$, corrupted by insertion error,
\begin{equation}
    \xbold_i(t_0) =
    \xbold_{\rm ref}(t_0) + \delta \xbold_{i,\rm des}^{\rm ref} + \delta \xbold_{i,\rm init},
    \quad 
    \delta \xbold_{i,\rm init} = 
    \begin{bmatrix}
        \delta \rbold_{i,\rm init} \\ \delta \vbold_{i,\rm init}
    \end{bmatrix}
    ,\quad
    \delta \rbold_{i,\rm init} \sim \mathcal{N}(\boldsymbol{0}_{3 \times 1}, \boldsymbol{\sigma}_{\rbold,\rm init}^2)
    ,\quad
    \delta \vbold_{i,\rm init} \sim \mathcal{N}(\boldsymbol{0}_{3 \times 1}, \boldsymbol{\sigma}_{\vbold,\rm init}^2)
    ,
\end{equation}
where $\boldsymbol{\sigma}_{\rbold,\rm init} \in \mathbb{R}^3$ and $\boldsymbol{\sigma}_{\vbold,\rm init} \in \mathbb{R}^3$ are the initial position and velocity insertion standard deviations, respectively.

\subsubsection{Dynamics Error}
The SRP term causes the most significant modeling error within the HFEM~\eqref{eq:HFEM_dynamics}.
Following~\cite{Davis2022}, we consider Gaussian variations on SRP parameters realized between each control maneuver,
\begin{subequations}
\begin{align}
    A/m &= {A_0/m_0} (1 + \delta A/m), \quad
    \delta A/m \sim \mathcal{N}(0, \sigma_{A/m}^2)
    \\
    C_r &= {C_{r,0}} (1 + \delta C_r), \quad
    \delta C_r \sim \mathcal{N}(0, \sigma_{C_r}^2)
    ,
\end{align}
\end{subequations}
where ${A_0/m_0} \in \mathbb{R}$ and ${C_{r,0}} \in \mathbb{R}$ are average values, and $\sigma_{A/m} \in \mathbb{R}$ and $\sigma_{C_r} \in \mathbb{R}$ are relative standard deviations.

\subsubsection{Navigation Error}
Navigation error is inherent to any space flight problem due to the existence of both modeling error in the dynamics and measurement error.
Let $\hat{\Xbold}$ denote the estimate of $\Xbold$.
In this work, we model navigation error in terms of a normal distribution,
\begin{equation}
    \hat{\Xbold} = \begin{bmatrix}
        \hat{\xbold}_0 \\ \vdots \\ \hat{\xbold}_{M-1}
    \end{bmatrix}
    = \begin{bmatrix}
        {\xbold}_0 + \delta \xbold_{i,\rm nav} 
        \\ \vdots \\
        {\xbold}_{M-1} + \delta \xbold_{i,\rm nav}
    \end{bmatrix}
    ,\quad
    \delta \xbold_{i,\rm nav} = 
    \begin{bmatrix}
        \delta \rbold_{i,\rm nav} \\ \delta \vbold_{i,\rm nav}
    \end{bmatrix}
    ,\quad
    \delta \rbold_{i,\rm nav} \sim \mathcal{N}(\boldsymbol{0}_{3 \times 1}, \boldsymbol{\sigma}_{\rbold,\rm nav}^2)
    ,\quad
    \delta \vbold_{i,\rm nav} \sim \mathcal{N}(\boldsymbol{0}_{3 \times 1}, \boldsymbol{\sigma}_{\vbold,\rm nav}^2)
    ,
\end{equation}
where $\boldsymbol{\sigma}_{\rbold,\rm nav} \in \mathbb{R}^3$ is the position standard deviation and $\boldsymbol{\sigma}_{\vbold,\rm nav} \in \mathbb{R}^3$ is the velocity standard deviation of the navigation error, respectively.

\subsubsection{Control Execution Error}
Control execution error exists due to the discrepancy between the control model within the control problem and the actual hardware on the spacecraft.
Following common practice~\cite{Davis2022,Shimane2025-by}, the control execution error is modeled based on the Gates model~\cite{Gates1963}, which combines absolute magnitude error, a relative magnitude error, and a direction error.
Let $\ubold_{i,k} \in \mathbb{R}^3$ denote the \textit{commanded} impulse by the controller for spacecraft $i$ at $t_k$.
The actual executed control, $\tilde{\ubold}_{i,k}$, is given by
\begin{equation}
    \tilde{\ubold}_{i,k}
    = \Tbold(\delta \varphi) \left( 
        \ubold_{i,k} + \delta \ubold_{\rm abs} + \delta \ubold_{\rm rel}
    \right)
    ,
\end{equation}
where $\delta \ubold_{\rm abs}$ is the absolute magnitude error with standard deviation $\sigma_{u_{\rm abs}}$,
\begin{equation}
    \delta \ubold_{\rm abs} = \delta u_{\rm abs} \dfrac{\ubold_{i,k}}{\| \ubold_{i,k} \|_2}
    ,\quad \delta u_{\rm abs} \sim \mathcal{N}(0,\sigma_{u_{\rm abs}}^2)
    ,
\end{equation} $\delta \ubold_{\rm rel}$ is the relative magnitude error with standard deviation $\sigma_{u_{\rm rel}}$, 
\begin{equation}
    \delta \ubold_{\rm rel} = \delta u_{\rm rel} \ubold_{i,k}
    ,\quad \delta u_{\rm rel} \sim \mathcal{N}(0,\sigma_{u_{\rm rel}}^2)
    ,
\end{equation}and $\Tbold(\delta \varphi)$ is the direction error with standard deviation $\sigma_{\delta \varphi}$,
\begin{equation}
    \Tbold(\delta \varphi)
    =
    \cos(\delta \varphi) \Ibold_3 + \sin(\delta \varphi)
    \boldsymbol{i}^{\times} +
    [1 - \cos(\delta \varphi)] \boldsymbol{i} \boldsymbol{i}^T
    ,\quad
    \delta \varphi \sim \mathcal{N}(0, \sigma_{\delta \varphi}^2)
    ,
\end{equation}
where $\boldsymbol{i} \in \mathbb{R}^{3}$ is a random unit vector, and $\boldsymbol{i}^{\times} \in \mathbb{R}^{3 \times 3}$ is the skew-symmetric form of $\boldsymbol{i}$.

% % -------------------------------------------------------------------- %
% \subsection{Station-Keeping Model Predictive Control for Single Vehicle}
% \label{sec:background_SKMPC}

% \begin{itemize}
%     \item Paragraph about recursive full-state tracking achieved via MPC with targeting constraint
%     \item Talk about reference trajectory
% \end{itemize}

% ===================================================================== %
\section{Multi-Vehicle Optimal Control Problem for Formation Flight}
\label{sec:mvocp}
This Section develops the MVOCP.
We first introduce the prediction horizon of the OCP.
We then define path constraints of interest in~\ref{sec:MVOCP_pathconstraints}.
These constraints are incorporated into the full OCP formulation, a nonconvex optimization problem with CT path constraints, developed in~\ref{sec:MVOCP_mvocp}.

% -------------------------------------------------------------------- %
\subsection{Prediction Horizon}
\label{sec:prediction_horizon}
A key requirement for operating spacecraft in NRHO is to maintain the orbit with a low station-keeping cadence; both the CAPSTONE~\cite{Cheetham2022} maneuvers and Gateway~\cite{Davis2022} plan maneuvers once per revolution at $\theta_{\rm ref} = 200^{\circ}$.
In the context of formation flight, Foss et al.~\cite{Foss2025-sm} considers two maneuvers per revolution, at $\theta_{\rm ref} = 160^{\circ}$ and $200^{\circ}$.
The maneuver times along the NRHO are chosen to be consistent with previous studies~\cite{Foss2025-sm} and operationally relevant: the two maneuver times mark the beginning and end of the apolune pass, where the dynamics are less sensitive and payload operations are expected to occur.
This cadence still leaves over $\sim\!4$ days between the two maneuvers around apolune, and allows for higher controllability of the formation introduced by the additional maneuver per revolution.

We consider a prediction horizon consisting of $N_{\rm rev}$ revolutions, with two maneuvers per revolution, placed at times $t$ when $\theta_{\rm ref}(t) = 160^{\circ}$ and $200^{\circ}$.
Since the MPC is re-instantiated at each maneuver, the initial time of the prediction horizon alternates between $t$ $\st$ $\theta_{\rm ref} = 160^{\circ}$ and $200^{\circ}$.
\red{Let $(\cdot)_{j|k}$ denote the quantity predicted at $j \in \mathcal{J} = \{ 0,\ldots,N-1 \}$ time increments ahead of the current state estimate $\xbold(t_k)$ at time $t_k$, $\st$ $N = 1 + 2N_{\rm rev}$.
The times along the prediction horizon are denoted by $t_{0|k},\ldots,t_{N-1|k}$.
Note that either $\theta_{\rm ref}(t_{0|k}) = \theta_{\rm ref}(t_{2|k}) = \ldots = \theta_{\rm ref}(t_{N-1|k}) = 160^{\circ}$ and $\theta_{\rm ref}(t_{1|k}) = \theta_{\rm ref}(t_{3|k}) = \ldots = \theta_{\rm ref}(t_{N-2|k}) = 200^{\circ}$ or vice-versa.}

% -------------------------------------------------------------------- %
\subsection{Path Constraints}
\label{sec:MVOCP_pathconstraints}
We consider two sets of path constraints that are relevant for a formation flight context between multiple spacecraft.
The first set consists of lower and upper bounds on the relative separation between the spacecraft; the lower bound ensures the formation is safe from collision, while the upper bound is motivated by operations, ensuring, for example, that the data rate between the spacecraft is above a prescribed level.
The second set of constraints ensures the relative Sun phase angle of each spacecraft is bounded; in this way, each spacecraft is optically observable by the other.

We enforce the minimum separation constraint throughout the entire prediction due to its direct \rev{effect} on mission safety.
\rev{Meanwhile,} the maximum separation constraint and the bounds on the relative Sun phase angle are enforced continuously during time windows where the true anomaly along the reference NRHO, $\theta_{\rm ref}(t)$, is between $160^{\circ}$ and $200^{\circ}$.
We hereafter refer to the trajectory portion at $t$ where $160^{\circ} \leq \theta_{\rm ref}(t) \leq 200^{\circ}$ as the \textit{apolune segment}, and the trajectory portion at $t$ where $-160^{\circ} \leq \theta_{\rm ref}(t) \leq 160^{\circ}$ as the \textit{perilune segment}.
\rev{Reasons behind omitting the upper bound on the maximum separation and the Sun phase angle bounds during the perilune segment are twofold.
First, continuously enforcing these conditions through perilune would potentially require an additional control maneuver around perilune,} where the dynamics are most sensitive and erroneous maneuvers may lead to severe deviation of the intended motion.
\rev{Second,} it is operationally unlikely to require communication between each spacecraft continuously during perilune passage anyway, as that would equate to requiring rapid slewing maneuvers.
% This decision is made based on the assumption that during the rapid perilune transition from $t$ where $\theta_{\rm ref}(t) = 200^{\circ}$ to $\theta_{\rm ref}(t) = 160^{\circ}$, it is 1) operationally unlikely to require communication between each spacecraft continuously, especially as that would equate to requiring rapid slewing maneuvers, and 2) continuously enforcing would potentially require a third control maneuver to occur near perilune, where the dynamics is most sensitive and erroneous maneuvers may lead to severe deviation of the intended motion.

\subsubsection{Bounded Separation}
The bounded separation constraints ensure the distance between any two spacecraft in the formation is bounded from below and above by $[\Delta r_{\min}, \Delta r_{\max}]$, where $0 < \Delta r_{\min} < \delta r_{\alpha}^{\beta} < \Delta r_{\max}$ for all $(\alpha,\beta) \in \mathcal{C}$, where $\mathcal{C}$ is the set of combinations of pairs out of $M$ spacecraft.
The lower bound ensures any two spacecraft maintain a safe distance from one another and avoid collision.
The upper bound ensures any two spacecraft maintain a distance that is suitable for inter-spacecraft communication and/or observation.
The bounded separation constraints are
\begin{subequations} \label{eq:MVOCP_boundsep}
\begin{align}
    g_{\Delta r_{\min},\alpha\beta} &:= 
    \Delta r_{\min} - 
    \| \rbold_{\alpha}(t) - \rbold_{\beta}(t) \|_2 \leq 0
        \quad \forall
        (\alpha,\beta) \in \mathcal{C},\,
        t\in [t_{0|k},t_{N-1|k}]
    ,
    \label{eq:MVOCP_minsep}
    \\
    g_{\Delta r_{\max},\alpha\beta} &:= 
    \| \rbold_{\alpha}(t) - \rbold_{\beta}(t) \|_2 - \Delta r_{\max}\leq 0
        \quad \forall 
        (\alpha,\beta) \in \mathcal{C},\,
        % t\in[t_{0|k},t_{N-1|k}]
        t\in \{ t_{0|k} \leq t \leq t_{N-1|k}
        \st
        160^{\circ} \leq \theta_{\rm ref}(t) \leq 200^{\circ}
        \}
    .
    \label{eq:MVOCP_maxsep}
\end{align}
\end{subequations}
Note that while constraint~\eqref{eq:MVOCP_minsep} is enforced continuously over the entire prediction horizon,~\eqref{eq:MVOCP_maxsep} is enforced continuously over apolune segments.

\subsubsection{Relative Sun Phase Angle}
\begin{figure}[t]
     \centering
     \begin{subfigure}[b]{0.46\textwidth}
         \centering
         \includegraphics[width=0.9\textwidth]{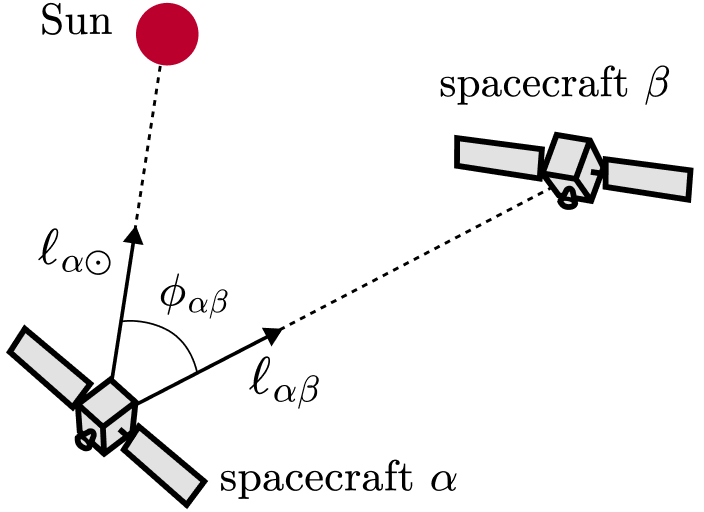}
        \caption{$\phi_{\alpha\beta}$, formed by spacecraft $\alpha$ relative to spacecraft $\beta$}
         \label{fig:RelativeSunPhaseAngle_ij}
     \end{subfigure}
     \hfill %\\
     \begin{subfigure}[b]{0.46\textwidth}
         \centering
         \includegraphics[width=0.9\textwidth]{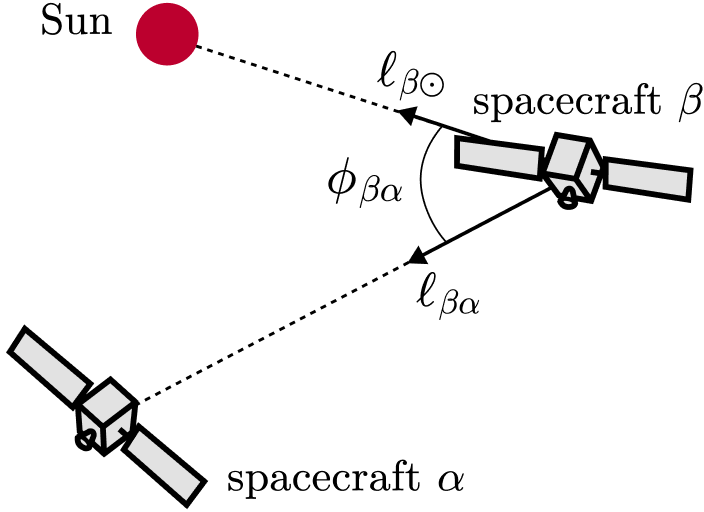}
        \caption{$\phi_{\beta \alpha}$, formed by spacecraft $\beta$ relative to spacecraft $\alpha$}
         \label{fig:RelativeSunPhaseAngle_ji}
     \end{subfigure}
    \caption{Relative Sun phase angles $\phi_{\alpha\beta}$ and $\phi_{\beta \alpha}$}
    \label{fig:RelativeSunPhaseAngle}
\end{figure}

We introduce a constraint on the phase angle of a spacecraft as observed by another spacecraft, \rev{ensuring adequate illumination condition between one another. Such a constraint may be relevant for formations utilizing optical systems for imaging or communication, and serve as an example of a path constraint that is not governed by the dominant dynamics mode along the NRHO.}
We define the \textit{relative Sun phase angle}, $\phi_{\alpha\beta}$, as the phase angle of spacecraft $\alpha$ as observed by spacecraft $\beta$,
\begin{equation}
    \phi_{\alpha\beta}(t) = \cos^{-1} \left( \ellbold_{\alpha\beta}(t) \cdot \ellbold_{\alpha\odot}(t) \right)
    ,
\end{equation}
where $\ellbold_{\alpha\beta}$ and $\ellbold_{\alpha \odot}$ are the line-of-sight vectors from spacecraft $\alpha$ to $\beta$ and from spacecraft $\alpha$ to the Sun, respectively,
\begin{equation}
    \ellbold_{\alpha\beta}(t) = \dfrac{\rbold_{\beta}(t) - \rbold_{\alpha}(t)}{\| \rbold_{\beta}(t) - \rbold_{\alpha}(t) \|_2},
    \quad
    \ellbold_{\alpha \odot}(t) = \dfrac{\rbold_{\odot}(t) - \rbold_{\alpha}(t)}{\| \rbold_{\odot}(t) - \rbold_{\alpha}(t) \|_2}
    .
\end{equation}
Figure~\ref{fig:RelativeSunPhaseAngle} illustrates $\phi_{\alpha\beta}$ as well as the reciprocal $\phi_{\beta \alpha}$, the latter being the phase angle of spacecraft $\beta$ as observed by spacecraft $\alpha$.
Note that since the distance between the two spacecraft is much smaller than the distance from either spacecraft to the Sun, $\phi_{\alpha \beta} + \phi_{\beta \alpha} \approx 180^{\circ}$.

% phase angle figure

To prevent the interference of the Sun with the inter-spacecraft link, we ensure that $\phi_{\alpha\beta}$ for each $(\alpha,\beta)$ pair is bounded from below and above by $[\phi_{\min}, \phi_{\max}]$, where $0 < \phi_{\min} < \phi_{\max} < \pi$.
We thus enforce
\begin{subequations}   \label{eq:constraint_boundphaseangle}
\begin{align}
    g_{\mathrm{\phi_{\min}},\alpha \beta} := 
    \phi_{\min} - \phi_{\alpha\beta}(t) &\leq 0
        \quad \forall
        (\alpha,\beta) \in \mathcal{C},\,
        t\in \{ t_{0|k} \leq t \leq t_{N-1|k}
        \st
        160^{\circ} \leq \theta_{\rm ref}(t) \leq 200^{\circ}
        \}
    ,   \label{eq:constraint_minphaseangle}
    \\
    g_{\mathrm{\phi_{\max}},\alpha \beta} := 
     \phi_{\alpha\beta}(t) - \phi_{\max} &\leq 0
        \quad \forall
        (\alpha,\beta) \in \mathcal{C},\,
        t\in \{ t_{0|k} \leq t \leq t_{N-1|k}
        \st
        160^{\circ} \leq \theta_{\rm ref}(t) \leq 200^{\circ}
        \}
    .
    \label{eq:constraint_maxphaseangle}
\end{align}
\end{subequations}
For the relative phase angle constraints, both the lower bound~\eqref{eq:constraint_minphaseangle} and upper bound~\eqref{eq:constraint_maxphaseangle} are enforced continuously over apolune segments.

% -------------------------------------------------------------------- %
\subsection{Problem Formulation}
\label{sec:MVOCP_mvocp}
We define the MVOCP, which seeks to find the optimal controls for maintaining the formation in the vicinity of the reference NRHO while enforcing the path constraints from Section~\ref{sec:MVOCP_pathconstraints} within the finite prediction horizon, using the minimum total propellant consumption.
%\subsubsection{Problem Formulation}
% Let $(\cdot)_{j|k}$ denote the quantity predicted at $j \in \mathcal{J} = \{ 0,\ldots,N-1 \}$ time increments ahead of the current state estimate $\xbold(t_k)$ at time $t_k$.
Let $\ubold_{0|k},\ldots,\ubold_{N-1|k}$ denote the control vectors at each control node.
Based on our choice of prediction horizon with $N_{\rm rev}$ revolutions and 2 maneuvers per revolution, $N = 2N_{\rm rev} + 1$. 
The finite-horizon MVOCP is 
\begin{subequations}    \label{eq:MVOCP}
\begin{align}
    \minimize{
        \substack{\Xbold_{0|k},\ldots,\Xbold_{N-1|k}\\\Ubold_{0|k},\ldots,\Ubold_{N-1|k}}
    }
    \quad& \sum_{i=0}^{M-1} \sum_{j=0}^{N-1} \| \ubold_{i,j|k} \|_2
        \label{eq:MVOCP_obj}
    \\\st\quad&
    \Xbold_{j+1|k} = \Xbold_{j|k} + \Bbold_{\rm concat} \Ubold_{j|k} 
    + \int_{t_{j|k}}^{t_{j+1|k}} \fbold_{\rm concat} (\Xbold_{j|k}(t),t) \mathrm{d}t
        \quad \forall j \in \mathcal{J} \setminus \{N-1\} 
    \label{eq:MVOCP_dynamics}
    \\&
    \Xbold_{0|k} = \hat{\Xbold}(t_k)
    \label{eq:MVOCP_initialconditions}
    \\&
    \| \rbold_{i,N-1|k} - \rbold_{\mathrm{ref},i}(t_{N-1|k}) \|_2 \leq \epsilon_r
        \quad \forall i \in \mathcal{I} 
    \label{eq:MVOCP_finalset_r}
    \\&
    \| \vbold_{i,N-1|k} + \ubold_{i,N-1|k} - \vbold_{\mathrm{ref},i}(t_{N-1|k}) \|_2 \leq \epsilon_v
        \quad \forall i \in \mathcal{I} 
    \label{eq:MVOCP_finalset_v}
    \\&
    \text{eqn.~\eqref{eq:MVOCP_boundsep},~\eqref{eq:constraint_boundphaseangle}}
    \nonumber
\end{align}
\end{subequations}
The objective~\eqref{eq:MVOCP_obj} aims to minimize the total propellant used by all spacecraft in the constellation.
Constraints~\eqref{eq:MVOCP_dynamics} enforce the dynamics via multiple shooting, discretized at each control node; constraint~\eqref{eq:MVOCP_initialconditions} fixes the initial state to the state estimate at $t_k$; constraints~\eqref{eq:MVOCP_finalset_r} and~\eqref{eq:MVOCP_finalset_v} ensure all spacecraft in the formation track their respective absolute reference trajectories; constraints~\eqref{eq:MVOCP_minsep}, and~\eqref{eq:MVOCP_maxsep} bound the inter-spacecraft separation; finally, constraints~\eqref{eq:constraint_minphaseangle} and~\eqref{eq:constraint_maxphaseangle} bound the inter-spacecraft relative Sun phase angle.
While we enforce no constraint on the maximum control magnitude, a typical station-keeping maneuver on the NRHO is on the order of at most a few \SI{}{cm/s} per maneuver~\cite{Davis2022,Shimane2025}, which can be executed effectively as an impulse by a typical lunar mission spacecraft~\cite{Cheetham2022}.

Two aspects of problem~\eqref{eq:MVOCP} still require our attention.
\rev{First}, since this OCP is solved as part of an MPC scheme, we must ensure recursive feasibility under uncertainty; specifically, once a solution at $t_k$ is obtained, a corrupted control $\Ubold_{0|k} + \delta \Ubold$ is executed, and the true state $\Xbold(t_k)$ is propagated until $t_{k+1}$, the instantiated OCP at $t_{k+1}$ with state estimate $\hat{\Xbold}(t_{k+1})$ must admit a feasible solution.
Second, constraints~\eqref{eq:MVOCP_boundsep} and~\eqref{eq:constraint_boundphaseangle} are defined in CT, which require either a discrete approximation, which risks inter-sample violation, or an isoperimetric reformulation.
These two aspects are treated in Sections~\ref{sec:empirical_recursive_feasible} and~\ref{sec:continuous_time_constraints}, respectively.

% ===================================================================== %
\section{Solution Approach for Receding Horizon Control}
\label{sec:solution_approach}
In this Section, we address issues encountered due to the recursive nature of the station-keeping problem, where the MVOCP must be solved repeatedly in the presence of uncertainty.

% -------------------------------------------------------------------- %
\subsection{Constraint Tightening}
\label{sec:empirical_recursive_feasible}
Problem~\eqref{eq:MVOCP} is infeasible if at $t_k$, the state estimate $\hat{\Xbold}(t_k)$ violates one of the path constraints~\eqref{eq:MVOCP_boundsep} and~\eqref{eq:constraint_boundphaseangle}.
Such an infeasibility is prone to occur in the presence of uncertainties.
One approach to remedy this infeasibility issue is to reformulate~\eqref{eq:MVOCP_boundsep} and~\eqref{eq:constraint_boundphaseangle} using chance constraints, incorporating uncertainties due to state estimation error, model mismatch, and control execution error, \rev{to achieve a feasible solution at some $1-\epsilon$ probability, where $\epsilon$ is a small positive user-defined constant.
Examples of spacecraft guidance problems incorporating chance constraints include~\cite{Kumagai2025-vu,Lew2020,Takubo2024-tf,Oguri2024,Elango2025RDV}.}
\rev{Implementing chance constraints across nonlinear functions such as the dynamics propagation~\eqref{eq:MVOCP_dynamics} results in arbitrary distributions, for which reformulation into a tractable form with techniques such as the Chebyshev-Cantelli inequality is known to introduce conservatism~\cite{Mesbah2016-cn,Heirung2018-ub}.
Meanwhile, direct calibration of the required constraint tightening offline has the advantage of reducing the conservatism~\cite{Lorenzen2017-se}.
Due to the high importance of control cost dictated by the scarcity of propellant onboard the spacecraft, we adopt a similar offline tightening process.}

Consider a generic CT state path constraint,
\begin{equation}    \label{eq:pathconstraint_generic}
    g(\Xbold(t),t) \leq 0 \quad t \in [t_{0|k},t_{N-1|k}]
    .
\end{equation}
We define a \textit{tightened constraint} $\tilde{g}(\Xbold(t), t)$,
\begin{equation}    \label{eq:pathconstraint_gradual_tighten}
    \tilde{g}(\Xbold(t), t) = g(\Xbold(t), t) + \zeta(t, t_{0|k}, t_{N-1|k}) \leq 0 \quad t \in [t_{0|k},t_{N-1|k}]
    ,
\end{equation}
where $\zeta$ is a monotonically increasing \textit{tightening function}, with $\zeta(t_{0|k}, t_{0|k}, t_{N-1|k}) = 0$.
The rationale behind enforcing~\eqref{eq:pathconstraint_gradual_tighten} in lieu of~\eqref{eq:pathconstraint_generic} is that for $t > t_{0|k}$ within the prediction horizon at $t_k$, an extra margin is built in to a feasible solution due to the term $\zeta$.
With an appropriately chosen $\zeta$, the solution to the OCP instantiated at $t_k$ satisfying~\eqref{eq:pathconstraint_gradual_tighten} includes a margin $\zeta(t_{1|k}, t_{0|k}, t_{N-1|k}) > 0$ at $t_{1|k}$, such that at the next recursion at $t_{k+1}$, the initial state $\Xbold_{0|k+1}$ satisfies the original constraint~\eqref{eq:pathconstraint_generic} despite the presence of uncertainties.

\begin{figure}[]
     \centering
     \begin{subfigure}[b]{0.46\textwidth}
         \centering
         \includegraphics[width=\textwidth]{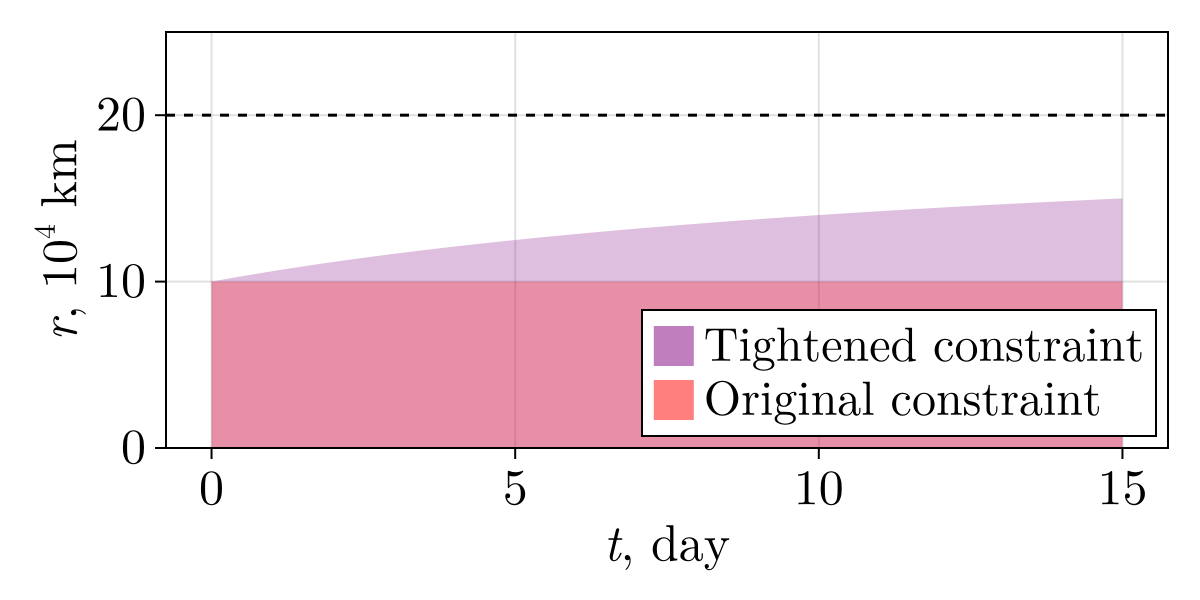}
        \caption{$\kappa = 10^3$}
         \label{fig:constraint_tighten_rmin_kappa1e3}
     \end{subfigure}
     \hfill
     \begin{subfigure}[b]{0.46\textwidth}
         \centering
         \includegraphics[width=\textwidth]{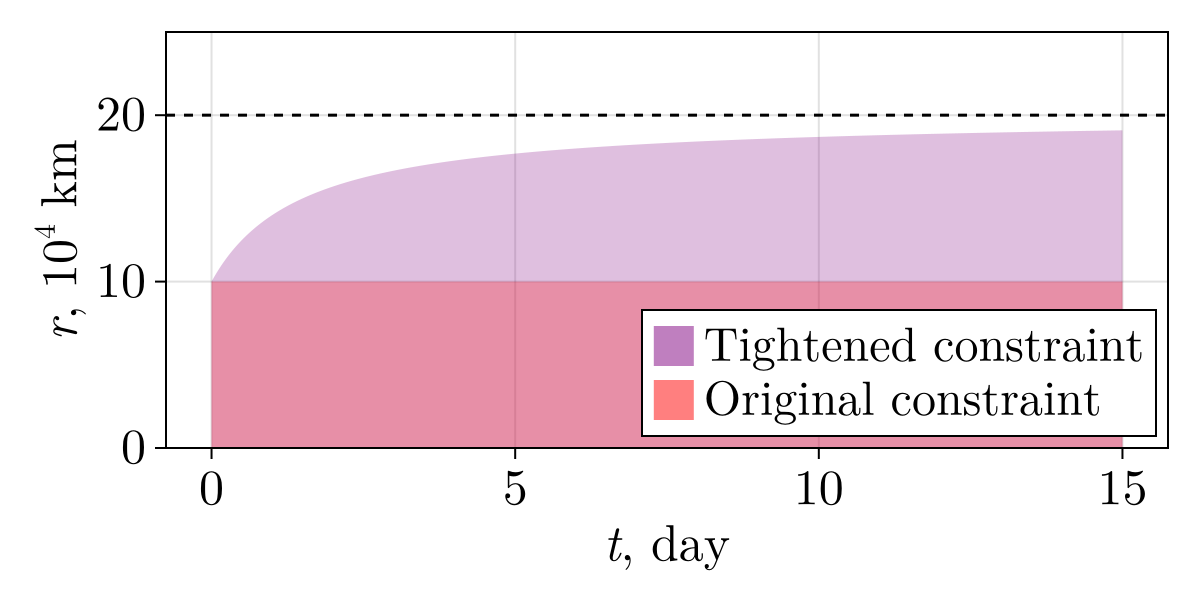}
        \caption{$\kappa = 10^4$}
         \label{fig:constraint_tighten_rmin_kappa1e4}
     \end{subfigure}
     \\
     \begin{subfigure}[b]{0.46\textwidth}
         \centering
         \includegraphics[width=\textwidth]{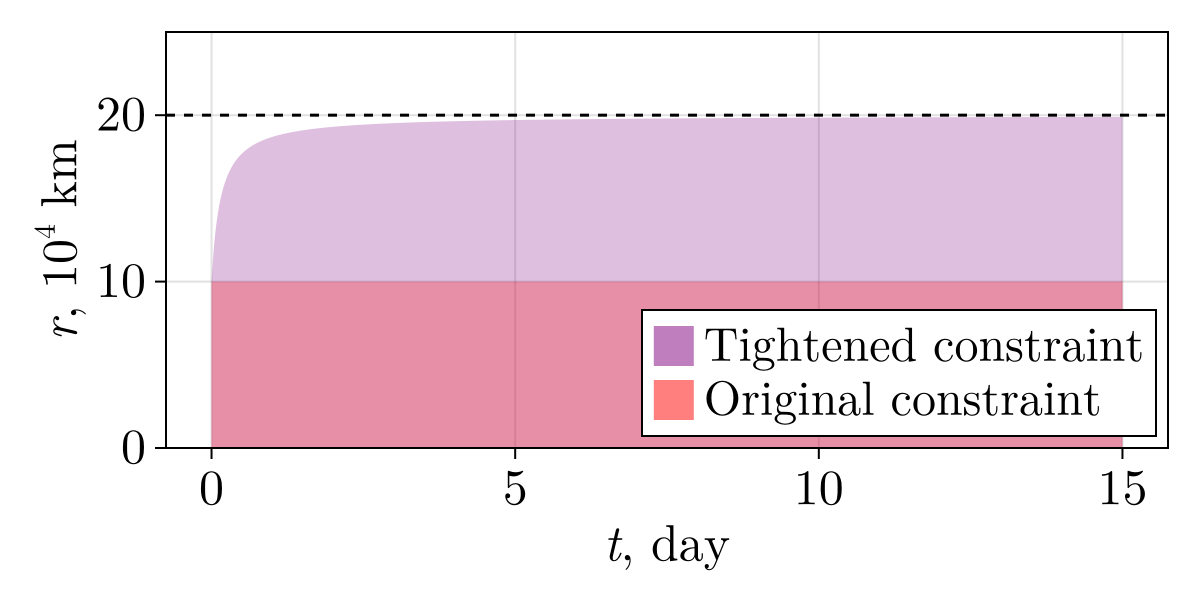}
        \caption{$\kappa = 10^5$}
         \label{fig:constraint_tighten_rmin_kappa1e5}
     \end{subfigure}
    \caption{Gradually tightened minimum separation constraint ($\Delta r_{\min} = 10$ \SI{}{km}, $\eta_{\Delta r_{\min}} = 10$ \SI{}{km})}
    %; the red region is the region constrained by the original path constraint, and the purple region is the additional forbidden region by the tightened path constraint}
    \label{fig:constraint_tighten_rmin_kappavar}
\end{figure}

\rev{Given the requirements on $\zeta$, we consider the empirical expression}
%The $\zeta$ function is designed to be as intuitive and to use as few parameters as possible,
\begin{equation}
    \zeta (t, t_{0|k}, t_{N-1|k}) = \eta - \dfrac{1}{\kappa t_{\rm ratio} + 1/\eta}
    ,\quad
    t_{\rm ratio} = \dfrac{t - t_{0|k}}{t_{N-1|k} - t_{0|k}}
    ,
\end{equation}
where $\eta$ is a margin with the same unit as the constraint $g$, and $\kappa$ is a tightening parameter, where a larger $\kappa$ results in faster tightening.
\rev{Assuming a priori knowledge of the uncertainty distributions in the problem described in Section~\ref{sec:background_uncertainties} as well as controller design considerations such as maneuver spacing and horizon length $N$, the parameters $\eta$ and $\kappa$ may be tuned offline in a way that the controlled trajectory under disturbance remains feasible with respect to the original constraint at a desired $1-\epsilon$ probability.}
To illustrate the effect of constraint tightening, consider the lower-bound inter-spacecraft separation constraint~\eqref{eq:MVOCP_minsep} as an example, with $\Delta r_{\min} = 10$~\SI{}{km} and margin $\eta_{\Delta r_{\min}} = 10$ \SI{}{km}.
Figure~\ref{fig:constraint_tighten_rmin_kappavar} shows forbidden inter-spacecraft regions as $\kappa$ is varied from $10^3$ to $10^5$, where the red-shaded region is restricted by the original path constraint, and the purple-shaded region is restricted by the additional tightening term.

% \subsubsection{Adaptive relaxation for non-critical constraints}
% Through preliminary experiments, we find the maximum separation constraint~\eqref{eq:MVOCP_maxsep} to be prone to violations following perilune passes (i.e., as the spacecraft approaches $\theta = 160^{\circ}$.
% \red{Explain why that is the case}
% To prevent the problem from becoming infeasible, one may choose conservative tunings for $M$ and $\kappa$ that results in the problem being recursively feasible under uncertainty.
% Meanwhile, deeming the maximum separation constraint as being \textit{non}-critical, we consider adaptively relaxing pre-defined upper bound $\Delta r_{\max}$ and margin $M$ on the gradually tightened~\eqref{eq:MVOCP_maxsep}.
% We define a pair-wise adaptively relaxed upper bound on the separation, $\Delta r_{\max,ij\,\mathrm{adapt}}$,
% \begin{equation}
%     \Delta r_{\max,ij\,\mathrm{adapt}} = 
%     \begin{cases}
%         \Delta r_{\max} & \| \rbold_i(t) - \rbold_k(t) \|_2 \leq \Delta r_{\max},
%         \\
%         \| \rbold_i(t) - \rbold_k(t) \|_2 & \text{otherwise},
%     \end{cases}
% \end{equation}
% with corresponding margin pair-wise adaptive $M_{ij\,\mathrm{adapt}}$
% \begin{equation}
%     M_{ij\,\mathrm{adapt}} = 
%     \begin{cases}
%         M  & \| \rbold_i(t) - \rbold_k(t) \|_2 \leq \Delta r_{\max},
%         \\
%         M - (\| \rbold_i(t) - \rbold_k(t) \|_2 - \Delta r_{\max}) & \text{otherwise}.
%     \end{cases}
% \end{equation}

% -------------------------------------------------------------------- %
\subsection{Continuous-Time Constraint Satisfaction}
\label{sec:continuous_time_constraints}
Satisfying path constraints~\eqref{eq:MVOCP_boundsep} and~\eqref{eq:constraint_boundphaseangle} requires either a discrete-time approximation or an isoperimetric reformulation.
The former involves simply enforcing 
\begin{equation}    \label{eq:discrete_approx_pathconstraints}
    g(\Xbold_{j|k}, t_{j|k}) \leq 0 \quad \forall j \in \mathcal{J}
    .
\end{equation}
However, at $t_k < t < t_{k+1}$, there are no mechanisms that ensure the path constraints are not violated.
Instead, in this work, we employ an isoperimetric reformulation to avoid the risk of inter-sample violation altogether.
A generic CT state path constraint,
\begin{equation}
    g(\Xbold(t),t) \leq 0 \quad t \in [t_{0|k},t_{N-1|k}]
    ,
\end{equation}
can be equivalently expressed in isoperimetric form as 
\begin{equation}    \label{eq:generic_isoperimetric_integral}
    \int_{t_{j|k}}^{t_{j+1|k}} \max{\left[ 0, g(\Xbold(t),t) \right]}^2 \mathrm{d}t = 0
    \quad \forall j \in \mathcal{J} \setminus \{N-1\} 
    ,
\end{equation}
where the integrand is made differentiable through the square operator.
In the case of OCP~\eqref{eq:MVOCP}, $g(\cdot)$ correspond to constraints~\eqref{eq:MVOCP_minsep},~\eqref{eq:MVOCP_maxsep},~\eqref{eq:constraint_minphaseangle}, and~\eqref{eq:constraint_maxphaseangle}.

Path constraints reformulated as~\eqref{eq:generic_isoperimetric_integral} can be enforced by introducing a slack state variable $y$ for each path constraint, with fictitious dynamics
\begin{equation}    \label{eq:ydot_generic}
    \dot{y} = \max{\left[ 0, g(\Xbold(t),t) \right]}^2 
    ,
\end{equation}
and boundary conditions
\begin{equation}~\label{eq:ct_slack_equality}
    y_{j+1|k} = y_{j|k}
    \quad \forall j \in \mathcal{J} \setminus \{N-1\} 
    .
\end{equation}
Notably,~\eqref{eq:ct_slack_equality} results in linearly dependent gradients with dynamics constraints for $y$, resulting in failure to meet linear independence constraint qualification (LICQ).
\rev{Following the approach in~\cite{Elango2024},} we introduce a small tolerance~$\epsilon_{\rm LICQ}$ and relax~\eqref{eq:ct_slack_equality} as
\begin{equation}~\label{eq:ct_slack_equality_relaxed}
    y_{j+1|k} - y_{j|k} \leq \epsilon_{\rm LICQ}
    \quad \forall j \in \mathcal{J} \setminus \{N-1\} 
    .
\end{equation}
\rev{Note that, from Theorem 10 and Remark 11 in~\cite{Elango2024}, it is possible to derive a constraint tightening term $\delta_{g}(\epsilon_{\rm LICQ})$ such that $\int_{t_{j|k}}^{t_{j+1|k}} \max{\left[ 0, g(\Xbold(t),t) + \delta_{g}(\epsilon_{\rm LICQ}) \right]}^2 \mathrm{d}t \leq \epsilon_{\rm LICQ}$ implies $g(\Xbold(t),t) \leq 0$ continuously over $t \in [t_{j|k}, t_{j+1|k}]$.
In numerical implementation, a sufficiently small $\epsilon_{\rm LICQ}$ can be chosen such that the relaxation is numerically significant relative to the optimization solver tolerance, while remaining physically negligible.
The chosen values for $\epsilon_{\rm LICQ}$ and the expected order of magnitude of worst-case constraint violation are reported in Section~\ref{sec:numerical_experiments}.
% With respect to the inter-spacecraft range constraints~\eqref{eq:constraint_boundphaseangle}, we set $\epsilon_{\rm LICQ} = 10^{-8}$\footnote{The slack's order of magnitude is with respect to the canonical scales used within the optimization problem.}, the order of magnitude of the resulting violation
% \begin{equation}
%     \mathcal{O} \left(
%         \underset{t \in [t_{j|k}, t_{j+1|k}]}{\esssup} 
%         \max [0, g(\Xbold(t),t) ]
%     \right)
%     \sim 
%     \sqrt{\dfrac{\epsilon_{\rm LICQ}}{t_{j+1|k} - t_{j|k}}}
% \end{equation}
% evaluates to $10^{-1}\,\mathrm{km}$.
% With the Sun phase angle constraints~\eqref{eq:constraint_boundphaseangle}, we set $\epsilon_{\rm LICQ} = 10^{-6}$, which results in order of magnitude of constraint violation of $10^{-2}\,\mathrm{deg}$.
}

% -------------------------------------------------------------------- %
\subsection{Reformulated Optimal Control Problem}
The OCP~\eqref{eq:MVOCP} is reformulated by taking into account the gradual path-constraint tightening from Section~\ref{sec:empirical_recursive_feasible} and the isoperimetric path-constraint reformulation from Section~\ref{sec:continuous_time_constraints}.
Let $\ybold \in \mathbb{R}^{L}$ denote the slack state variables corresponding to $L$ CT path constraints, and $\Zbold = \begin{bmatrix} \Xbold^T(t) & \ybold^T(t) \end{bmatrix}^T \in \mathbb{R}^{6M + L}$ denote the augmented state variables, with corresponding augmented dynamics $\fbold_{\rm aug}$, composed of the concatenated dynamics $\fbold_{\rm concat}$ from~\eqref{eq:f_concat} and slack dynamics $\fbold_{\rm slack}$,
\begin{equation}    \label{eq:f_aug}
    \dot{\Zbold}(t) = 
    \begin{bmatrix}
        \dot{\Xbold}(t) \\ \dot{\ybold}(t)   
    \end{bmatrix} 
    =
    \fbold_{\rm aug}(\Zbold(t),t) = \begin{bmatrix}
        \fbold_{\rm concat}(\Xbold(t), t) \\ 
        \fbold_{\rm slack}(\Xbold(t), t)
    \end{bmatrix} 
    ,\quad
    \fbold_{\rm slack}(\Xbold(t), t) =
    \begin{bmatrix}
        W_0 \max{\left[ 0, g_0(\Xbold(t),t) \right]}^2 \\ \vdots \\ W_{L-1} \max{\left[ 0, g_{L-1}(\Xbold(t),t) \right]}^2
    \end{bmatrix}
    % \begin{bmatrix}
    %     W_0 \dot{y}_{0}(t) \\ \vdots \\ W_{L-1} \dot{y}_{L-1}(t)
    % \end{bmatrix}
    ,
\end{equation}
where $W_0,\ldots,W_{L-1}$ are scaling parameters used to adjust the scale of the state around unity to improve numerical conditions when integrating $\fbold_{\rm aug}$.
Note that since derivatives of $\ybold$ defined according to~\eqref{eq:ydot_generic} do not depend on $\ybold$ themselves, the argument of $\fbold_{\rm slack}$ can be reduced to $\Xbold$ instead of $\Zbold$.
While, as a consequence, the argument of $\fbold_{\rm aug}$ should also be $\Xbold$ instead of $\Zbold$, we retain $\Zbold$ to clarify that $\fbold_{\rm aug}$ define the dynamics of $\Zbold$.
With the bounded separation constraints~\eqref{eq:MVOCP_minsep} and~\eqref{eq:MVOCP_maxsep} and the relative phase angle constraints~\eqref{eq:constraint_minphaseangle} and~\eqref{eq:constraint_maxphaseangle} for a problem involving $M$ spacecraft, $L = 4 \times \comb{M}{2}$.
The reformulated MVOCP is given by
\begin{subequations}    \label{eq:MVOCP_ct}
\begin{align}
    \minimize{
        \substack{\Zbold_{0|k},\ldots,\Zbold_{N-1|k}\\\Ubold_{0|k},\ldots,\Ubold_{N-1|k}}
    }
    \quad& \sum_{i=0}^{M-1} \sum_{j=0}^{N-1} \| \ubold_{i,j|k} \|_2
        \label{eq:MVOCP_ct_obj}
    \\\st\quad&
    \Zbold_{j+1|k} = \Zbold_{j|k} 
        + \begin{bmatrix}
            \Bbold_{\rm concat} \\ \boldsymbol{0}_{L \times 3M}
        \end{bmatrix} \Ubold_{j|k}
        + \int_{t_{j|k}}^{t_{j+1|k}} \fbold_{\rm aug}(\Zbold(t),t)\mathrm{d}t
            \quad \forall j \in \mathcal{J} \setminus \{N-1\}
    \label{eq:MVOCP_ct_dynamics}
    \\&
    \ybold_{j+1|k} - \ybold_{j|k} \leq \epsilon_{\rm LICQ} \boldsymbol{1}_{L \times 1}
        \quad \forall j \in \mathcal{J} \setminus \{N-1\}
    \label{eq:MVOCP_ct_slack_eq_relaxed}
    \\&
        \nonumber
        \text{eqn.~\eqref{eq:MVOCP_initialconditions},~\eqref{eq:MVOCP_finalset_r},~\eqref{eq:MVOCP_finalset_v}}
\end{align}
\end{subequations}

\subsection{Sequential Convex Programming}
Problem~\eqref{eq:MVOCP_ct} is solved via SCP~\cite{Mao2019-wp,Malyuta2022-ye}.
In an SCP scheme, a convexified subproblem is solved iteratively by linearizing and/or convexifying the original OCP with respect to a reference solution\footnote{The \textit{reference solution} within the SCP is the iteratively updated trajectory of all spacecraft in the formation, and is unrelated to the tracked \textit{reference NRHO}.}.
The reference solution is updated by an appropriate step-acceptance criterion, and subproblems are successively solved until a local optimal solution to the original nonconvex problem is found.
In case of~\eqref{eq:MVOCP_ct}, the only nonconvexity occurs in the dynamics constraints~\eqref{eq:MVOCP_ct_dynamics}.

We adopt the augmented Lagrangian SCP from~\cite{Oguri2023} due to its successful adoption in cislunar trajectory design problems~\cite{Kumagai2024,Oguri2024,Shimane2025}.
Two caveats when implementing an SCP scheme are \textit{artificial unboundedness} and \textit{artificial infeasibility}~\cite{Malyuta2022-ye}.
To avoid these issues, the augmented Lagrangian SCP incorporates slack variables on the nonconvex constraints, which are penalized in an augmented objective function, and trust-region constraints, which limit the search space of the decision variables to a neighborhood of the reference solution, ensuring the linearization provides a reliable prediction of the nonconvex behavior.
Let $(\cdot)^{(l)}$ denote quantities within the SCP algorithm during its $l^{\rm th}$ iteration, and let $\bar{(\cdot)}^{(l)}$ denote the reference solution at the $l^{\rm th}$ iteration.
Let $\xibold_{j|k} \in \mathbb{R}^{6M + L}$ denote the slack variables associated with the $j^{\rm th}$ dynamics constraint~\eqref{eq:MVOCP_ct_dynamics}, with corresponding Lagrange multipliers $\boldsymbol{\lambda}_{j|k}^{(l)}  \in \mathbb{R}^{6M + L}$, let $\boldsymbol{\Delta}^{(l)} \in \mathbb{R}^{6M + L}_+$ denote the vector of trust-regions on the augmented state vector components, and let $w^{(l)} \in \mathbb{R}_+$ denote the penalty weight parameter.
The convex subproblem corresponding to~\eqref{eq:MVOCP_ct} is 
\begin{subequations}    \label{eq:MVOCP_ct_convex_subproblem}
\begin{align}
    \minimize{
        \substack{
            \Zbold_{0|k},\ldots,\Zbold_{N-1|k} \\
            \Ubold_{0|k},\ldots,\Ubold_{N-1|k} \\
            \xibold_{0|k},\ldots,\xibold_{N-2|k}
        }
    }
    \quad& \sum_{i=0}^{M-1} \sum_{j=0}^{N-1} \| \ubold_{i,j|k} \|_2
    + 
    \sum_{j=0}^{N-2}
    \left(
        \boldsymbol{\lambda}_{j|k}^T \xibold_{j|k} + \dfrac{w^{(l)}}{2} \xibold_{j|k}^T \xibold_{j|k}
    \right)
    \\\st\quad&
    \Zbold_{j+1|k} =
        \boldsymbol{\Psi}_{j+1,j|k}
        \left(
        \Zbold_{j|k} 
        + \begin{bmatrix}
            \Bbold_{\rm concat} \\ \boldsymbol{0}_{L \times 3M}
        \end{bmatrix} \Ubold_{j|k}
        \right)
        + \boldsymbol{C}_{j|k}
        + \xibold_{j|k}
        % + \int_{t_{j|k}}^{t_{j+1|k}} \fbold_{\rm aug}(\Zbold(t),t)\mathrm{d}t
            \quad \forall j \in \mathcal{J} \setminus \{N-1\} %k = 0,\ldots,N-2
    \label{eq:MVOCP_ct_convex_subproblem_dynamics}
    \\&
        \left| \Zbold_{j|k} - \bar{\Zbold}_{j|k}^{(l)} \right| \leq \boldsymbol{\Delta}^{(l)}
            \quad \forall j \in \mathcal{J} \setminus \{N-1\}
    \label{eq:MVOCP_ct_convex_subproblem_trustregion}
    \\&
        \nonumber
        \text{eqn.~\eqref{eq:MVOCP_initialconditions},~\eqref{eq:MVOCP_finalset_r},~\eqref{eq:MVOCP_finalset_v},~\eqref{eq:MVOCP_ct_slack_eq_relaxed}}
\end{align}
\end{subequations}
\rev{Note that in~\eqref{eq:MVOCP_ct_convex_subproblem}, all nonconvex constraints, namely the dynamics and the nonconvex path constraints, are linearized and incorporated in the linearized augmented dynamics constraint~\eqref{eq:MVOCP_ct_convex_subproblem_dynamics}.}
In~\eqref{eq:MVOCP_ct_convex_subproblem_dynamics}, $\boldsymbol{\Psi}_{j+1,j|k}$ is the state-transition matrix (STM) for the augmented state $\Zbold$, and $\boldsymbol{C}_{j|k}$ contains the dynamics terms along the current reference solution,
\begin{equation}
    \boldsymbol{C}_{j|k} = 
    \left(
        \bar{\Zbold}_{j|k} + \int_{t_{j|k}}^{t_{j+1|k}} \fbold_{\rm aug}(\bar{\Zbold}(t), t) \mathrm{d}t
    \right)
    -
    \boldsymbol{\Psi}_{j+1,j|k}
    \left(
    \bar{\Zbold}_{j|k} 
    + \begin{bmatrix}
        \Bbold_{\rm concat} \\ \boldsymbol{0}_{L \times 3M}
    \end{bmatrix} \bar{\Ubold}_{j|k}
    \right)
    ,
\end{equation}
and constraints~\eqref{eq:MVOCP_ct_convex_subproblem_trustregion} enforce the trust-region, where the inequality is applied element-wise.
\red{The STMs $\boldsymbol{\Psi}_{j+1,j|k}$ are computed by integrating the state $\Zbold$ along with the STM.
Expressions for the Jacobian of the augmented system $\fbold_{\rm aug}$ are presented in Appendix~\ref{appendix:jacobian}}.

The convex subproblem~\eqref{eq:MVOCP_ct_convex_subproblem} is almost exactly an example of the class of problems considered in~\cite{Oguri2023} with the exception of~\eqref{eq:MVOCP_ct_convex_subproblem_trustregion}, where we impose the element-wise inequality with $\boldsymbol{\Delta}^{(l)}$, whereas~\cite{Oguri2023} uses a scalar $\Delta$ and imposes $\| \Zbold_{j|k} - \bar{\Zbold}_{j|k}^{(l)} \|_{\infty} \leq \Delta^{(l)}$ instead.
At a practical level, we find the definition of trust region for each variable to be particularly useful with problem~\eqref{eq:MVOCP_ct_convex_subproblem}, where the nonlinearity of slack variables $\ybold$ can be largely different from the nonlinearity of the dynamics.
An even more sophisticated approach would involve a full state-dependent trust-region, where an independent trust-region vector $\boldsymbol{\Delta}_{j|k}$ is defined independently for each time-step $(\cdot)_{j|k}$ and tuned based on the local nonlinearity, as demonstrated in~\cite{Bernardini2024-hu}.
We however find defining $\boldsymbol{\Delta} = \boldsymbol{\Delta}_{0|k} = \ldots = \boldsymbol{\Delta}_{N-1|k}$ to be sufficient for the present application.
Between successive steps $l$ and $l+1$, we adopt the step-acceptance, Lagrange multiplier update, and weight update criteria from~\cite{Oguri2023}.

\rev{Due to the nonconvex property of~\eqref{eq:MVOCP_ct}, providing a good initial reference solution, i.e., $\bar{\Zbold}_{j|k}^{(0)}$ and $\bar{\Ubold}_{j|k}^{(0)}$, is essential for the SCP algorithm to converge to a local optimal solution reliably and with as few iterations as possible.
In the station-keeping context, an intuitive initial guess can be constructed based on the reference NRHO, since the state of each spacecraft in the formation lies in the vicinity of the reference.
Specifically, we set the initial guess on the state $\bar{\Zbold}_{j|k}^{(0)}$ as follows:
\begin{equation}    \label{eq:initial_guess}
    \bar{\Zbold}_{j|k}^{(0)}
    = \begin{bmatrix}
        \bar{\Xbold}_{j|k}^{(0)} \\
        \ybold_{j|k}^{(0)}
    \end{bmatrix}
    ,\quad
    \bar{\Xbold}_{j|k}^{(0)}
    =
    \begin{bmatrix}
        \xbold_{\rm ref}(t_{j|k}) + \epsilon_{\rm init} \boldsymbol{1}_{6 \times 1}
        \\
        \xbold_{\rm ref}(t_{j|k}) + 2 \epsilon_{\rm init} \boldsymbol{1}_{6 \times 1}
        \\
        \vdots 
        \\
        \xbold_{\rm ref}(t_{j|k}) + M \epsilon_{\rm init} \boldsymbol{1}_{6 \times 1}
    \end{bmatrix}
    ,\quad
    \ybold_{j|k}^{(0)} = \boldsymbol{0}_{L \times 1}
    ,
\end{equation}
where $\epsilon_{\rm init} = 10^{-6}$ is included to ensure the Jacobian $\partial \fbold_{\rm aug}/\partial \Zbold$ along $\bar{\Zbold}_{j|k}^{(0)}$ is well-defined.
We set $\bar{\Ubold}_{j|k}^{(0)} = \boldsymbol{0}_{3M \times 1}$ in the expectation that the required control maneuver is small.
In the context of the MPC, one may also use solutions from previous recursions as a warm-starting solution to generate $\bar{\Zbold}_{j|k}^{(0)}$ and $\bar{\Ubold}_{j|k}^{(0)}$.
While such an approach could reduce the number of iterations required to solve the MVOCP, we find the initial guess based on~\eqref{eq:initial_guess} to be sufficiently good for the SCP algorithm to converge reliably.}

% -------------------------------------------------------------------- %
\subsection{Model Predictive Control Framework}
The proposed MPC framework consists of recursively solving the MVOCP~\eqref{eq:MVOCP_ct}. Figure~\ref{fig:recursive_simulation} illustrates the recursive process.
The red boxes indicate processes where the errors described in Section~\ref{sec:background_uncertainties} enter the system.
Based on the prediction horizon from Section~\ref{sec:prediction_horizon} with controls located at $\theta_{\rm ref}$ of $160^{\circ}$ and $200^{\circ}$, \red{the initial time of the MPC}, $t_{0|k}$, alternates between the two true anomaly regions along the NRHO during the recursion.

\rev{As with any MPC framework, the online implementability of a solution algorithm to the internal OCP must be taken into consideration.
The MVOCP~\eqref{eq:MVOCP_ct} is a nonconvex optimization problem, which together with the computation of the STMs for linearization at each SCP iteration, amounts to computational requirements that are beyond today's practical onboard compute resources for robotic spacecraft.
Instead, station-keeping maneuvers are typically computed on the ground using nonlinear/nonconvex techniques~\cite{Cheetham2022,Davis2022}, and maneuver commands are sent back to the spacecraft.
Such an operational scheme is possible due to the sparse control cadence occurring only once every few days, and justified by the typically high value of cislunar assets.
The proposed MPC framework is to be implemented on ground-based computational resources, where computational resources are abundant in comparison to what is required to solve~\eqref{eq:MVOCP_ct}.}

% flow chart
\begin{figure}
    \centering
    \begin{tikzpicture}[node distance=1.5cm]
    
    \node (start) [process] {Initialize $\Xbold(t_0)$};
    \node (estimate) [process, below of=start] {Estimate current state $\hat{\Xbold}(t_k)$};
    \node (mvocp) [process, below of=estimate] {Solve MVOCP~\eqref{eq:MVOCP_ct} with SCP};
    \node (execute) [process, below of=mvocp] {Execute control $\Ubold_k$};
    \node (propagate) [process, below of=execute] {Propagate until next control node at $t_{k+1}$};

    \node (insertcorrupt) [processerror, right of=start, xshift=4.5cm] {Initial orbit insertion error $\delta \xbold_{i,\rm init} \, \forall i$};
    \node (odcorrupt) [processerror, right of=estimate, xshift=4.5cm] {Navigation error $\delta \xbold_{i,\rm nav} \, \forall i$};
    \node (controlcorrupt) [processerror, right of=execute, xshift=4.5cm] {Control execution error $\delta \Ubold$};
    \node (dyncorrupt) [processerror, right of=propagate, xshift=4.5cm] {SRP error $\delta A/m$, $\delta C_r$};

    \draw [arrow] (start) -- (estimate);
    \draw [arrow] (estimate) -- (mvocp);
    \draw [arrow] (mvocp) -- (execute);
    \draw [arrow] (execute) -- (propagate);
    \coordinate (left propagate) at ($(propagate.west) - (1.8cm,0)$);
    \draw[arrow]  
        (propagate.west) -- node[anchor=south] {$k \gets k+1$} (left propagate)
        |- (estimate.west);
    
    \draw [arrow, red] (insertcorrupt) -- (start);
    \draw [arrow, red] (odcorrupt) -- (estimate);
    \draw [arrow, red] (controlcorrupt) -- (execute);
    \draw [arrow, red] (dyncorrupt) -- (propagate);
    
    \end{tikzpicture}
    \caption{Model predictive control framework for recursive station-keeping of spacecraft formation}
    \label{fig:recursive_simulation}
\end{figure}

% ===================================================================== %
\section{Numerical Experiments}
\label{sec:numerical_experiments}
We demonstrate the proposed approach for a two- \rev{and three-}spacecraft formation tracking a common reference NRHO.
\rev{We begin by analyzing the solution of single instances of the MVOCP~\eqref{eq:MVOCP_ct}, reporting the effect of the LICQ relaxation with $\epsilon_{\rm LICQ}$, the behavior of constraint tightening, and the computational effort required to solve these instances.}
\rev{We then conduct Monte Carlo experiments demonstrating the MPC scheme solving the MVOCP recursively.}
Each experiment consists of 100 Monte Carlo samples, each lasting 10 revolutions, which corresponds to $\sim\!\! 65.5$~\SI{}{days}. This duration ensures that the formation-flying spacecraft completes a full quasi-periodic cycle along its 9:2 resonance, which lasts approximately two synodic months. 
With controls located at $\theta_{\rm ref}(t) = 160^{\circ}$ and $200^{\circ}$, there are two MVOCP instances solved per revolution, resulting in 20 instances solved during the 10-revolution experiment.
\rev{Other experiment variants, such as with different control cadence or along other LPO, would also be of interest and are left for future work.}

We conduct four Monte Carlo experiments, summarized in Table~\ref{tab:summary_mc_experiments}: experiments I and II consider the MVOCP~\eqref{eq:MVOCP} without the relative Sun phase angle constraint~\eqref{eq:constraint_boundphaseangle}, i.e., with only the bounded separation as path constraints; meanwhile, experiments III and IV consider~\eqref{eq:MVOCP} as is, i.e., with both bounded separation and relative Sun phase angle path constraints.
Experiments I and III enforce continuous path constraints through the isoperimetric reformulation, while experiments II and IV only enforce the path constraints at the control nodes.
With II and IV, the nonconvex path constraints, that is, the lower bounds on separation~\eqref{eq:MVOCP_minsep} as well as the lower and upper bounds on the relative Sun phase angle~\eqref{eq:constraint_boundphaseangle} are linearized and enforced gradually through the augmented Lagrangian scheme from~\cite{Oguri2023}, as is done for the nonconvex dynamics constraints~\eqref{eq:MVOCP_dynamics}.

In all experiments, the reference is tracked at tolerances $\epsilon_r = 20\,\mathrm{km}$ and $\epsilon_v = 5\,\mathrm{m/s}$ by the end of a prediction horizon of $N_{\rm rev} = 5$ revolutions, as enforced by constraints~\eqref{eq:MVOCP_finalset_r} and~\eqref{eq:MVOCP_finalset_v}.
The parameters pertaining to the uncertainties, largely based on~\cite{Davis2022}, are provided in Table~\ref{tab:uncertainty_params}.
\rev{We assume navigation errors of each spacecraft in the formation are uncorrelated, resulting in the most conservative case; in practice, the consistency of dynamical effects across spacecraft and the existence of relative sensing improve the state estimation quality, which would consequently improve the guidance and control scheme performance.}
Table~\ref{tab:constraint_tightening_params} summarizes the constraint tightening parameters associated with~\eqref{eq:MVOCP_boundsep} and~\eqref{eq:constraint_boundphaseangle}.
As mission requirements, we consider a minimum separation of $\Delta r_{\min} = 10$ \SI{}{km} at all times and a maximum separation of $\Delta r_{\max} = 100$ \SI{}{km} in apolune segments; in addition, with experiments III and IV, we consider a phase angle bounded between $\phi_{\min} = 30^{\circ}$ and $\phi_{\max} = 150^{\circ}$ at apolune segments.

\rev{For experiments I and III, LICQ relaxation slacks must be chosen; with respect to the inter-spacecraft range constraints~\eqref{eq:constraint_boundphaseangle}, we set $\epsilon_{\rm LICQ} = 10^{-8}$\footnote{The slack's order of magnitude is with respect to the canonical scales used within the optimization problem.}, the order of magnitude of the resulting violation
\begin{equation}
    \mathcal{O} \left(
        \underset{t \in [t_{j|k}, t_{j+1|k}]}{\esssup} 
        \max [0, W \cdot g(\Xbold(t),t) ]
    \right)
    \sim 
    \sqrt{\dfrac{\epsilon_{\rm LICQ}}{t_{j+1|k} - t_{j|k}}}
    ,
\end{equation}
where $W$ is the scaling weight on the constraint, evaluates to $10^{-1}\,\mathrm{km}$.
With the Sun phase angle constraints~\eqref{eq:constraint_boundphaseangle}, we set $\epsilon_{\rm LICQ} = 10^{-6}$, which results in order of magnitude of constraint violation of $\sim 2.4\,\mathrm{deg}$.}
 
% Table of experiment summary
\begin{table}[]
\centering
\caption{Summary of Monte Carlo experiments}
\begin{tabular}{@{}lll@{}}
    \toprule
    Experiment & Path constraints & Path constraints type \\ \midrule
    I & \multirow{2}{*}{Rel. separation~\eqref{eq:MVOCP_boundsep}}                 
    & Continuous 
    \\
    II &                                                     
    & Discrete approximation (control nodes only)
    \\
    III & \multirow{2}{*}{Rel. separation~\eqref{eq:MVOCP_boundsep} \& rel. Sun phase angle~\eqref{eq:constraint_boundphaseangle}}
    & Continuous
    \\
    IV         &                                                     
    & Discrete approximation (control nodes only) \\
    \bottomrule
\end{tabular}
\label{tab:summary_mc_experiments}
\end{table}

% table of uncertainty parameters
\begin{table}[]
\centering  
\caption{Uncertainty parameters in Monte Carlo experiments}
\begin{tabular}{ll}
    \hline
    Parameter & Value \\ \hline
    Initial orbit insertion position error $3$-$\sigma_{\rbold 0}$, \SI{}{km}  & $5$ \\
    Initial orbit insertion velocity error $3$-$\sigma_{\vbold 0}$, \SI{}{m/s} & $0.1$ \\
    SRP $A/m$ relative error $3$-$\sigma_{A/m}$, \%                & $30$ \\
    SRP $C_r$ relative error $3$-$\sigma_{C_r}$, \%                & $15$ \\
    Navigation position error $3$-$\sigma_{\rbold}$, \SI{}{km}     & $1$ \\
    Navigation velocity error $3$-$\sigma_{\vbold}$, \SI{}{cm/s}    & $0.8$ \\
    Control absolute error $3$-$\sigma_{u_{\rm abs}}$, \SI{}{mm/s} & $1$ \\
    Control relative error $3$-$\sigma_{u_{\rm rel}}$, \%          & $1.5$ \\
    Control direction error $3$-$\sigma_{\delta \varphi}$, \SI{}{deg}      & $0.5$ \\
    \hline
\end{tabular}
\label{tab:uncertainty_params}
\end{table}

% table of constraint tightening parameters
\begin{table}[h]
    \centering
    \caption{Constraint tightening parameters}
    \begin{subtable}[h]{0.48\textwidth}
        \centering
        \caption{Separation constraints~\eqref{eq:MVOCP_boundsep}}
        \label{tab:OCP_parameters_separation_constraints}
        \begin{tabular}{@{}ll@{}}
        \toprule
        Parameter & Value \\ \midrule
        Scaling  weight $W$ & $1$ \\
        Bounds $(\Delta r_{\min}, \Delta r_{\max})$, \SI{}{km} & $(10,100)$ \\
        Tightening margins $(\eta_{\Delta r_{\min}}, \eta_{\Delta r_{\max}})$, \SI{}{km} & $(10,10)$ \\
        Tightening parameters $(\kappa_{\Delta r_{\min}}, \kappa_{\Delta r_{\max}})$ & $(10^4, 10^4)$ \\
        \bottomrule
        \end{tabular}
    \end{subtable}
    \hfill
    \begin{subtable}[h]{0.48\textwidth}
        \centering
        \caption{Relative Sun phase angle constraints~\eqref{eq:constraint_boundphaseangle}}
        \label{tab:OCP_parameters_phaseangle_constraints}
        \begin{tabular}{@{}ll@{}}
        \toprule
        Parameter & Value \\ \midrule
        Scaling  weight $W$ & $10^{-3}$ \\
        Bounds $(\phi_{\min}, \phi_{\max})$, \SI{}{deg} & $(30^{\circ}, 150^{\circ})$ \\
        Tightening margins $(\eta_{\phi_{\min}}, \eta_{\phi_{\max}})$, \SI{}{deg} & $(30^{\circ}, 30^{\circ})$ \\
        Tightening parameters $(\kappa_{\phi_{\min}}, \kappa_{\phi_{\max}})$ & $(10^4, 10^4)$ \\
        \bottomrule
        \end{tabular}
    \end{subtable}
    \label{tab:constraint_tightening_params}
\end{table}

The dynamics, MPC algorithm, and simulation environment are implemented in Julia.
The dynamics are non-dimensionalized by choosing a distance unit of $10^4$ \SI{}{km} while ensuring a non-dimensional $\mu = 1$ to improve the numerical conditions of the MVOCP.
The dynamics are numerically integrated with \verb|DifferentialEquations.jl|~\cite{rackauckas2017differentialequations}, using Verner's 7/6 and 8/7 Runge-Kutta methods~\cite{Verner2010-it} within the SCP and during the recursion, respectively, with relative and absolute tolerances set to $10^{-12}$ within the SCP, and a relative tolerance of $10^{-12}$ and an absolute tolerance of $10^{-14}$ during the recursion.
Initial value problem solvers of separate order and varying tolerances are used to (i) speed up the computation within the SCP while (ii) maintaining accurate integration between one control action and another for the simulation to be representative of an actual mission.
The convex subproblem within the SCP is modeled with \verb|JuMP|~\cite{Lubin2023} and solved using the interior-point conic solver Clarabel~\cite{Goulart2024}, using default parameters.
SCP parameters are given in Table~\ref{tab:scp_hyperparams}; the algorithm ends when both the optimality tolerance $\epsilon_{\rm opt}$ and the feasibility tolerance $\epsilon_{\rm feas}$, as defined in~\cite{Oguri2023}, are satisfied.
Parameters not reported in Table~\ref{tab:scp_hyperparams} have been set to their default settings as provided in~\cite{Oguri2023}.
\rev{All experiments are conducted on an Apple M3 Ultra CPU computer; multi-thread parallelism is used for Monte Carlo experiments, allocating threads to independent random seeds.}

% table of SCP parameters
\begin{table}[]
\centering  
\caption{Non-default SCP parameters}
\begin{tabular}{ll}
    \hline
    Parameter & Value \\ \hline
    Initial trust-region radius $\Delta^{(0)}$ on components of state $\xbold$  & $0.05$ \\
    Initial trust-region radius $\Delta^{(0)}$ on slack $y$ for separation constraints & $0.5$ \\
    Initial trust-region radius $\Delta^{(0)}$ on slack $y$ for relative Sun phase angle constraints & $5$ \\
    Initial augmented Lagrangian penalty weight $w^{(0)}$ & $100$ \\
    Lower and upper bounds on trust-region radius $\Delta$ & $(10^{-8}, 10)$ \\
    Optimality tolerance $\epsilon_{\rm opt}$ & $10^{-3}$ \\
    Feasibility tolerance $\epsilon_{\rm feas}$ & $10^{-8}$ \\
    \hline
\end{tabular}
\label{tab:scp_hyperparams}
\end{table}

% \subsection{Demonstration of Multi-Vehicle Optimal Control Problem Solution}
% \label{sec:numexp_demo_mvocp}
% We first present demonstrations of MVOCP solutions. 
% \red{TBD whether to keep}

\subsection{\rev{Analysis over Single Instance of the MVOCP}}
\label{sec:results_single_instance}
\rev{We consider the solution of individual MVOCP instances corresponding to experiments I through IV in Table~\ref{tab:summary_mc_experiments} with $M=2$.
Table~\ref{tab:single_instance_performance}~shows the performance of the SCP algorithm solving a single instance from each experiment case.
All instances are solved on a single thread and process.
A larger number of spacecraft, as well as the introduction of CT path constraints, renders the nonconvex constraints of the MVOCP harder to resolve, requiring the SCP to take longer iterations.
Nevertheless, the MVOCP's solve times are well within what a ground-based compute resource can handle for online implementation in the station-keeping problem.}

\begin{figure}[t]
     \centering
     \begin{subfigure}[b]{0.49\textwidth}
         \centering
         \includegraphics[width=\textwidth]{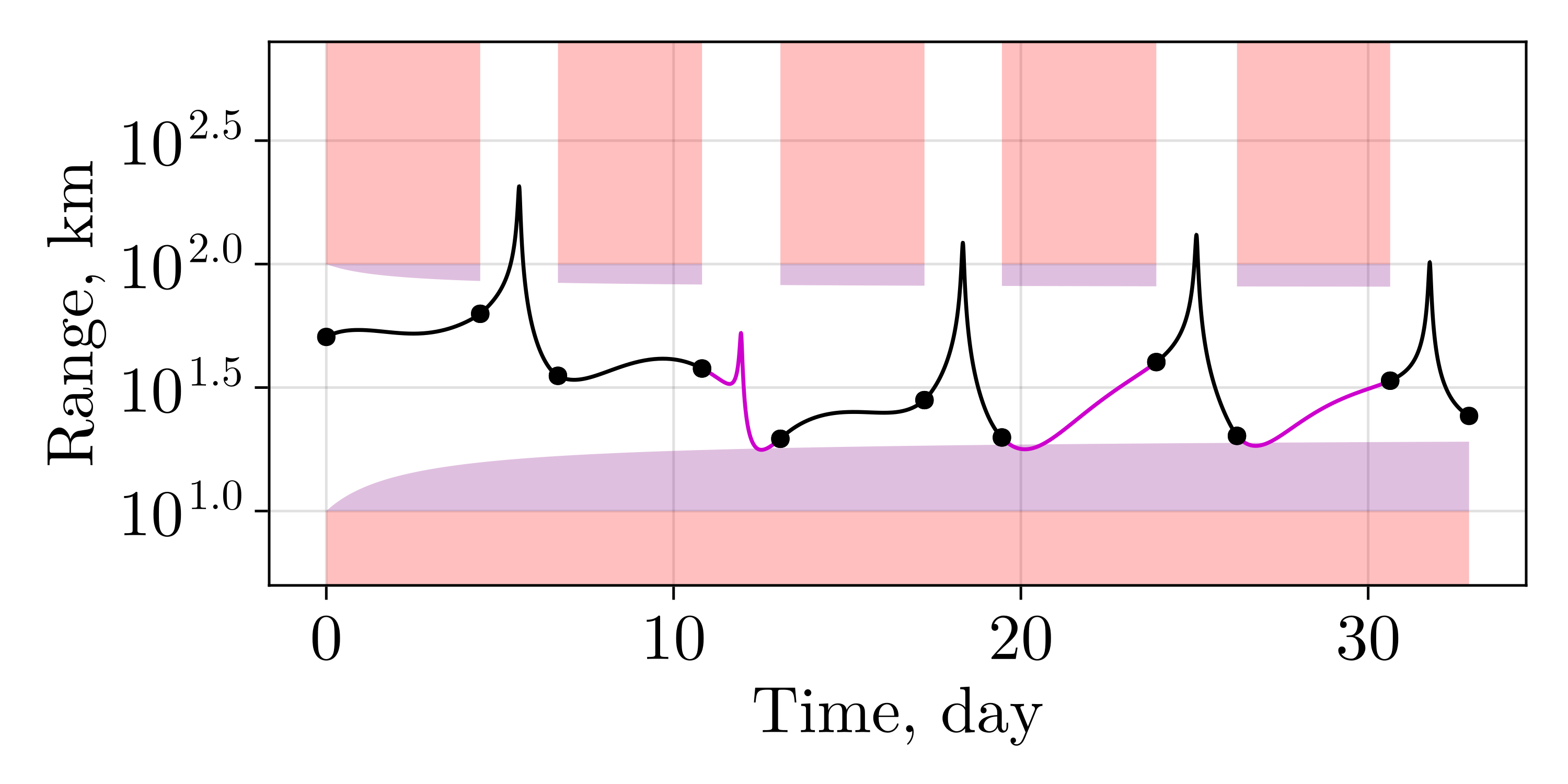}
        \caption{Inter-spacecraft range history}
         \label{fig:instance_isoperimetric_phase_X2_N2_range}
     \end{subfigure}
     \hfill %\\
     \begin{subfigure}[b]{0.49\textwidth}
         \centering
         \includegraphics[width=\textwidth]{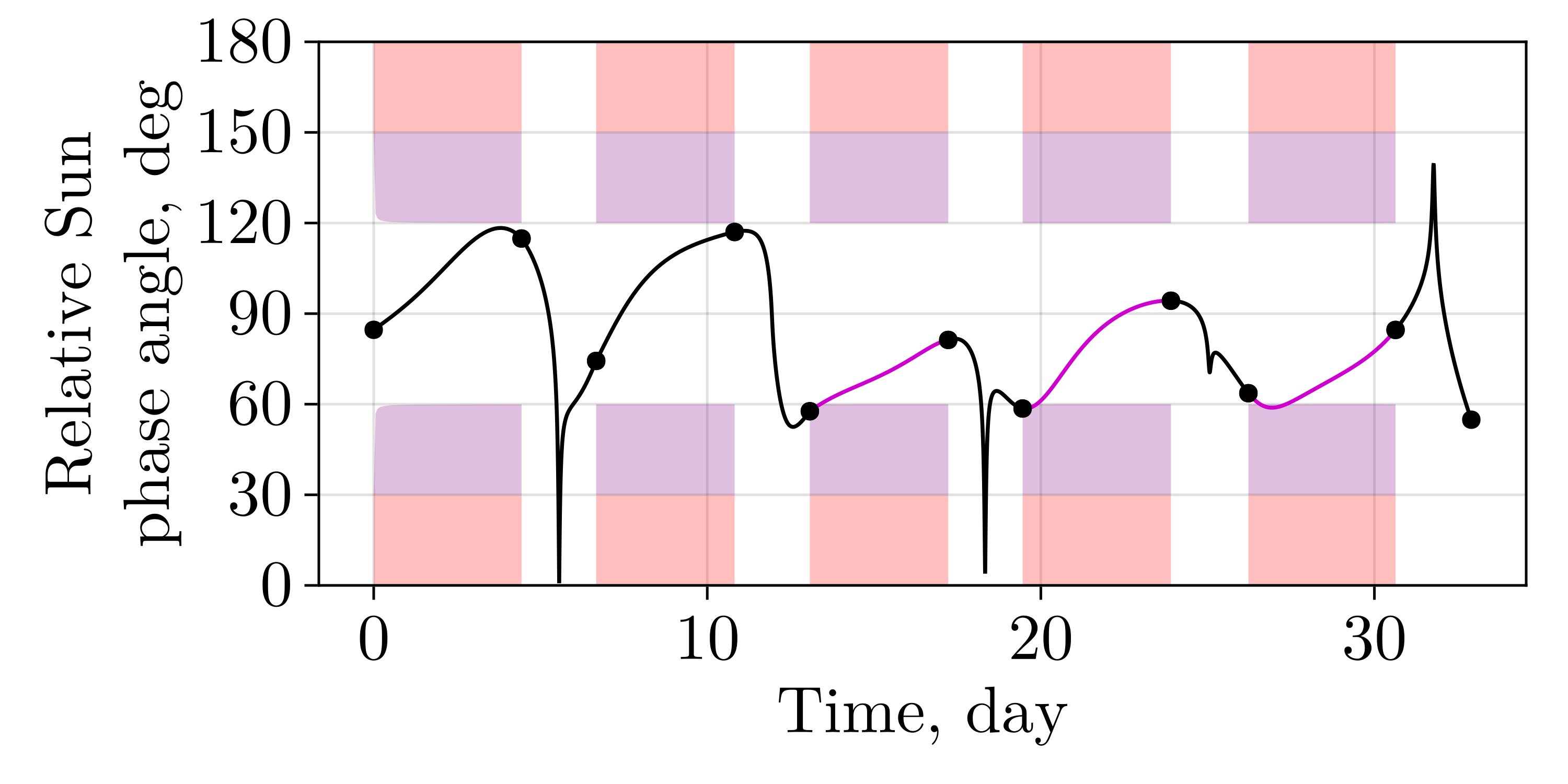}
        \caption{Relative Sun phase angle history}
         \label{fig:instance_isoperimetric_phase_X2_N2_phase}
     \end{subfigure}
    \caption{MVOCP solution with continuous-time path constraints}
    \label{fig:single_instance_isoperimetric}
\end{figure}

\begin{figure}[t]
     \centering
     \begin{subfigure}[b]{0.49\textwidth}
         \centering
         \includegraphics[width=\textwidth]{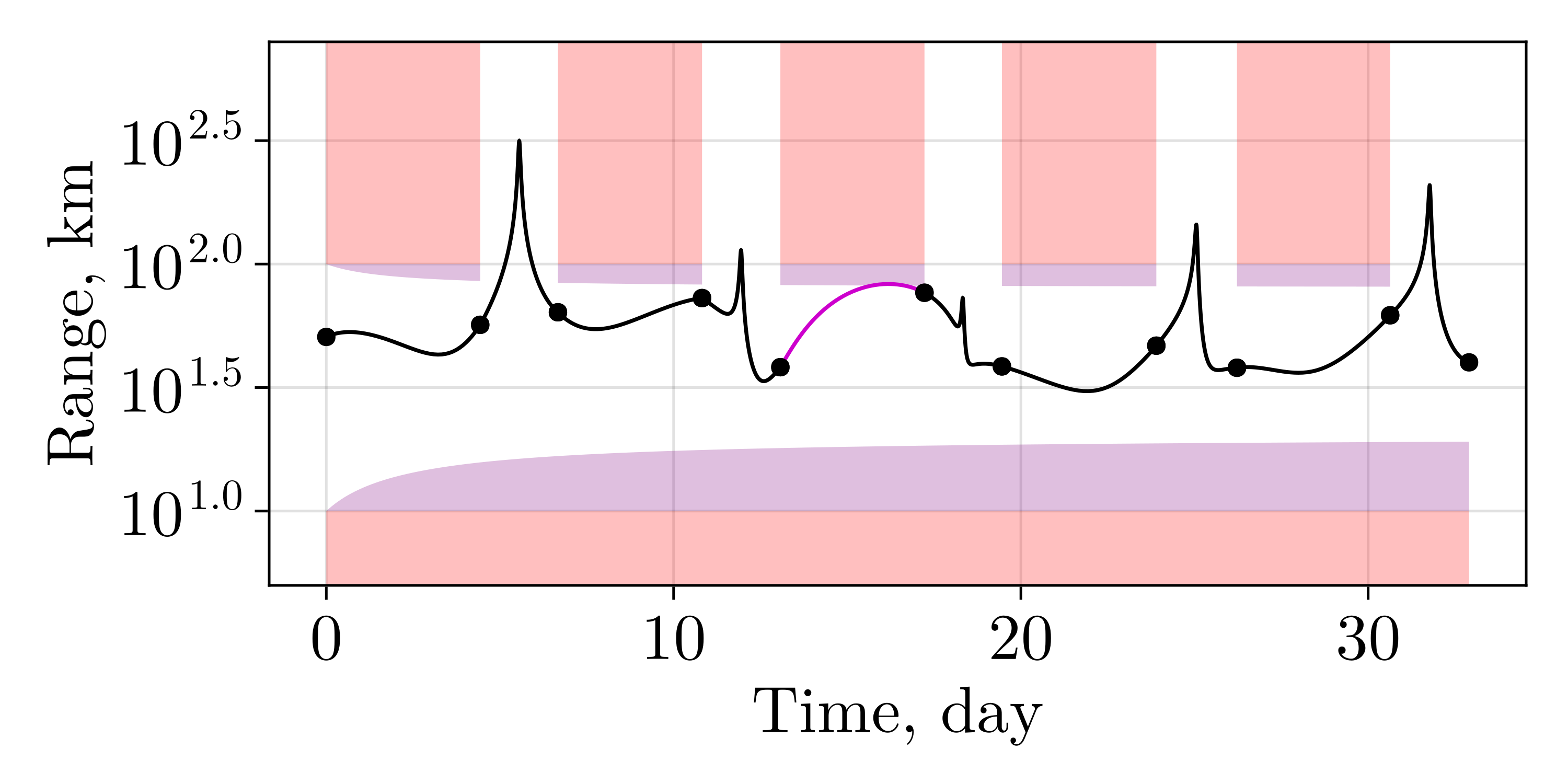}
        \caption{Inter-spacecraft range history}
         \label{fig:instance_nodes_phase_X2_N2_range}
     \end{subfigure}
     \hfill %\\
     \begin{subfigure}[b]{0.49\textwidth}
         \centering
         \includegraphics[width=\textwidth]{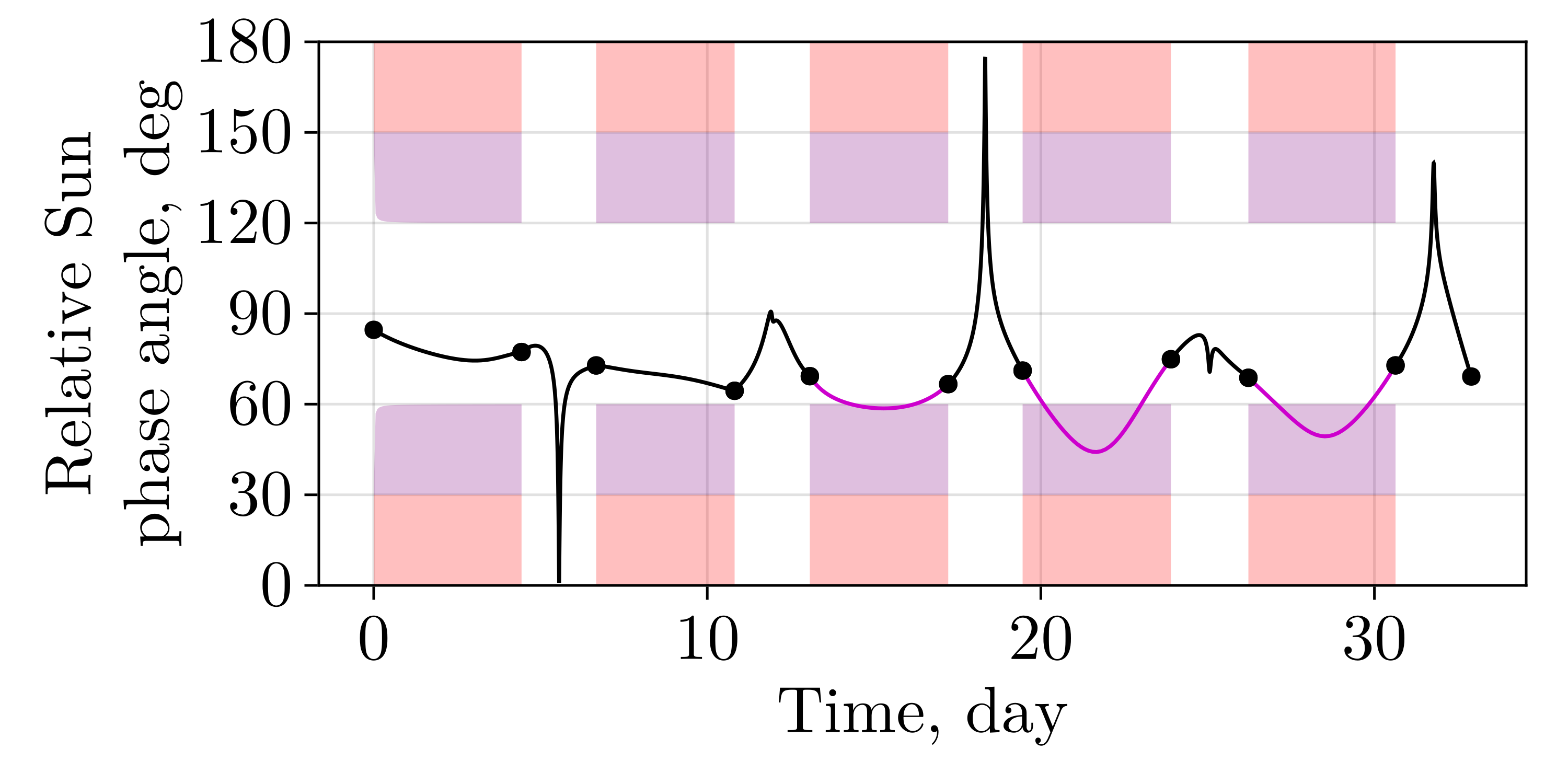}
        \caption{Relative Sun phase angle history}
         \label{fig:instance_nodes_phase_X2_N2_phase}
     \end{subfigure}
    \caption{MVOCP solution with path constraints enforced at nodes}
    \label{fig:single_instance_nodes}
\end{figure}

\rev{
Using the CT path constraints, as in experiments I and III, results in lower control cost than using discretized path constraints, as in experiments II and IV.
We visualize this behavior for the more complex experiments III and IV by looking at the time histories of the inter-spacecraft range $\|\rbold_1(t) - \rbold_2(t)\|_2$ and relative Sun phase angle $\phi_{12}(t)$ along the optimized solution, shown in Figures~\ref{fig:single_instance_isoperimetric} and~\ref{fig:single_instance_nodes}.
Because the MVOCP is nonconvex, the converged solutions in either problem are local optima; we find that the CT path constraints, while being more complicated to work out, provide information that steers the trajectory to a better local optimum than the node-constrained case.
}

\rev{
In Figures~\ref{fig:single_instance_isoperimetric} and~\ref{fig:single_instance_nodes}, the magenta segments correspond to cases with numerical violation of the tightened path constraints, checked a posteriori along the optimized solutions.
Because of the introduction of the LICQ relaxation~\eqref{eq:ct_slack_equality_relaxed}, we observe violations, albeit small, of the CT path constraints, even with the isoperimetric reformulation.
The largest violations are $0.2$~\SI{}{km} on the range and $2.3$~\SI{}{deg} on the Sun phase angle, which are within the expected worst-case violation due to LICQ relaxation.
Meanwhile, enforcing the path constraints on the nodes alone imposes no check on inter-sample violations; in this particular instance, the solution has a $1.2$~\SI{}{km} violation on the range and $15.8$~\SI{}{deg} on the Sun phase angle.
}

% single instance SCP solve
\begin{table}[]
    \centering
    \caption{SCP Performance solving individual instances of the MVOCP with $M=2$}
    \label{tab:single_instance_performance}
    \begin{tabular}{@{}llll@{}}
    \toprule
    Exp. & CPU time,~\SI{}{s} & Iterations & Objective,~\SI{}{m/s} \\ \midrule
    I          & 57.33  & 54 & $0.1940$  \\
    II         & 29.94  & 16 & $0.4640$  \\
    III        & 101.72 & 71 & $0.2248$  \\
    IV         & 28.54  & 15 & $0.7735$  \\ \bottomrule
    \end{tabular}
\end{table}

% Single instance plots for isoperimetric
% single instance plots for nodes only

\subsection{Monte Carlo Experiments with Two-Spacecraft Formation}
\rev{The Monte Carlo experiments demonstrate the performance of the MVOCP solved recursively within an MPC setting.}
The mean, standard deviation (Stdv.), and $3$-$\sigma$ of the cumulative cost of each spacecraft in the formation are summarized in Table~\ref{tab:cost_statistics}.
\rev{The ``MVOCP feasible'' column reports the number of cases where the MVOCP admits a feasible solution, and the ``CT feasible'' column reports the cases with CT path constraint satisfaction.
The columns $\mathrm{err}_{\Delta r}$ and $\mathrm{err}_{\phi}$ report the largest CT path constraint violation, if any.
As expected, a large number of cases from experiments II and IV, enforcing constraints at the nodes alone, suffer from inter-sample violations.
Since the recursion is continued as long as the MVOCP admits a feasible solution, the control cost statistics are computed based on all 100 Monte Carlo samples for all experiments.}
The equivalent yearly cost is approximately $\times 5.6$ the values reported in this Table.
\rev{A one-to-one comparison with yearly costs on the order of $~10^1$ \SI{}{cm/s} in~\cite{Foss2025-sm} is inapplicable due to 1) the omission of relative navigation information in our simulations, and 2) the difference in path constraints, where~\cite{Foss2025-sm} only enforces} a lower bound on the minimum separation, i.e., only~\eqref{eq:MVOCP_minsep}.
Despite the adversarial navigation conditions assumed in our work, we successfully demonstrate station-keeping for the formation with $3$-$\sigma$ cost in the order of $70 \times (365.5 / 65.5) \approx 390$~\SI{}{cm/s} to $100 \times (365.5 / 65.5) \approx 560$~\SI{}{cm/s} per spacecraft per year while enforcing path constraints continuously, as demonstrated by experiments I and III.
Further comparison of costs between experiments is discussed in subsequent subsections.

% Table of costs
\begin{table}[]
\centering
\footnotesize
\caption{MPC performance summary over 10 revolutions ($\sim 65.5$ \SI{}{days}) for 2-spacecraft formation}
\begin{tabular}{@{}lllllllllll@{}}
\toprule
\multirow{2}{*}{Exp.} &
\multirow{2}{*}{\begin{tabular}[c]{@{}l@{}}MVOCP\\feasible\end{tabular}} &
\multirow{2}{*}{\begin{tabular}[c]{@{}l@{}}CT\\feasible\end{tabular}} &
\multirow{2}{*}{\begin{tabular}[c]{@{}l@{}}$\mathrm{err}_{\Delta r}$,\\\SI{}{km}\end{tabular}}  & 
\multirow{2}{*}{\begin{tabular}[c]{@{}l@{}}$\mathrm{err}_{\phi}$,\\\SI{}{deg}\end{tabular}}  & 
\multicolumn{2}{c}{Spacecraft 1} & 
\multicolumn{2}{c}{Spacecraft 2} & 
\multicolumn{2}{c}{Formation Total} 
\\ 
& & & & 
  & Mean,~\SI{}{cm/s}& $3$-$\sigma$,~\SI{}{cm/s}
  & Mean,~\SI{}{cm/s}& $3$-$\sigma$,~\SI{}{cm/s}
  & Mean,~\SI{}{cm/s}& $3$-$\sigma$,~\SI{}{cm/s}
\\ \midrule
I &  100/100 &  100/100 & 0.00 & - & 18.53 & 36.71 & 19.97 & 38.76 & 38.50 & 67.35 \\
II &  100/100 &  (31/100) & 25.51 & - & 51.61 & 82.56 & 50.82 & 80.99 & 102.43 & 148.80 \\
III &  100/100 &  93/100 & 1.29 & 0.00 & 29.60 & 50.55 & 31.13 & 55.67 & 60.73 & 96.24 \\
IV &  98/100 &  (2/100) & 2.70 & 29.57 & 91.83 & 170.66 & 89.48 & 193.84 & 181.31 & 206.83 \\
\bottomrule
\end{tabular}
\label{tab:cost_statistics}
\end{table}

% -------------------------------------------------------------------- %
\subsubsection{Monte Carlo Results with Relative Separation Path Constraints}
\label{sec:mc_sep}
\begin{figure}[t]
     \centering
     \begin{subfigure}[b]{0.49\textwidth}
         \centering
         \includegraphics[width=\textwidth]{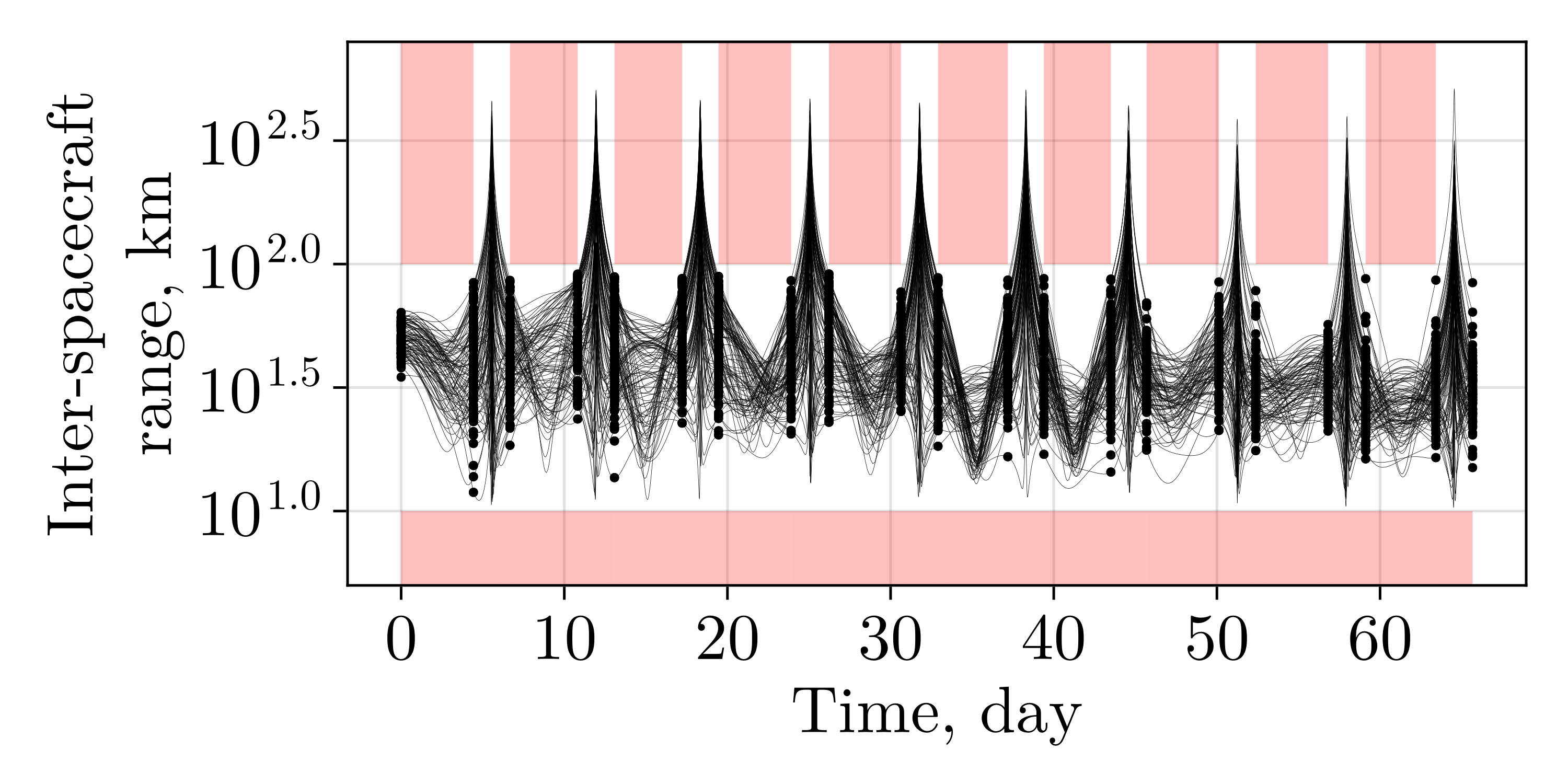}
        \caption{Inter-spacecraft range history}
         \label{fig:isoperimetric_sep_constrained_range_history_n2}
     \end{subfigure}
     \hfill %\\
     \begin{subfigure}[b]{0.49\textwidth}
         \centering
         \includegraphics[width=\textwidth]{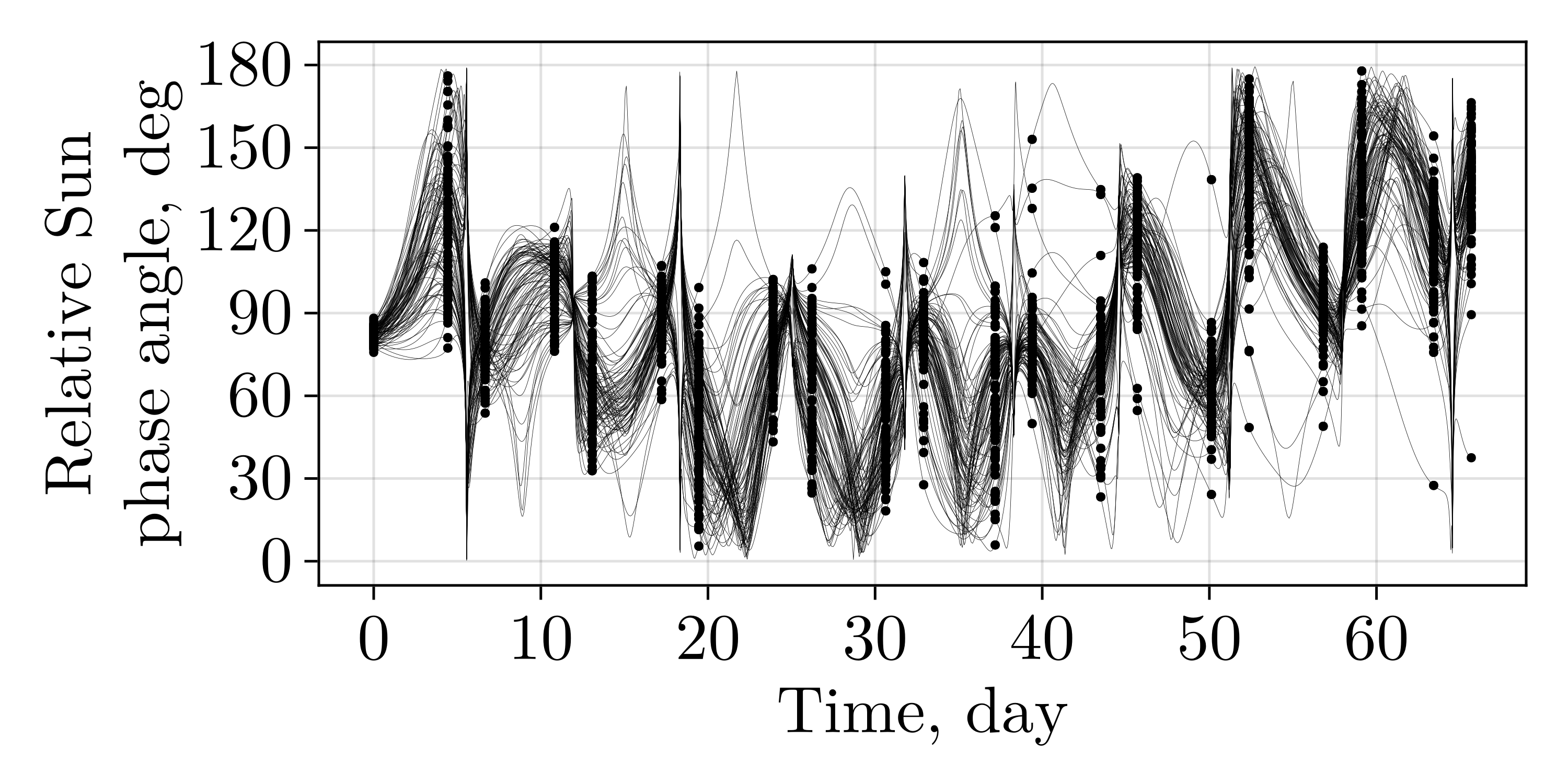}
        \caption{Relative Sun phase angle history (unconstrained)}
         \label{fig:isoperimetric_sep_constrained_phase_history_n2}
     \end{subfigure}
    \caption{Experiment I ($M=2$): inter-spacecraft range continuously enforced}
    % inter-spacecraft range and relative Sun phase angle histories along 100 Monte Carlo samples with continuously enforced path constraints}
    \label{fig:isoperimetric_sep_constrained}
\end{figure}

\begin{figure}[t]
     \centering
     \begin{subfigure}[b]{0.49\textwidth}
         \centering
         \includegraphics[width=\textwidth]{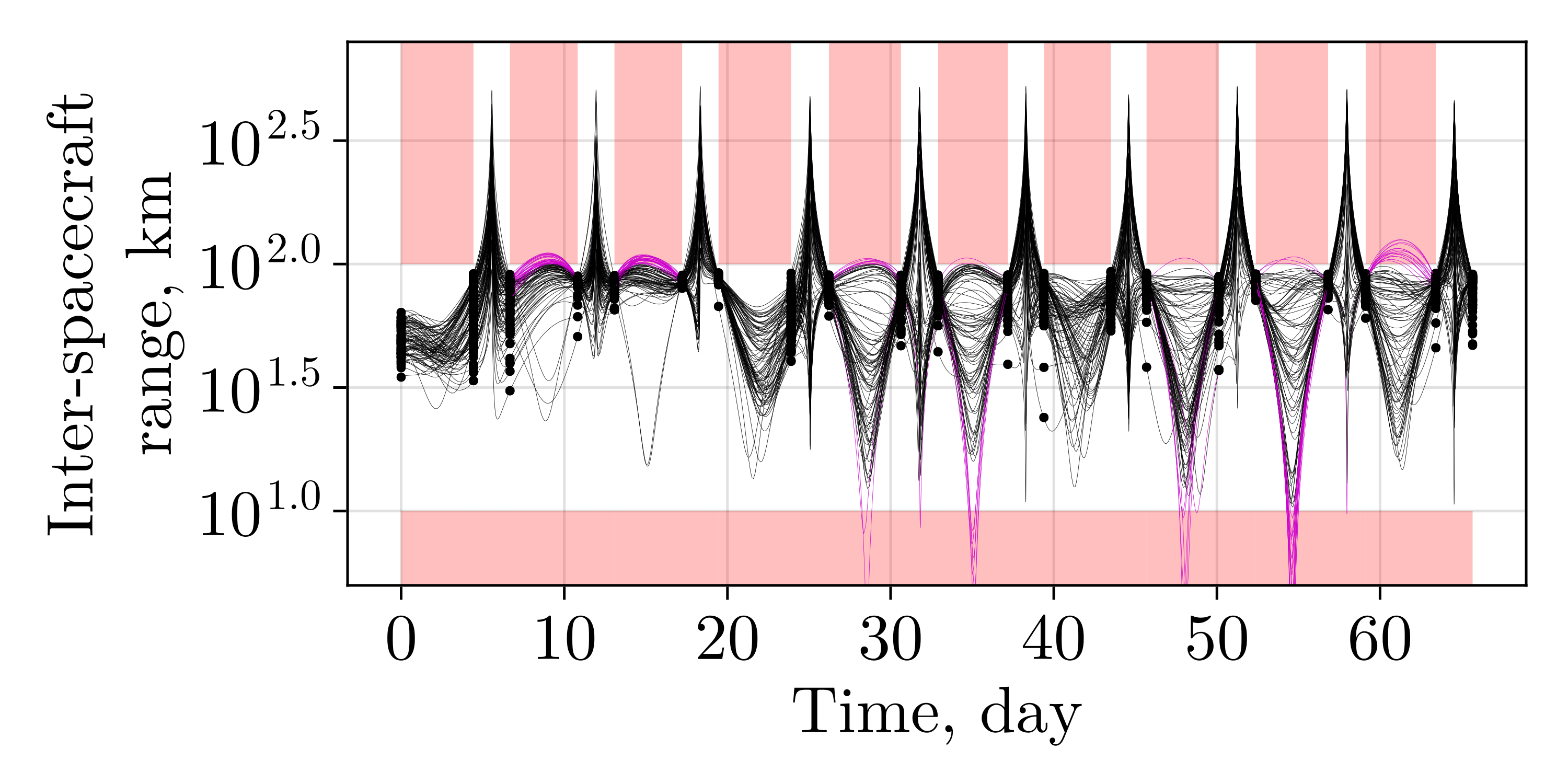}
        \caption{Inter-spacecraft range history}
         \label{fig:nodes_sep_constrained_range_history_n2}
     \end{subfigure}
     \hfill %\\
     \begin{subfigure}[b]{0.49\textwidth}
         \centering
         \includegraphics[width=\textwidth]{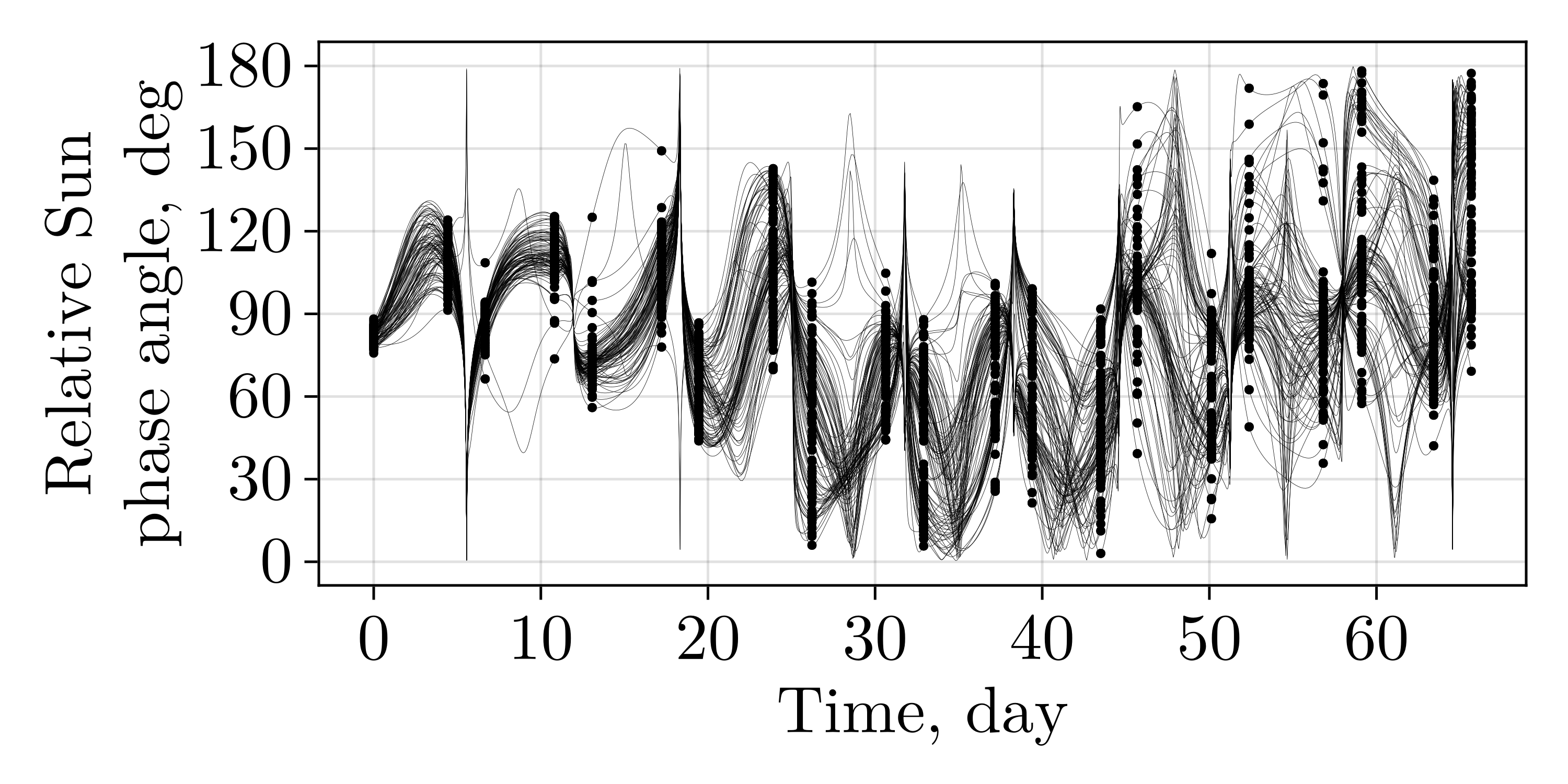}
        \caption{Relative Sun phase angle history (unconstrained)}
         \label{fig:nodes_sep_constrained_phase_history_n2}
     \end{subfigure}
    \caption{Experiment II ($M=2$): inter-spacecraft range enforced at nodes}
    % inter-spacecraft range and relative Sun phase angle histories along 100 Monte Carlo samples with path constraints enforced at nodes}
    \label{fig:node_sep_constrained}
\end{figure}

We first consider experiments I and II, where the relative separation path constraints~\eqref{eq:MVOCP_boundsep} are included.
Figure~\ref{fig:isoperimetric_sep_constrained} and Figure~\ref{fig:node_sep_constrained} show the time histories of $\|\rbold_1(t) - \rbold_2(t)\|_2$ and $\phi_{12}(t)$ from all Monte Carlo samples in experiments I and II, respectively. 
The red regions in the Figures indicate inter-spacecraft ranges that should be avoided according to mission requirements.
Note that while the minimum separation is enforced at all times, the maximum separation is only enforced at apolune segments, as defined in~\eqref{eq:MVOCP_maxsep}.
The traces in magenta indicate segments where the path constraint is violated.

With experiment I, the isoperimetric reformulation allows for the minimum separation constraint to be enforced continuously at all times, and the maximum separation constraint to be enforced continuously at apolune segments, as dictated by mission requirements.
Meanwhile, enforcing the path constraints simply at control nodes alone, as is done in experiment II, results in frequent violation of the minimum separation requirement, both during perilune and apolune segments.
Overcoming the inter-sample violation by adding nodes is not obvious, as the exact number and placement of nodes required to ensure continuous satisfaction of path constraints is not obvious ahead of time and cannot be guaranteed.
% Also, we observe that with experiment I, enforcing both minimum and maximum separations continuously results in larger variations in inter-spacecraft range at the control nodes, while with experiment II, the inter-spacecraft range lies within a much narrower variation.

With no relative Sun phase angle constraint in place, $\phi_{12}(t)$ is found to vary freely across the $\sim 65.5$ \SI{}{days} duration, as shown in Figures~\ref{fig:isoperimetric_sep_constrained_phase_history_n2} and~\ref{fig:nodes_sep_constrained_phase_history_n2}. 
Across the Monte Carlo samples, we find a common structure in the relative Sun phase angle history, dictated by the geometry of the NRHO in the Sun-Moon rotating frame, as shown in Figure~\ref{fig:NRHO_traj_SMrot}.

\begin{figure}[t]
     \centering
     \begin{subfigure}[b]{0.49\textwidth}
         \centering
         \includegraphics[width=\textwidth]{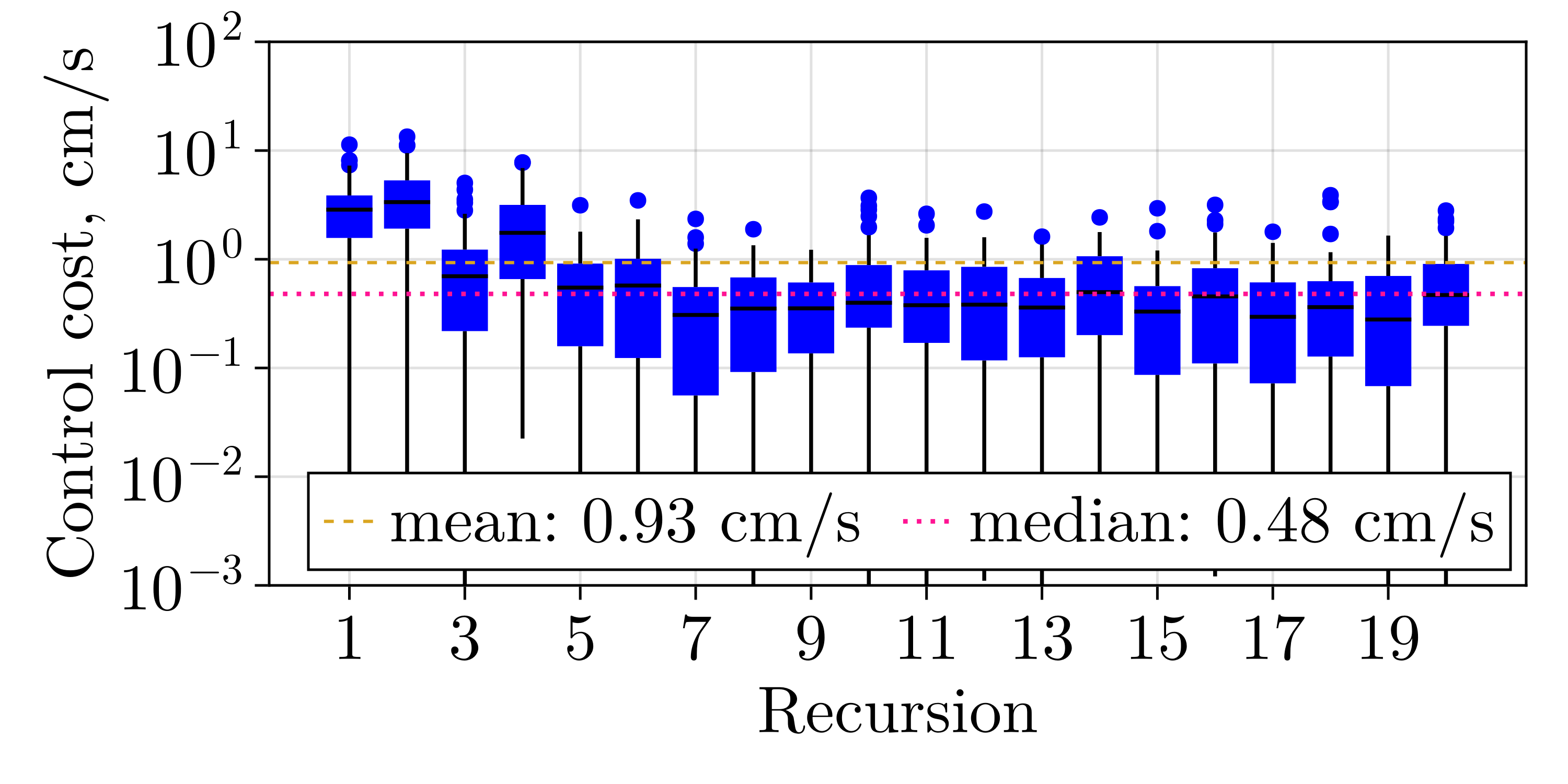}
        \caption{Spacecraft 1}
         \label{fig:control_boxplot_sc1_sep_isoperimetric_constrained_X2}
     \end{subfigure}
     \hfill %\\
     \begin{subfigure}[b]{0.49\textwidth}
         \centering
         \includegraphics[width=\textwidth]{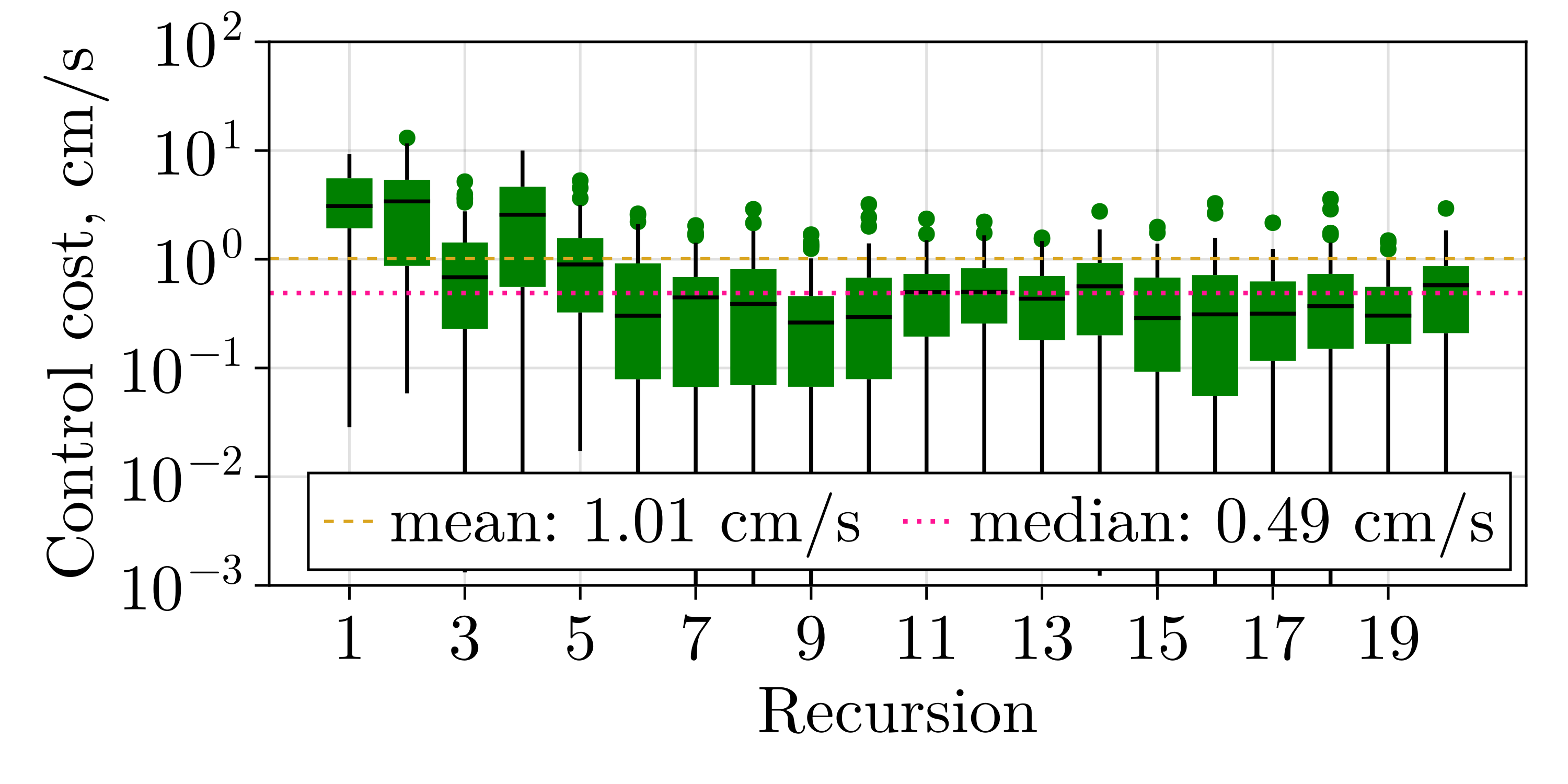}
        \caption{Spacecraft 2}
         \label{fig:control_boxplot_sc2_sep_isoperimetric_constrained_X2}
     \end{subfigure}
    \caption{Experiment I ($M=2$): distribution of control cost per recursion}
    \label{fig:control_boxplot_sep_isoperimetric_constrained_X2}
\end{figure}

\begin{figure}[t]
     \centering
     \begin{subfigure}[b]{0.49\textwidth}
         \centering
         \includegraphics[width=\textwidth]{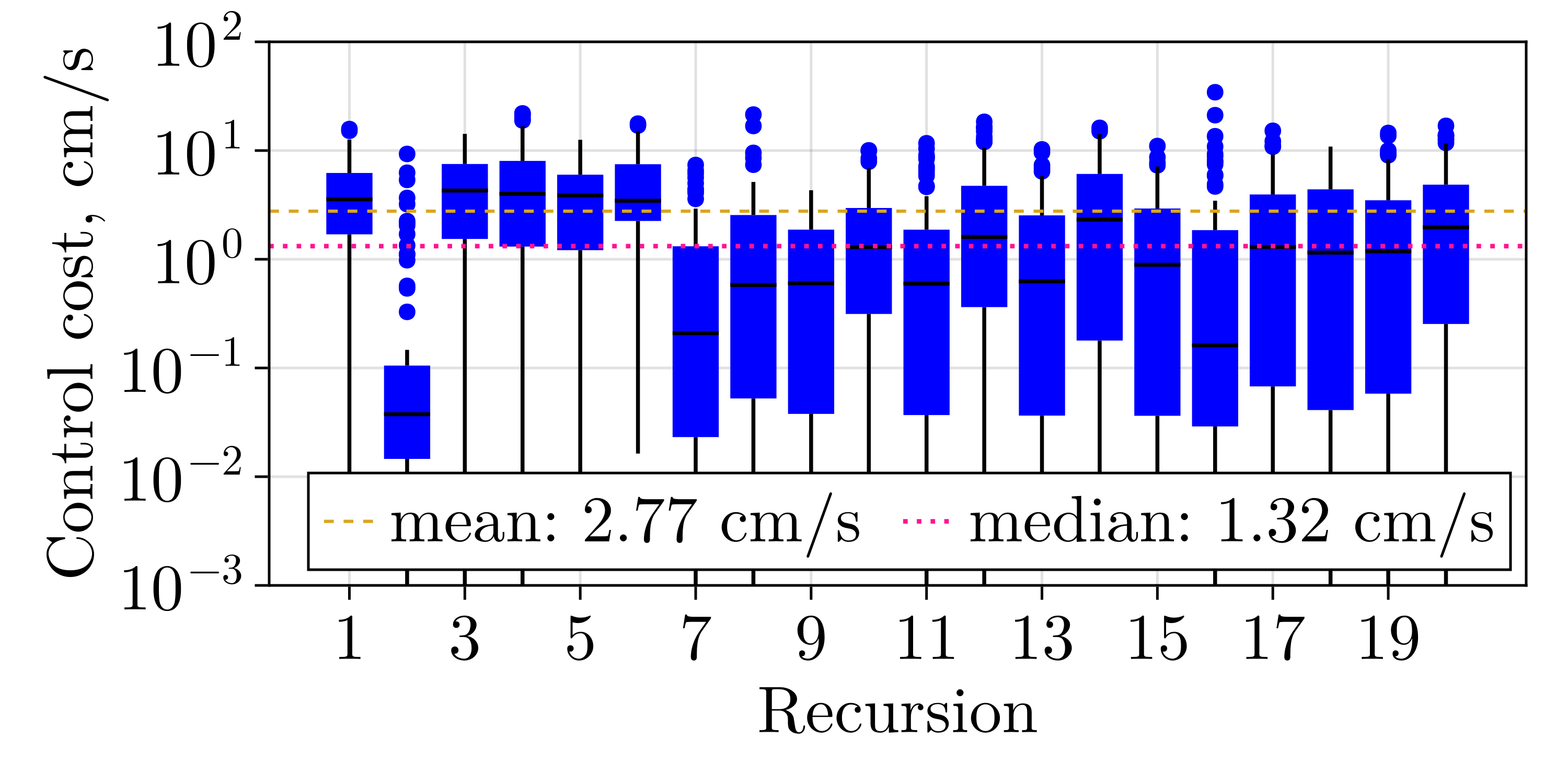}
        \caption{Spacecraft 1}
         \label{fig:control_boxplot_sc1_sep_nodes_constrained_X2}
     \end{subfigure}
     \hfill %\\
     \begin{subfigure}[b]{0.49\textwidth}
         \centering
         \includegraphics[width=\textwidth]{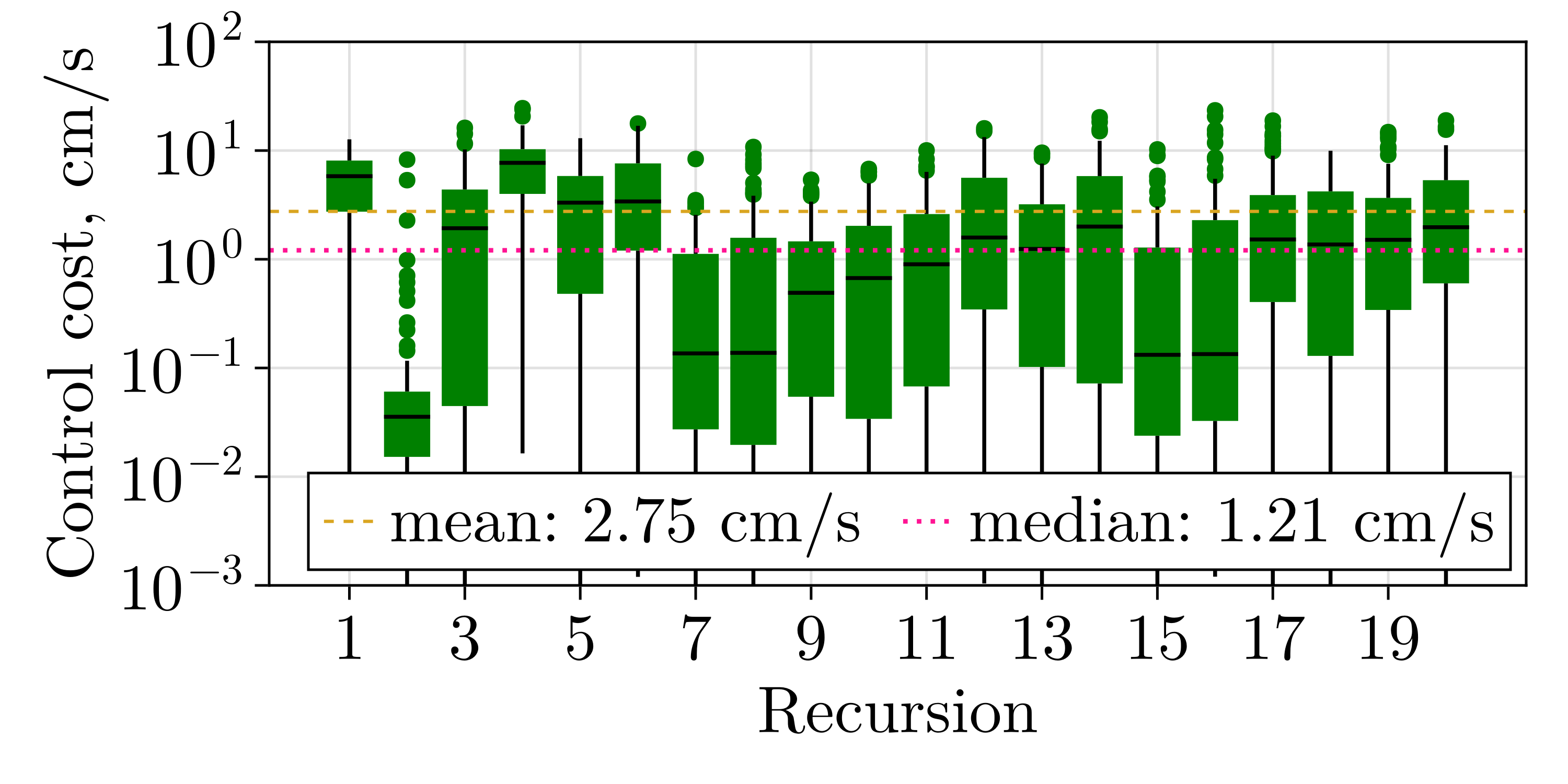}
        \caption{Spacecraft 2}
         \label{fig:control_boxplot_sc2_sep_nodes_constrained_X2}
     \end{subfigure}
    \caption{Experiment II ($M=2$): distribution of control cost per recursion}
    \label{fig:control_boxplot_sep_nodes_constrained_X2}
\end{figure}

While the MVOCP from experiment II is a more relaxed problem than that of experiment I, we find the cumulative mean and $3$-$\sigma$ costs of II to be $\times 1.6$ and $\times 2.3$ higher than I, respectively.
The distribution of the executed control magnitude by each spacecraft across the 20 recursions is shown in Figure~\ref{fig:control_boxplot_sep_isoperimetric_constrained_X2} and Figure~\ref{fig:control_boxplot_sep_nodes_constrained_X2} for experiments I and II, respectively; the higher cumulative cost in experiment II is a direct result of higher cost at each recursion as well.
Two effects, namely 1) the nonconvexity\rev{, as observed in Section~\ref{sec:results_single_instance},} and 2) the finite-horizon approximation of the MVOCP, are behind this cost difference.

With respect to the nonconvexity, using the same initial guess may lead to convergence to different local minima depending on the formulation; even if the solution of a given MVOCP instance from experiment I results in a lower objective value, the MVOCP instance from experiment II may converge to a separate local minimum solution with a higher objective.
Evidently, the inter-spacecraft range resulting from the recursive control in experiment II is tighter than in experiment I, which may well require a higher cumulative cost.

The finite-horizon approximation means that while the objective from the formulation in experiment II may be lower than experiment I at some time $t_k$, the resulting trajectory could lead to subsequent MVOCP instances with higher costs at later times $t_{k+1}$, $t_{k+2}$, etc., particularly in the presence of uncertainties.
While both I and II consider a finite-horizon approximation, enforcing continuous path constraints as in I results in a stronger incentive for steering the spacecraft geometry into a continuously feasible formation orientation (with minimal actuation) that may last for some finite time even beyond the prediction horizon, resulting in reserved control cost at subsequent recursions.
Meanwhile, with II, a solution that ``just'' satisfies the path constraints at the control node may provide insufficient information, and therefore incentive, to steer the spacecraft into such a geometry.
    
% \begin{enumerate}
%     \item With respect to the nonconvexity, using the same initial guess may lead to convergence to different local minima depending on the formulation; even if the solution of a given MVOCP instance from experiment I results in a lower objective value, the MVOCP instance from experiment II may converge to a separate local minimum solution with a higher objective.
%     Evidently, the inter-spacecraft range resulting from the recursive control in experiment II is tighter than in experiment I, which may well require a higher cumulative cost.
%     %
%     \item The finite-horizon approximation means that while the objective from the formulation in experiment II may be lower than experiment I at some time $t_k$, the resulting trajectory could lead to subsequent MVOCP instances with higher costs at later times $t_{k+1}$, $t_{k+2}$, etc., particularly in the presence of uncertainties.
%     While both I and II consider a finite-horizon approximation, enforcing continuous path constraints as in I results in a stronger incentive for steering the spacecraft geometry into a continuously feasible formation orientation (with minimal actuation) that may last for some finite time even beyond the prediction horizon, resulting in reserved control cost at subsequent recursions.
%     Meanwhile, with II, a solution that ``just'' satisfies the path constraints at the control node may provide insufficient information, and therefore incentive, to steer the spacecraft into such a geometry.
% \end{enumerate}

% phase constrained, isoperimetric

% phase constrained, node only

% control: sep constrained, isoperimetric

% control: sep constrained, nodes only

% -------------------------------------------------------------------- %
\subsubsection{Monte Carlo Results with Relative Separation \& Sun Phase Angle Path Constraints}
\label{sec:mc_phase}
\begin{figure}[t]
     \centering
     \begin{subfigure}[b]{0.49\textwidth}
         \centering
         \includegraphics[width=\textwidth]{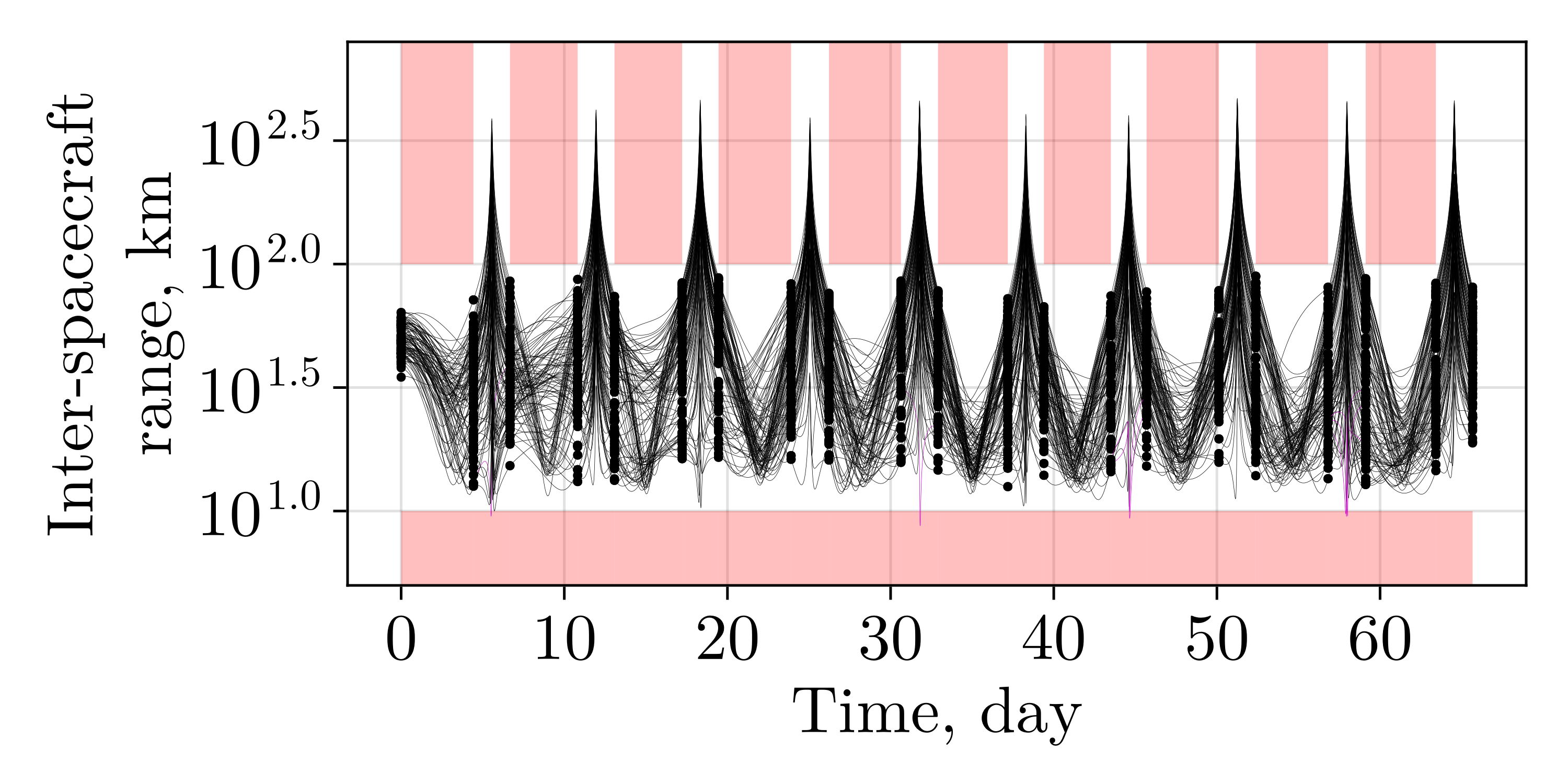}
        \caption{Inter-spacecraft range history}
         \label{fig:isoperimetric_phase_constrained_range_history_n2}
     \end{subfigure}
     \hfill %\\
     \begin{subfigure}[b]{0.49\textwidth}
         \centering
         \includegraphics[width=\textwidth]{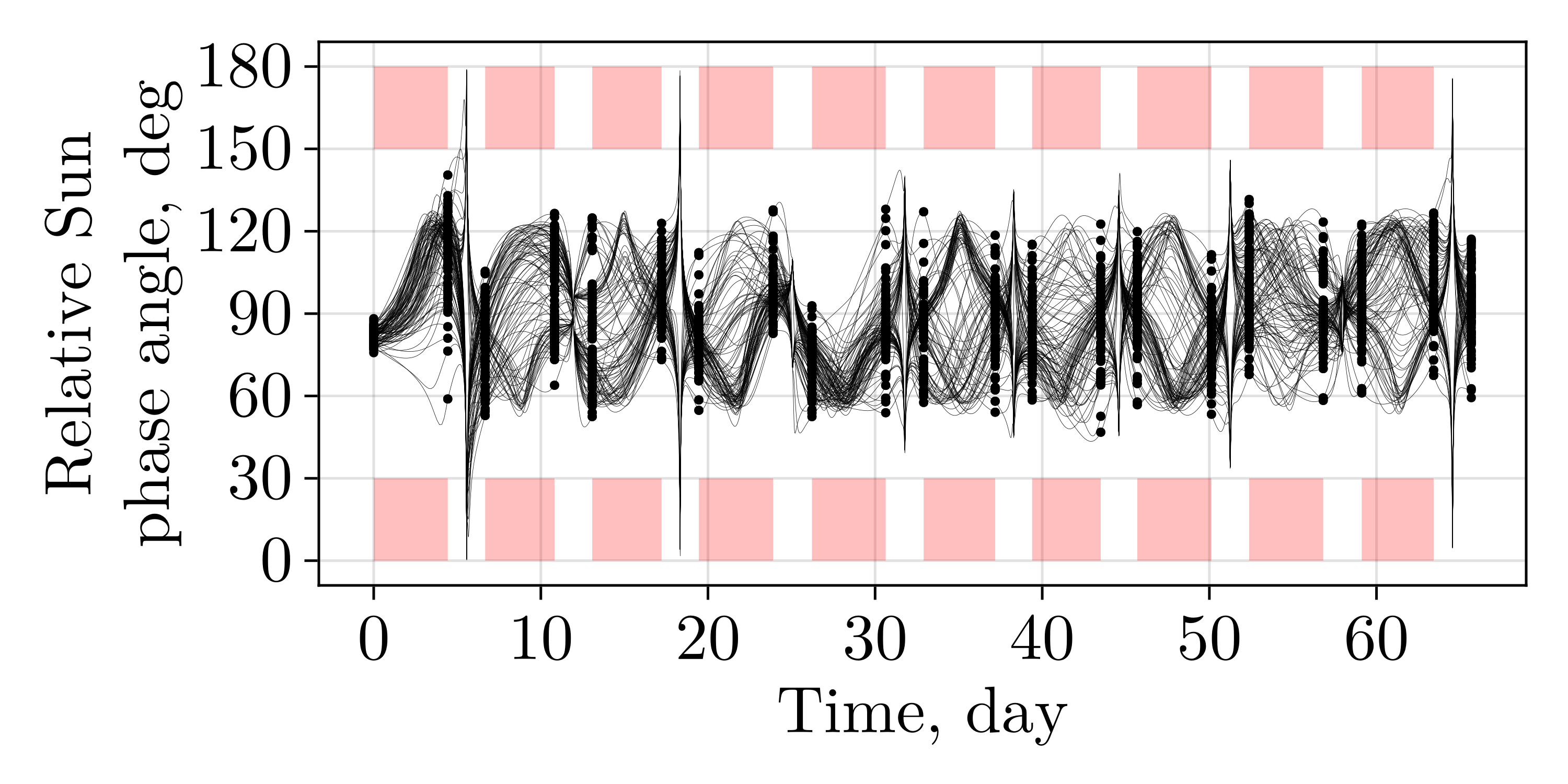}
        \caption{Relative Sun phase angle history}
         \label{fig:isoperimetric_phase_constrained_phase_history_n2}
     \end{subfigure}
    \caption{Experiment III ($M=2$): inter-spacecraft range and relative Sun phase angle continuously enforced}
    %: inter-spacecraft range and relative Sun phase angle histories along 100 Monte Carlo samples with continuously enforced path constraints}
    \label{fig:isoperimetric_phase_constrained}
\end{figure}

\begin{figure}[t]
     \centering
     \begin{subfigure}[b]{0.49\textwidth}
         \centering
         \includegraphics[width=\textwidth]{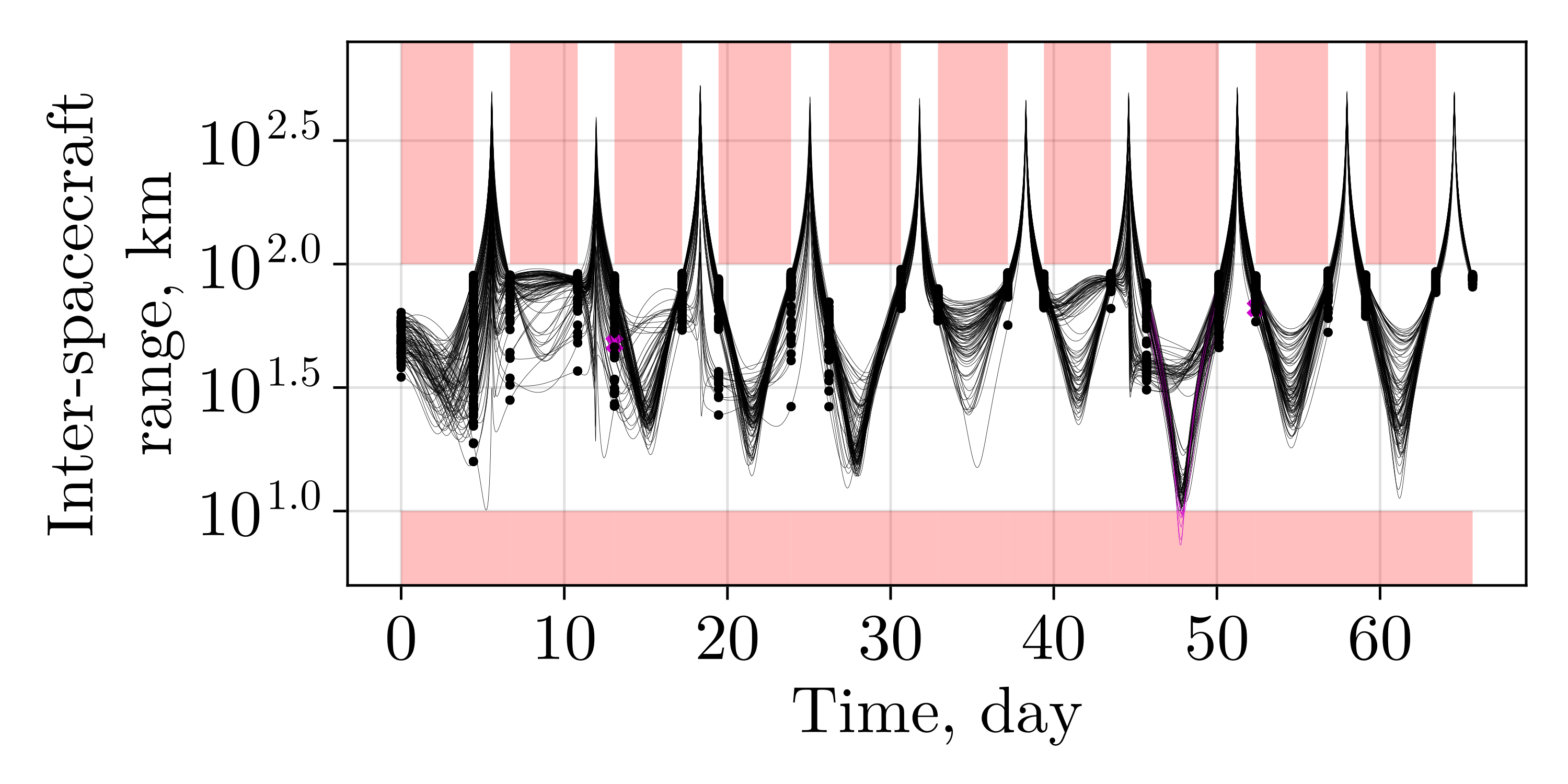}
        \caption{Inter-spacecraft range history}
         \label{fig:nodes_phase_constrained_range_history_n2}
     \end{subfigure}
     \hfill %\\
     \begin{subfigure}[b]{0.49\textwidth}
         \centering
         \includegraphics[width=\textwidth]{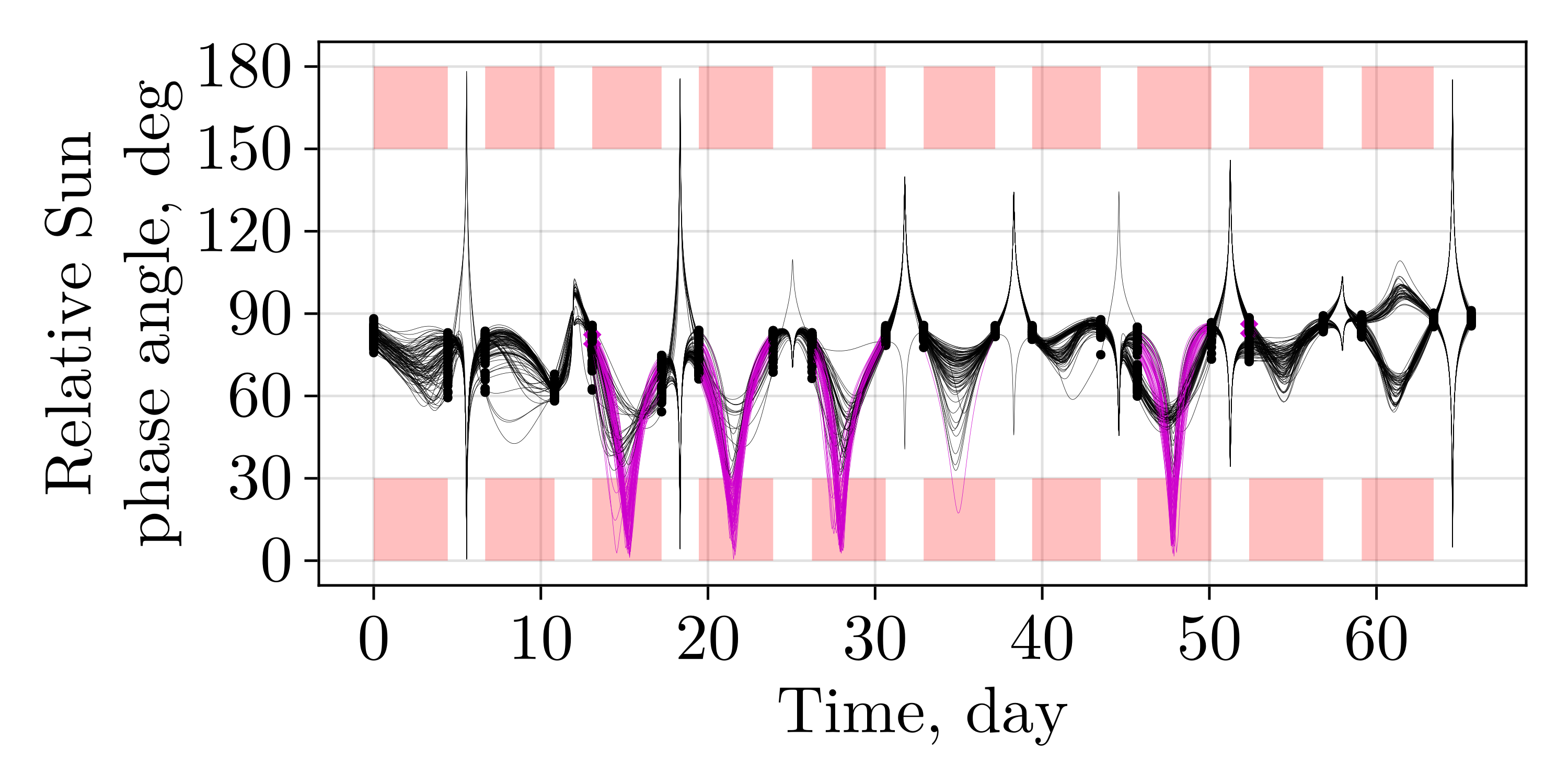}
        \caption{Relative Sun phase angle history}
         \label{fig:nodes_phase_constrained_phase_history_n2}
     \end{subfigure}
    \caption{Experiment IV ($M=2$): inter-spacecraft range and relative Sun phase angle enforced at nodes}
    %: inter-spacecraft range and relative Sun phase angle histories along 100 Monte Carlo samples with path constraints enforced at nodes}
    \label{fig:nodes_phase_constrained}
\end{figure}

We now consider experiments III and IV, where both the relative separation and Sun phase angle constraints are included.
Figure~\ref{fig:isoperimetric_phase_constrained} and Figure~\ref{fig:nodes_phase_constrained} show the time histories of the two path constraints from all Monte Carlo samples in experiments III and IV, respectively.
Here again, the red shaded regions indicate forbidden configurations; while the minimum separation is enforced at all times, the maximum separation as well as the minimum and maximum phase angle are only enforced during apolune segments.

The isoperimetric reformulation in experiment III successfully ensures the \rev{spacecraft is CT constraint feasible within the expected violation due to LICQ relaxation} for all samples. 
Enforcing only at the control nodes, as is done in experiment IV, results in frequent inter-sample violation of the minimum separation as well as the minimum relative Sun phase angle. Note that a violation of the lower bound $\phi_{\min}$ by $\phi_{\alpha\beta}$ means there is a simultaneous violation of the upper bound $\phi_{\max}$ by $\phi_{\beta \alpha}$.

\begin{figure}[t]
     \centering
     \begin{subfigure}[b]{0.49\textwidth}
         \centering
         \includegraphics[width=\textwidth]{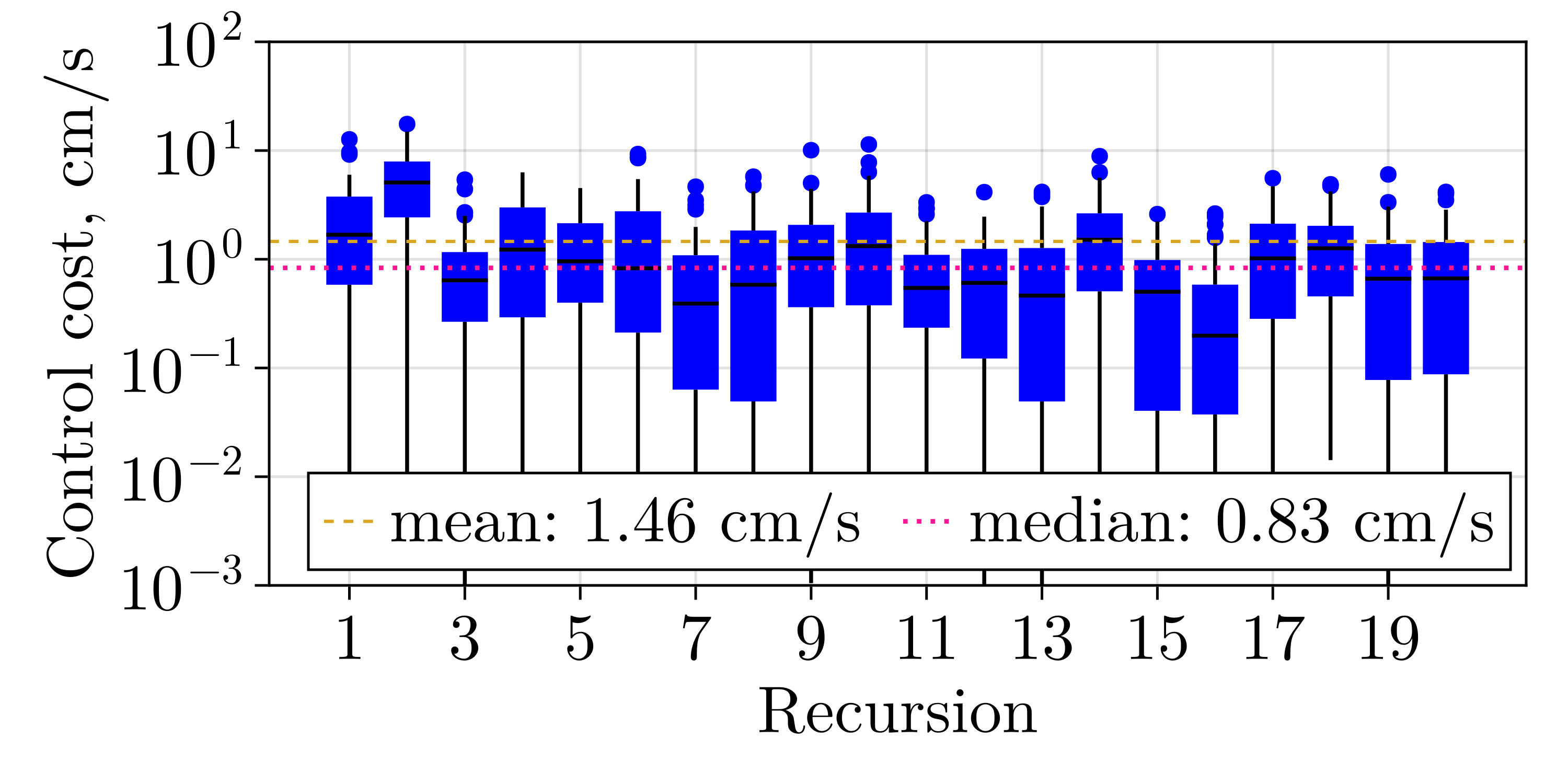}
        \caption{Spacecraft 1}
         \label{fig:control_boxplot_sc1_phase_isoperimetric_constrained_X2}
     \end{subfigure}
     \hfill %\\
     \begin{subfigure}[b]{0.49\textwidth}
         \centering
         \includegraphics[width=\textwidth]{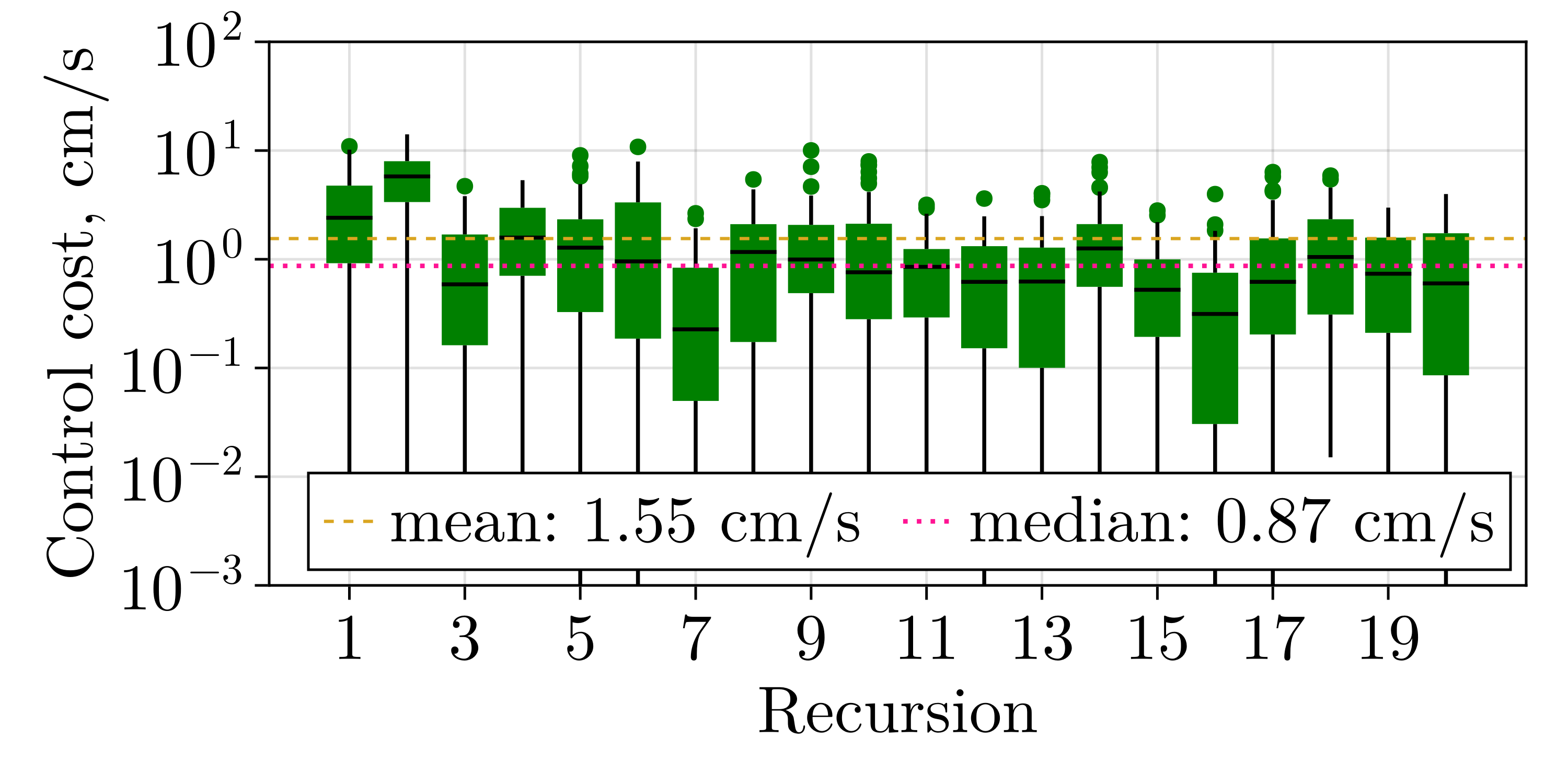}
        \caption{Spacecraft 2}
         \label{fig:control_boxplot_sc2_phase_isoperimetric_constrained_X2}
     \end{subfigure}
    \caption{Experiment III ($M=2$): distribution of control cost per recursion}
    \label{fig:control_boxplot_phase_isoperimetric_constrained_X2}
\end{figure}

\begin{figure}[t]
     \centering
     \begin{subfigure}[b]{0.49\textwidth}
         \centering
         \includegraphics[width=\textwidth]{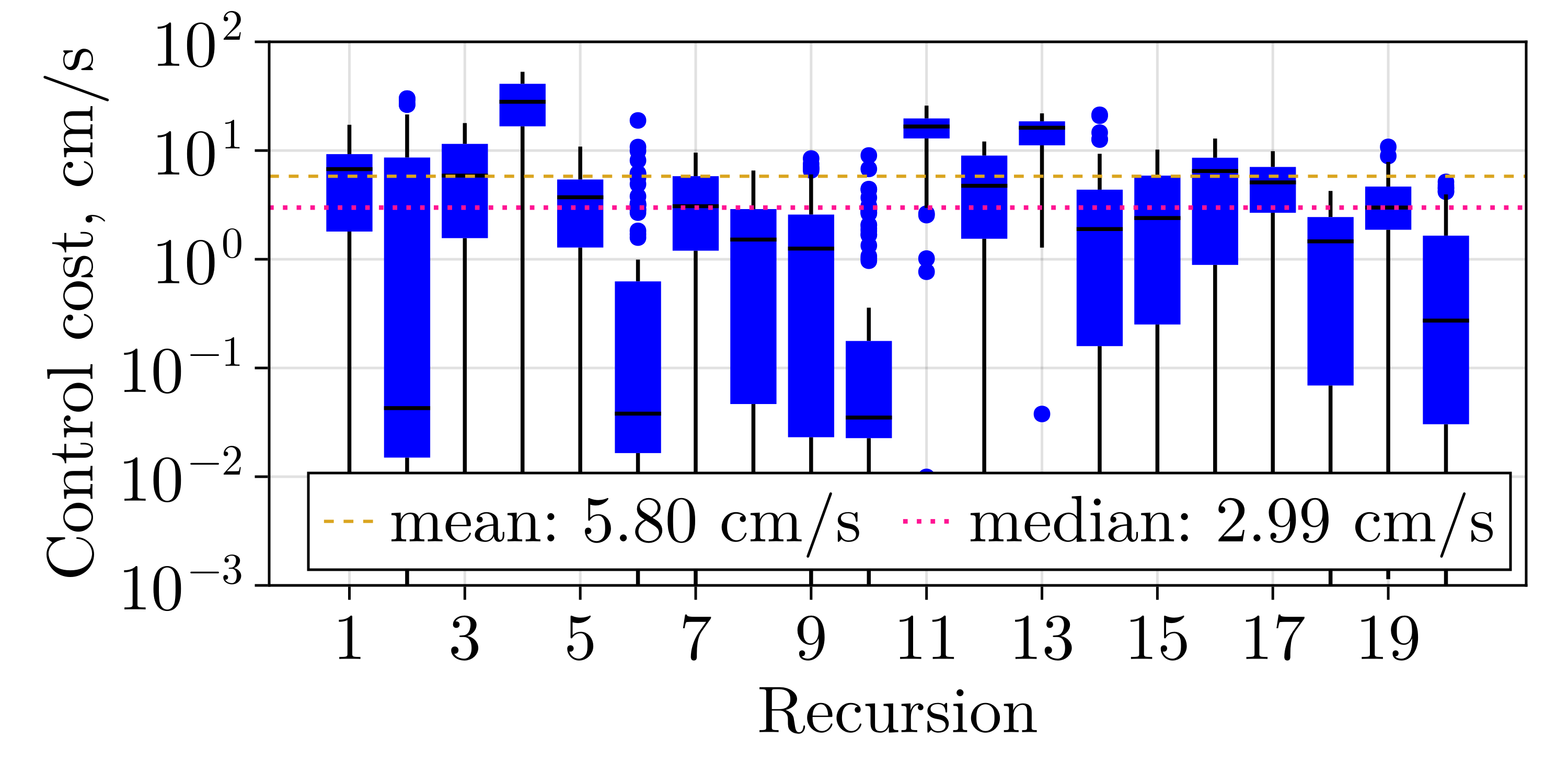}
        \caption{Spacecraft 1}
         \label{fig:control_boxplot_sc1_phase_nodes_constrained_X2}
     \end{subfigure}
     \hfill %\\
     \begin{subfigure}[b]{0.49\textwidth}
         \centering
         \includegraphics[width=\textwidth]{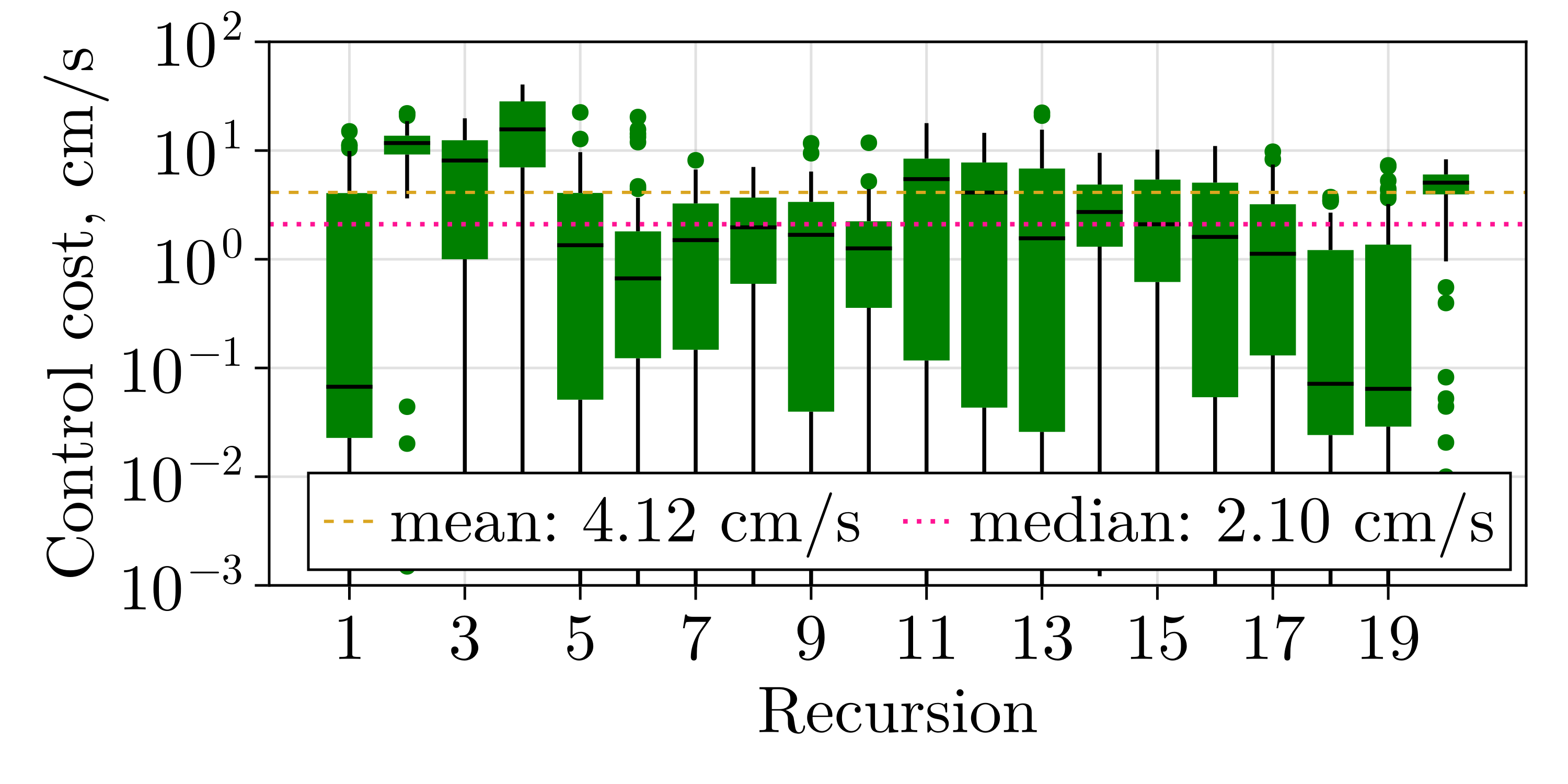}
        \caption{Spacecraft 2}
         \label{fig:control_boxplot_sc2_phase_nodes_constrained_X2}
     \end{subfigure}
    \caption{Experiment IV ($M=2$): distribution of control cost per recursion}
    \label{fig:control_boxplot_phase_nodes_constrained_X2}
\end{figure}

As presented in Table~\ref{tab:cost_statistics}, the mean and $3$-$\sigma$ cumulative costs are found to be lower for experiment III than experiment IV, even though the solution to an MVOCP instance in III is always a feasible solution to IV.
The distribution of the executed control magnitude by each spacecraft across the 20 recursions is shown in Figure~\ref{fig:control_boxplot_phase_isoperimetric_constrained_X2} and Figure~\ref{fig:control_boxplot_phase_nodes_constrained_X2} for experiments III and IV, respectively; similarly to experiments I and II, the lower cumulative cost with III is due to a lower per-recursion cost as well.

We also observe that the cost in IV is $\sim 53\%$ higher than in III, which is a significantly larger discrepancy than the cost difference between I and II.
The worst cost in IV is attributed to a more pronounced effect of the nonconvexity and finite-horizon approximation pointed out in Section~\ref{sec:mc_sep}, as the additional phase angle path constraints are introduced.
Our results also demonstrate that in a 2-spacecraft formation, the relative Sun phase angle can be achieved with a moderate increase in propellant budget, commonly accounted for based on worst-case values, e.g., $3$-$\sigma$ cost; under the assumed levels of uncertainty and spacecraft control cadence, a formation that satisfies both the bounded separation and relative phase angle constraints, as in III, can be achieved at a $\sim 57\%$ higher mean cumulative cost and $\sim 40\%$ higher $3$-$\sigma$ cost than a formation only satisfying bounded separation, as in I.

% phase constrained, isoperimetric

% phase constrained, node only

% control: sep+phase constrained, isoperimetric

% control: sep+phase constrained, nodes only

% -------------------------------------------------------------------- %
\subsubsection{Sensitivity on Constraint Tightening Parameters}
\red{We now consider the sensitivity of the MPC algorithm to constraint tightening parameters.
Specifically, we repeat experiments \rev{I through IV using $0.5\times\eta$ of the values from Table~\ref{tab:constraint_tightening_params}}.
With each experiment variant, we conduct Monte Carlo experiments with 100 samples over 10 revolutions, or 20 recursions; at a given recursion, if the MVOCP is infeasible, we prematurely end the simulation.}

Table~\ref{tab:cost_statistics_reduced} \rev{summarizes the cost statistics for the four experiments with reduced $\eta$.
As expected, the number of cases where the MVOCP instance is feasible at each recursion, as well as the number of cases that satisfy the CT path constraints, are drastically reduced.
Due to the added complexity of the relative Sun phase angle constraint, we find MVOCP instances from experiment III and IV to result in higher failure rates than from experiments I and II.
}

\begin{figure}[t]
     \centering
     \begin{subfigure}[b]{0.49\textwidth}
         \centering
         \includegraphics[width=\textwidth]{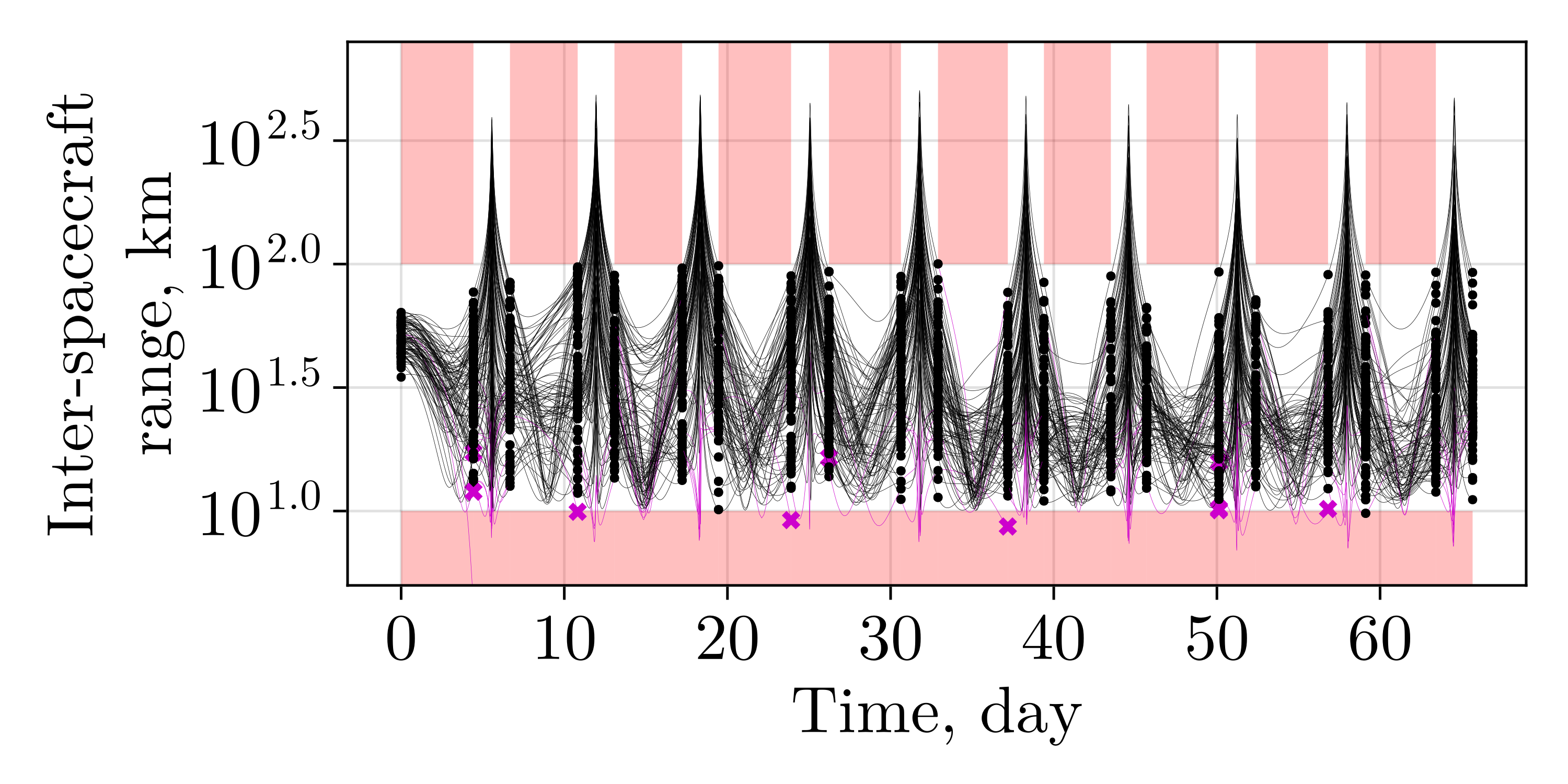}
        \caption{Inter-spacecraft range history}
         \label{fig:isoperimetric_phase_constrained_range_history_n2_X2-50}
     \end{subfigure}
     \hfill %\\
     \begin{subfigure}[b]{0.49\textwidth}
         \centering
         \includegraphics[width=\textwidth]{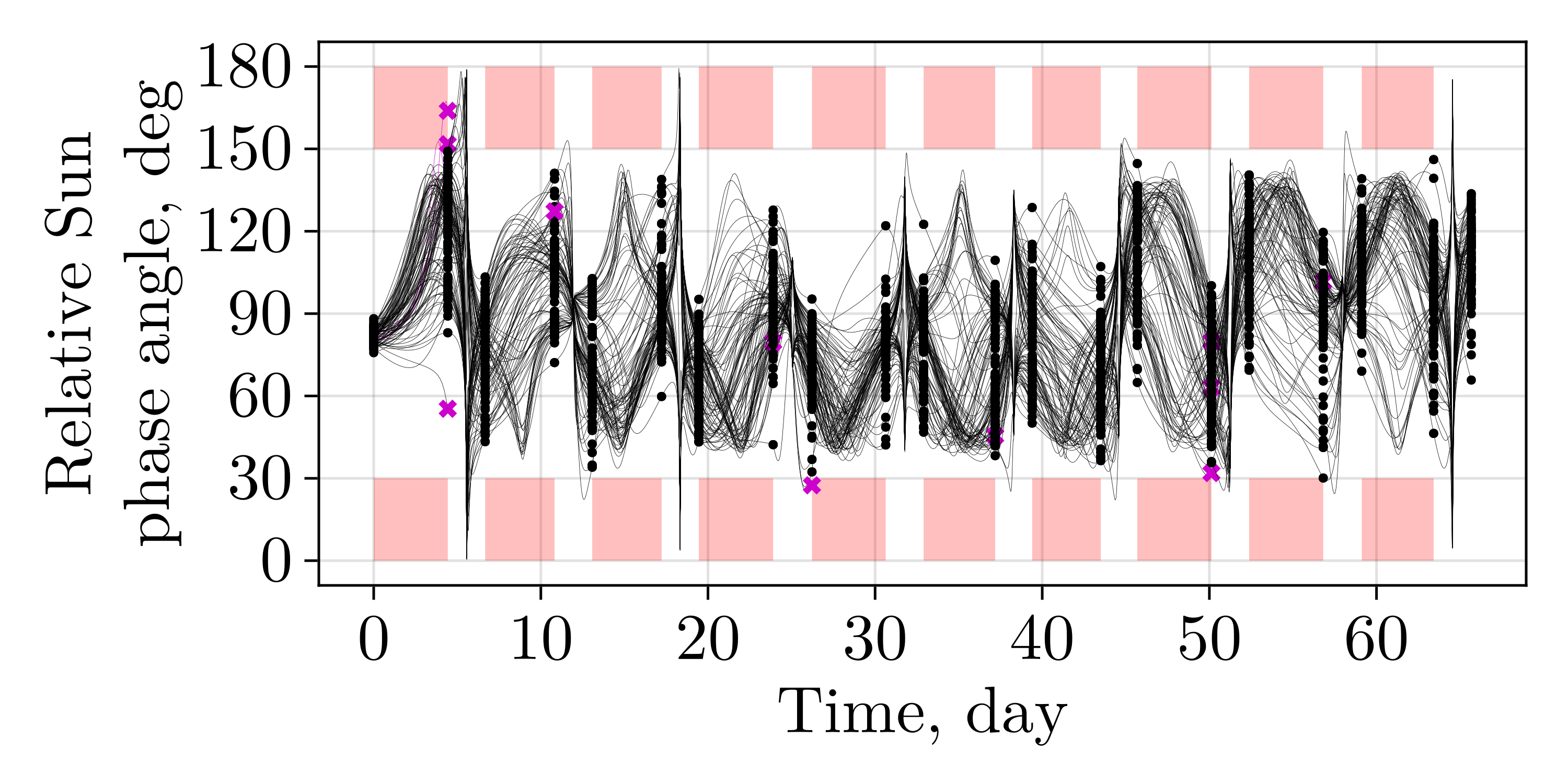}
        \caption{Relative Sun phase angle history}
         \label{fig:isoperimetric_phase_constrained_phase_history_n2_X2-50}
     \end{subfigure}
    \caption{Experiment III ($M=2$): constraints enforced at nodes with 50\% tightening}
    \label{fig:reduced_tightening_isoperimetric}
\end{figure}

\begin{figure}[t]
     \centering
     \begin{subfigure}[b]{0.49\textwidth}
         \centering
         \includegraphics[width=\textwidth]{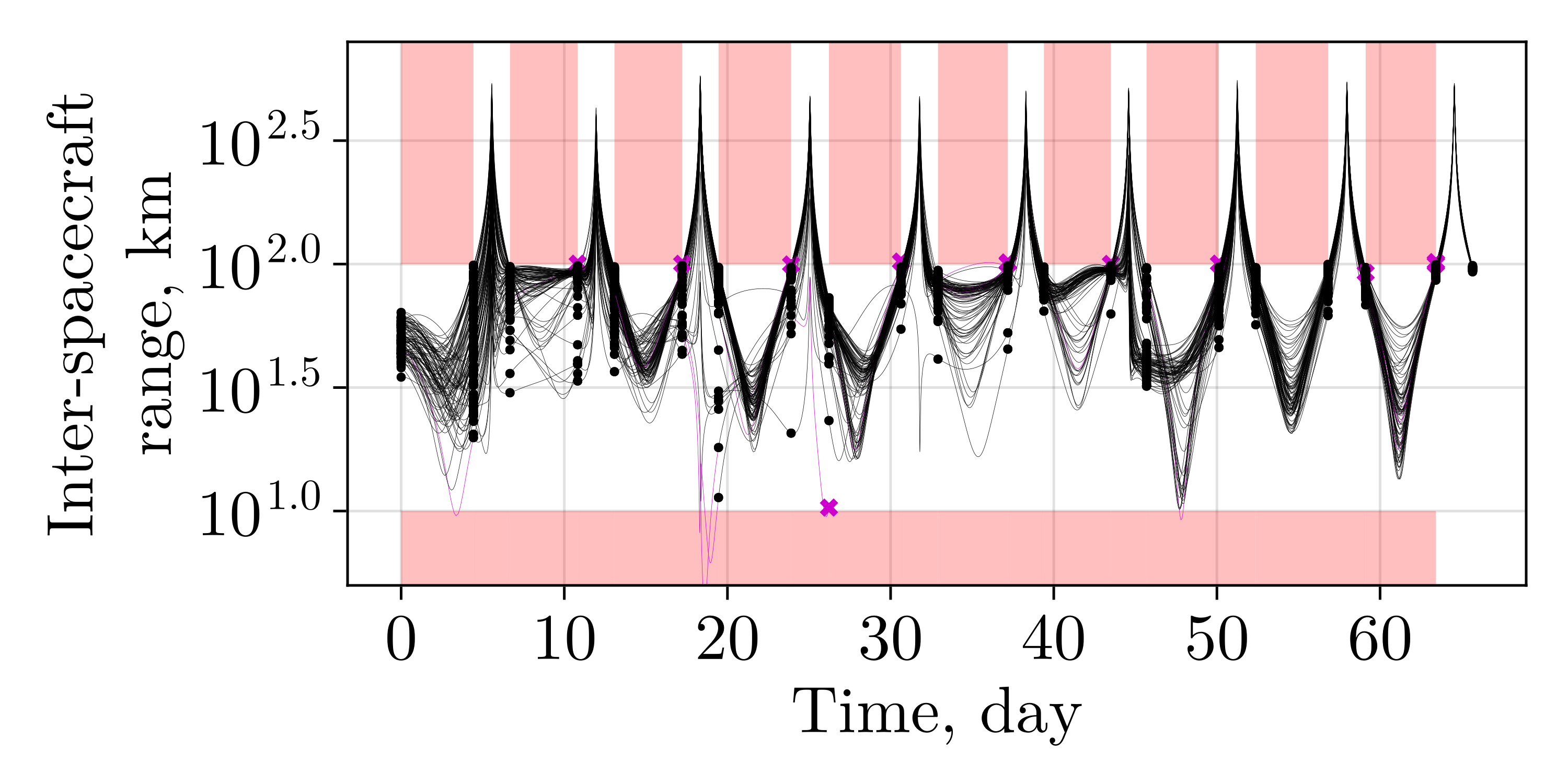}
        \caption{Inter-spacecraft range history}
         \label{fig:nodes_phase_constrained_range_history_n2_X2-50}
     \end{subfigure}
     \hfill %\\
     \begin{subfigure}[b]{0.49\textwidth}
         \centering
         \includegraphics[width=\textwidth]{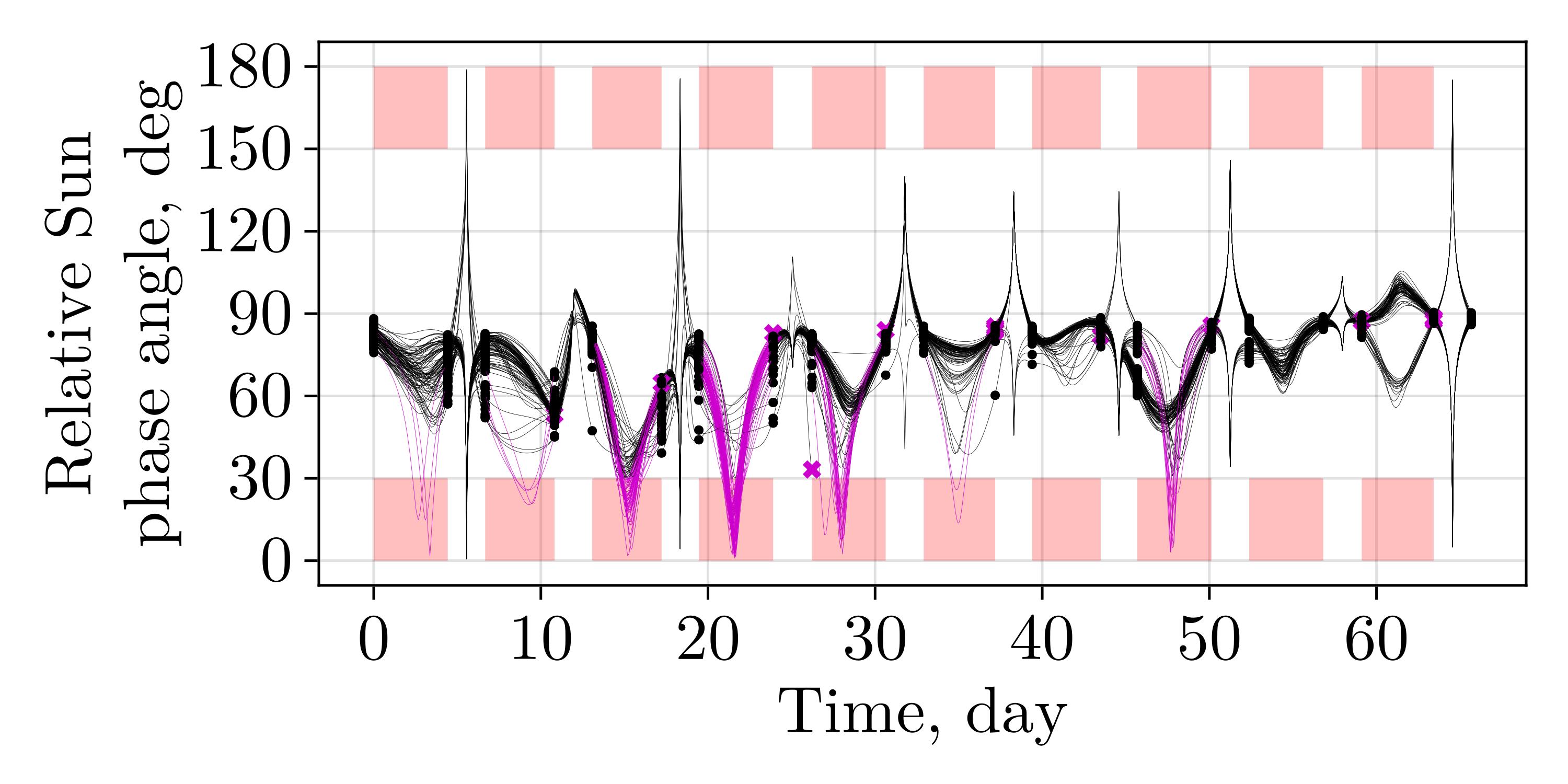}
        \caption{Relative Sun phase angle history}
         \label{fig:nodes_phase_constrained_phase_history_n2_X2-50}
     \end{subfigure}
    \caption{Experiment IV ($M=2$): constraints enforced at nodes with 50\% tightening}
    \label{fig:reduced_tightening_nodes}
\end{figure}

\rev{Figures~\ref{fig:reduced_tightening_isoperimetric} and~\ref{fig:reduced_tightening_nodes} show the time histories of the path constraints for experiments III and IV with reduced $\eta$.
Infeasible MVOCP instances, for which the Monte Carlo case is prematurely ended, are indicated by cross markers.
Comparing against Figures~\ref{fig:isoperimetric_phase_constrained} and~\ref{fig:nodes_phase_constrained}, we see the Monte Carlo samples spread over a wider inter-spacecraft range, resulting in more frequent constraint violations.
The relative Sun phase angle from experiment III, where the CT constraint is enforced, also sees a wider spread with reduced $\eta$, resulting in a few infeasible MVOCP instances.
In contrast, for experiments IV, the relative Sun phase angle constraint is satisfied at the nodes in most but one case even with a reduced $\eta$, and thus the majority of infeasible instances come from violations of inter-spacecraft range.
}

% Table of costs with reduced
\begin{table}[]
\centering
\footnotesize
\caption{MPC performance summary for 2-spacecraft formation with tightening at $0.5 \times \eta$}
\begin{tabular}{@{}lllllllllll@{}}
\toprule
\multirow{2}{*}{Exp.} &
\multirow{2}{*}{\begin{tabular}[c]{@{}l@{}}MVOCP\\feasible\end{tabular}} &
\multirow{2}{*}{\begin{tabular}[c]{@{}l@{}}CT\\feasible\end{tabular}} &
\multirow{2}{*}{\begin{tabular}[c]{@{}l@{}}$\mathrm{err}_{\Delta r}$,\\\SI{}{km}\end{tabular}}  & 
\multirow{2}{*}{\begin{tabular}[c]{@{}l@{}}$\mathrm{err}_{\phi}$,\\\SI{}{deg}\end{tabular}}  & 
\multicolumn{2}{c}{Spacecraft 1} & 
\multicolumn{2}{c}{Spacecraft 2} & 
\multicolumn{2}{c}{Formation Total} 
\\ 
& & & & 
  & Mean,~\SI{}{cm/s}& $3$-$\sigma$,~\SI{}{cm/s}
  & Mean,~\SI{}{cm/s}& $3$-$\sigma$,~\SI{}{cm/s}
  & Mean,~\SI{}{cm/s}& $3$-$\sigma$,~\SI{}{cm/s}
\\ \midrule
I &  98/100 &  79/100 & 1.46 & - & 17.13 & 34.62 & 18.34 & 36.87 & 35.48 & 63.11 \\
II &  89/100 &  (18/100) & 35.37 & - & 53.93 & 94.57 & 54.05 & 119.53 & 107.98 & 179.20 \\
III &  89/100 &  51/100 & 5.92 & 17.47 & 22.88 & 45.48 & 25.12 & 47.74 & 48.00 & 86.45 \\
IV &  84/100 &  (2/100) & 6.16 & 29.11 & 97.04 & 119.30 & 100.77 & 113.11 & 197.81 & 232.40 \\
\bottomrule
\end{tabular}
\label{tab:cost_statistics_reduced}
\end{table}

% experiment III 50% tightening

% experiment IV 50% tightening

% -------------------------------------------------------------------- %
\subsubsection{Formation Trajectory Satisfying Relative Separation \& Sun Phase Constraints}
\begin{figure}[]
     \centering
     \begin{subfigure}[b]{0.49\textwidth}
         \centering
         \includegraphics[width=\textwidth]{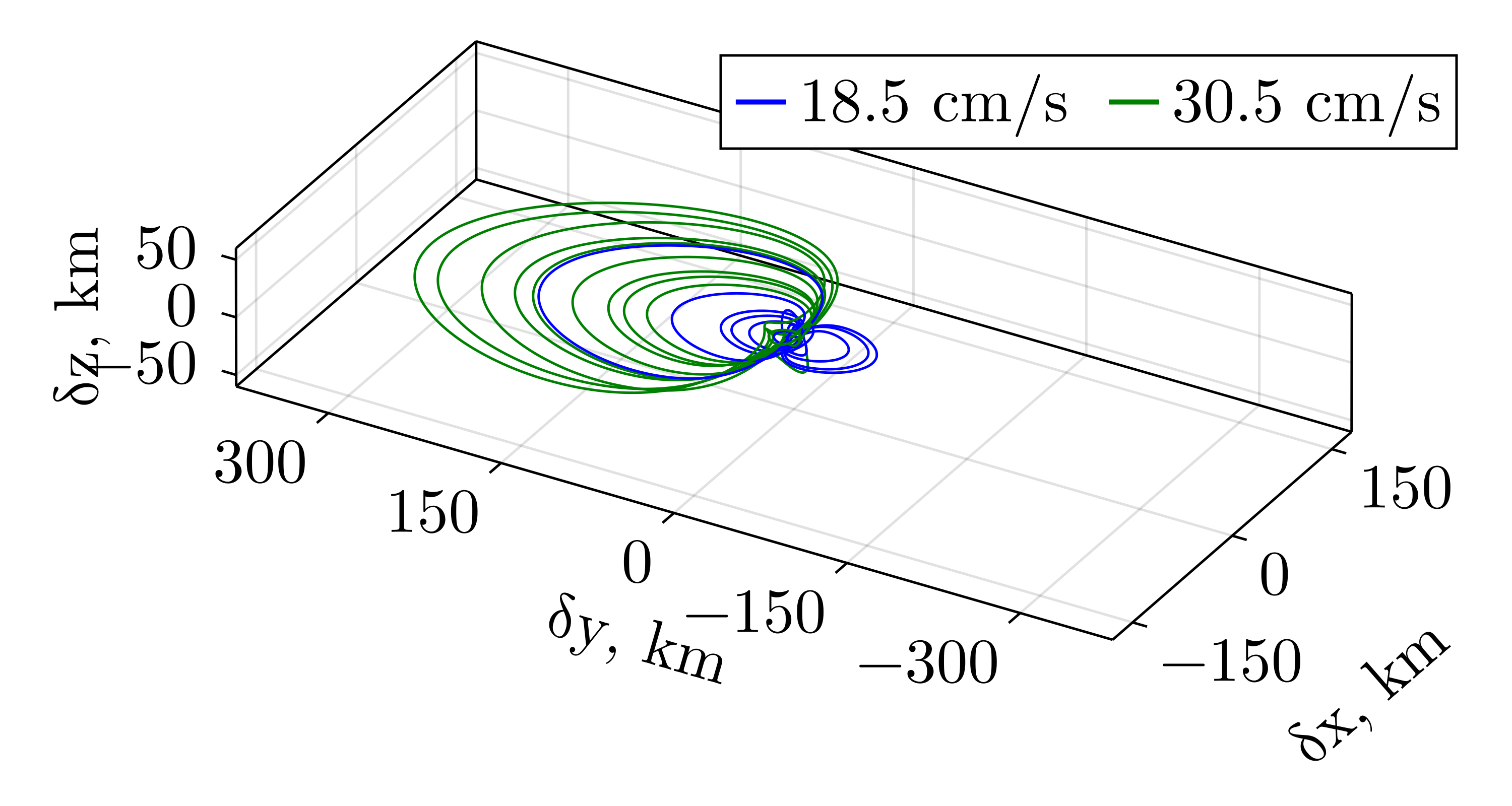}
        \caption{Perilune segments separated by RW-plane, varying amplitudes}
         \label{fig:relative_trajectory_RTN_mc22_n2}
     \end{subfigure}
     \hfill %\\
     \begin{subfigure}[b]{0.49\textwidth}
         \centering
         \includegraphics[width=\textwidth]{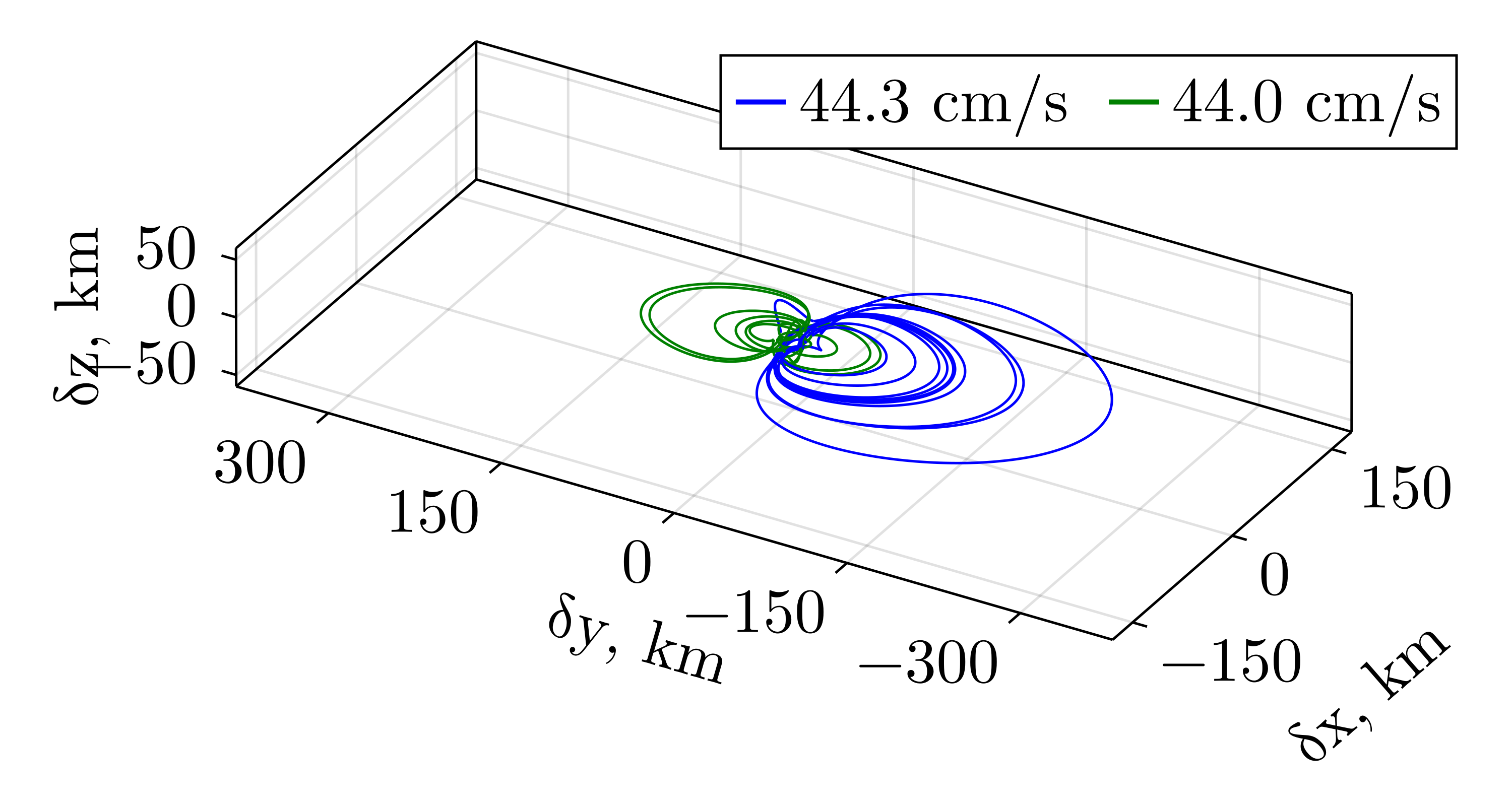}
        \caption{Perilune segments separated by RW-plane, similar amplitudes}
         \label{fig:relative_trajectory_RTN_mc10_n2}
     \end{subfigure}
     \\
     \begin{subfigure}[b]{0.49\textwidth}
         \centering
         \includegraphics[width=\textwidth]{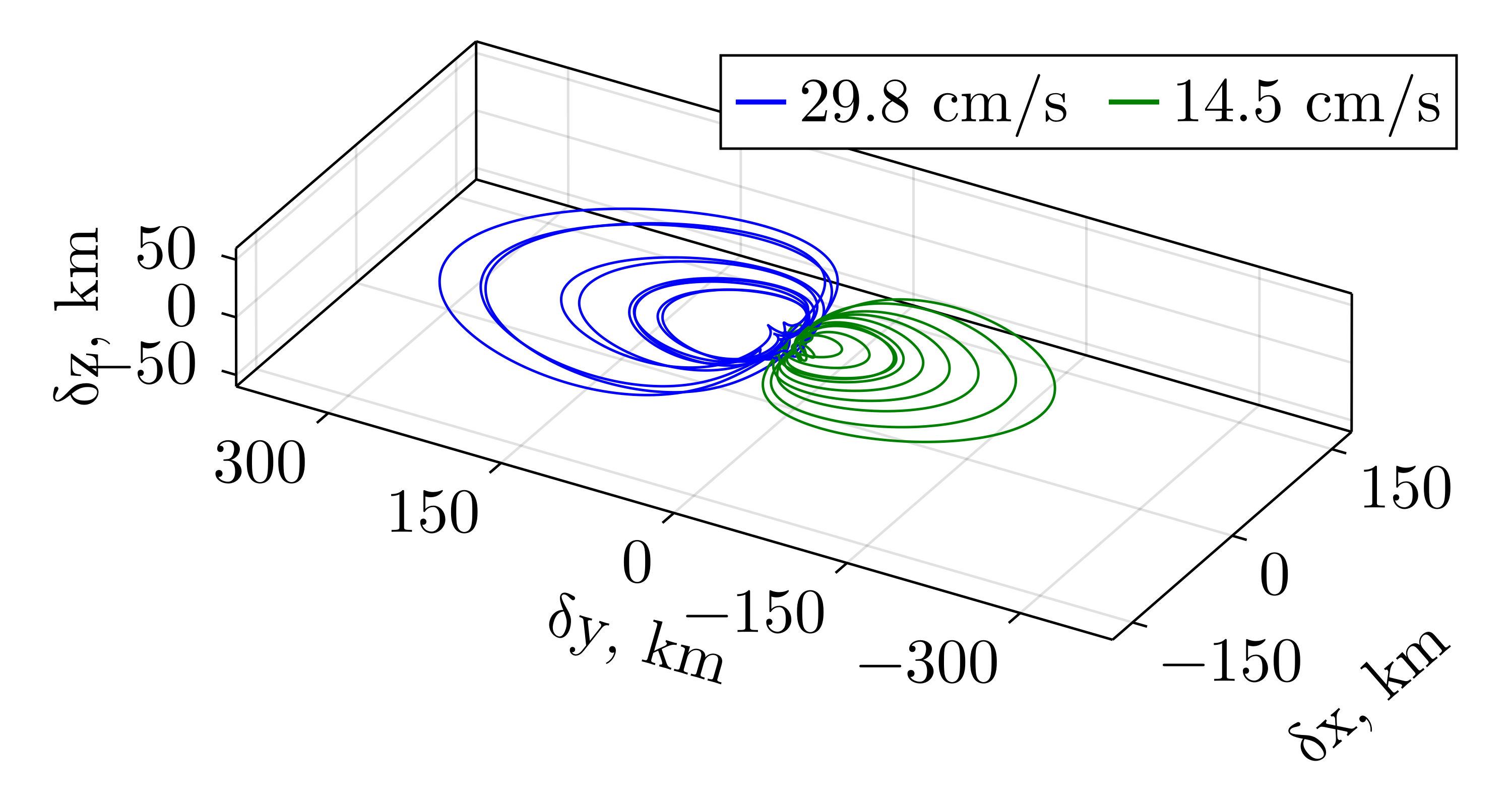}
        \caption{Perilune segments predominantly on one side of RW-plane}
         \label{fig:relative_trajectory_RTN_mc15_n2}
     \end{subfigure}
     \hfill %\\
     \begin{subfigure}[b]{0.49\textwidth}
         \centering
         \includegraphics[width=\textwidth]{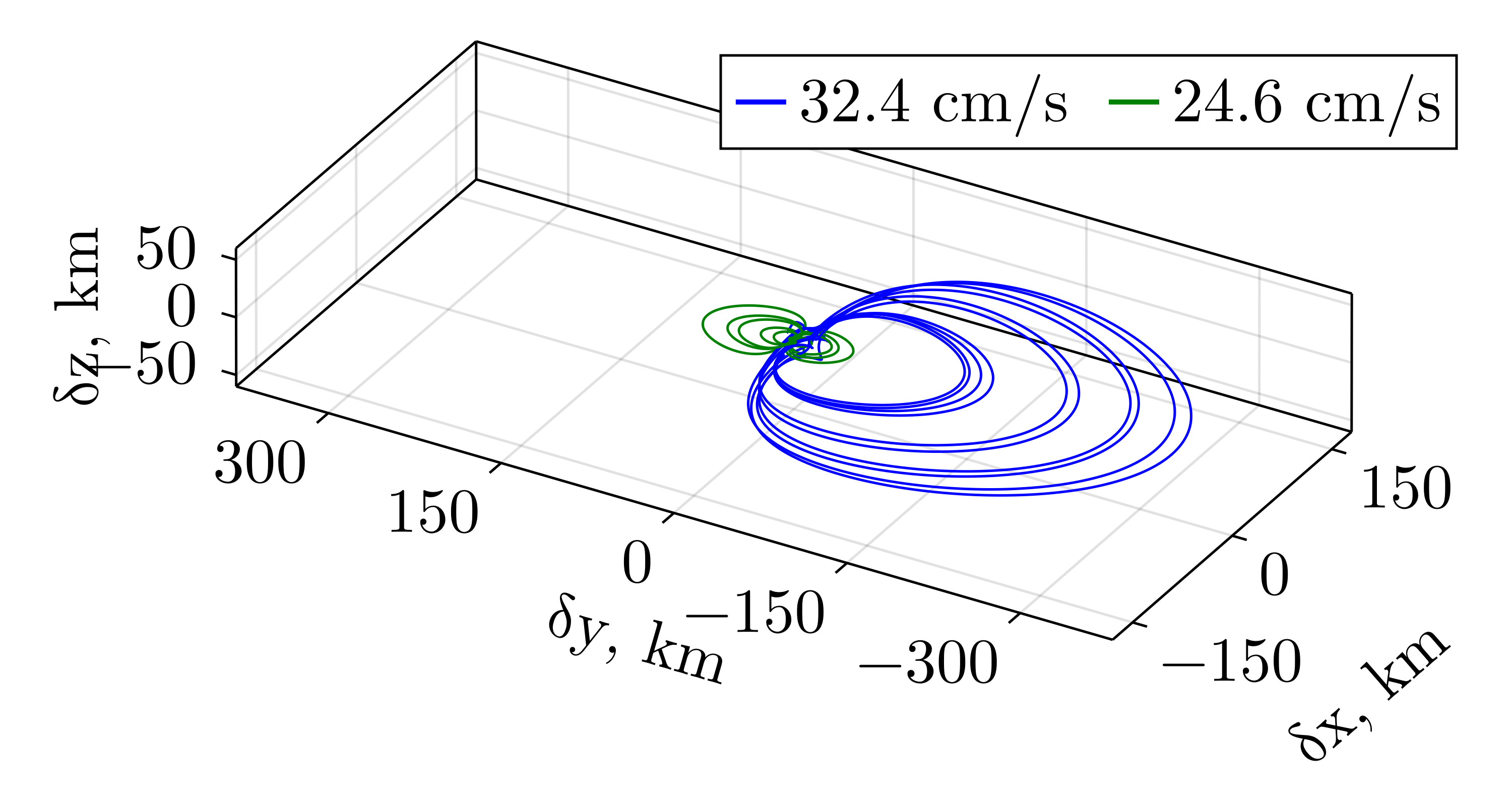}
        \caption{Perilune segments residing on both sides of RW-plane}
         \label{fig:relative_trajectory_RTN_mc23_n2}
     \end{subfigure}
    \caption{Sample reference-relative trajectories in 2-spacecraft formation in RSW frame over 10 revolutions}
    \label{fig:relative_trajectory_RTN}
\end{figure}

We provide an analysis of the relative trajectory of the formation from Section~\ref{sec:mc_phase}, where both bounded relative separation and Sun phase angle constraints are satisfied.
Figure~\ref{fig:relative_trajectory_RTN} shows sample reference-relative trajectories in~$\mathcal{F}_{\rm RSW}$ from four Monte Carlo samples in experiment III.
Across all relative trajectories, apolune segments are clustered around the origin due to the upper bound constraint on the relative separation, and perilune segments draw loops predominantly lying on the RT-plane, i.e., the $\delta x$-$\delta y$ plane.
We find a variety of relative trajectory orientations that are feasible results of recursive station-keeping under uncertainty. 
Notably, along the tangential direction, i.e., the $\delta y$ direction, the relative trajectories in Figures~\ref{fig:relative_trajectory_RTN_mc22_n2} and~\ref{fig:relative_trajectory_RTN_mc10_n2} have perilune segments on either side of the RW-plane, while the ones in Figures~\ref{fig:relative_trajectory_RTN_mc15_n2} and~\ref{fig:relative_trajectory_RTN_mc23_n2} have perilune segments residing on common sides; we also find variable amplitudes of perilune segments.
While all four trajectory orientations are feasible for the set of considered constraints, if a certain orientation property, e.g., perilune segments residing on either side of the RW-plane, is desired, then an additional constraint at the perilunes of the reference such as 
\begin{equation}
    \begin{bmatrix}
        0 \\ 1 \\ 0
    \end{bmatrix}^T
    \Tbold^{\rm Inr}_{\rm RSW}(t_{\ell}) \rbold_{1,\ell|k} \leq 0,
    \quad
    \begin{bmatrix}
        0 \\ 1 \\ 0
    \end{bmatrix}^T
    \Tbold^{\rm Inr}_{\rm RSW}(t_{\ell}) \rbold_{2,\ell|k} \geq 0,
    \, \text{ where } \,
    \theta_{\rm ref} (t_{\ell}) = 0^{\circ}
    ,
\end{equation}
may be incorporated into~\eqref{eq:MVOCP_ct}.

% trajectory

\begin{figure}[]
     \centering
     \begin{subfigure}[b]{0.375\textwidth}
         \centering
         \includegraphics[width=\textwidth]{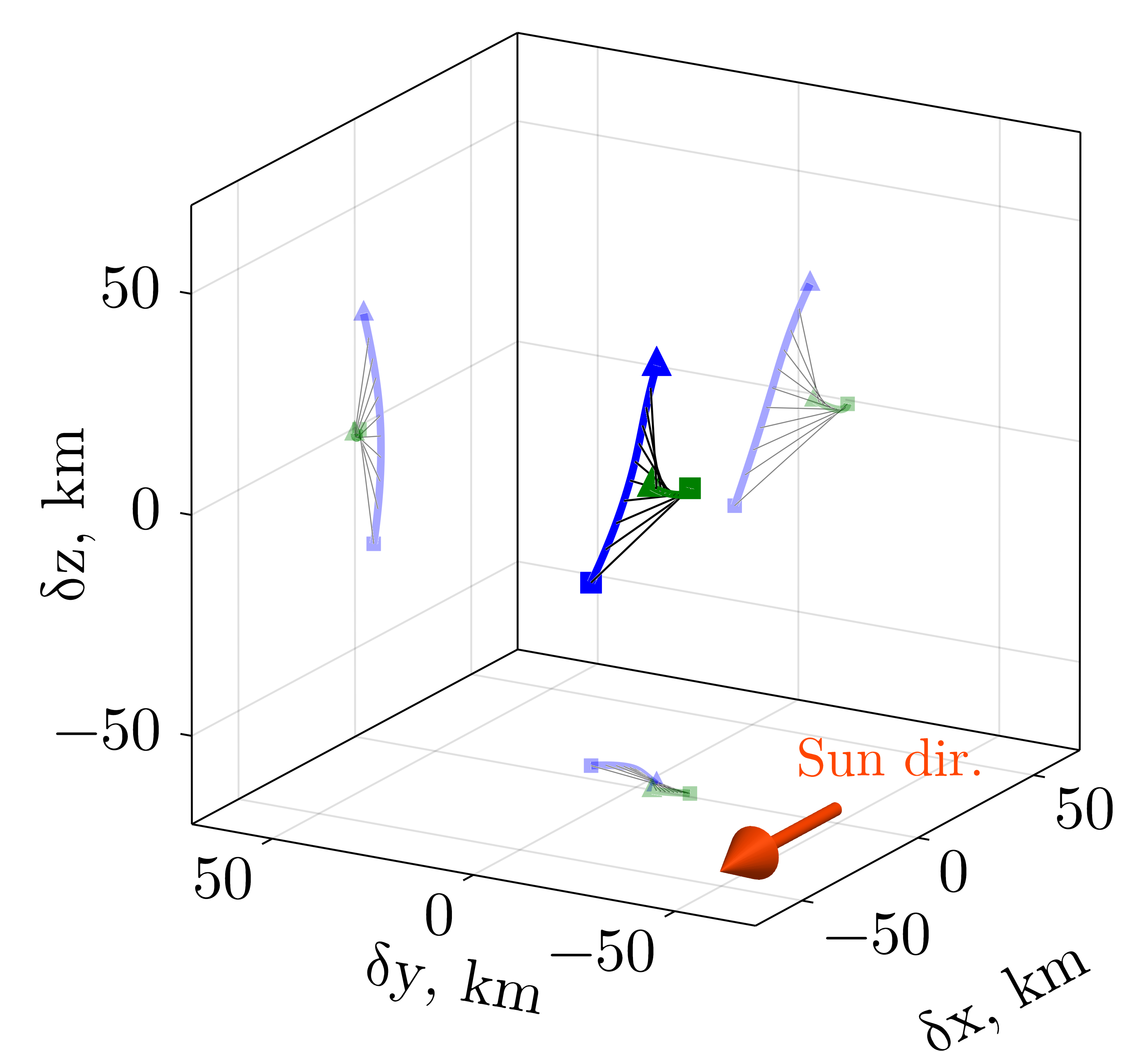}
        \caption{Trajectory around $1^{\rm st}$ apolune}
         \label{fig:traj_sun_avoidance_mc10_idx11}
     \end{subfigure}
     \hfill %\\
     \begin{subfigure}[b]{0.375\textwidth}
         \centering
         \includegraphics[width=\textwidth]{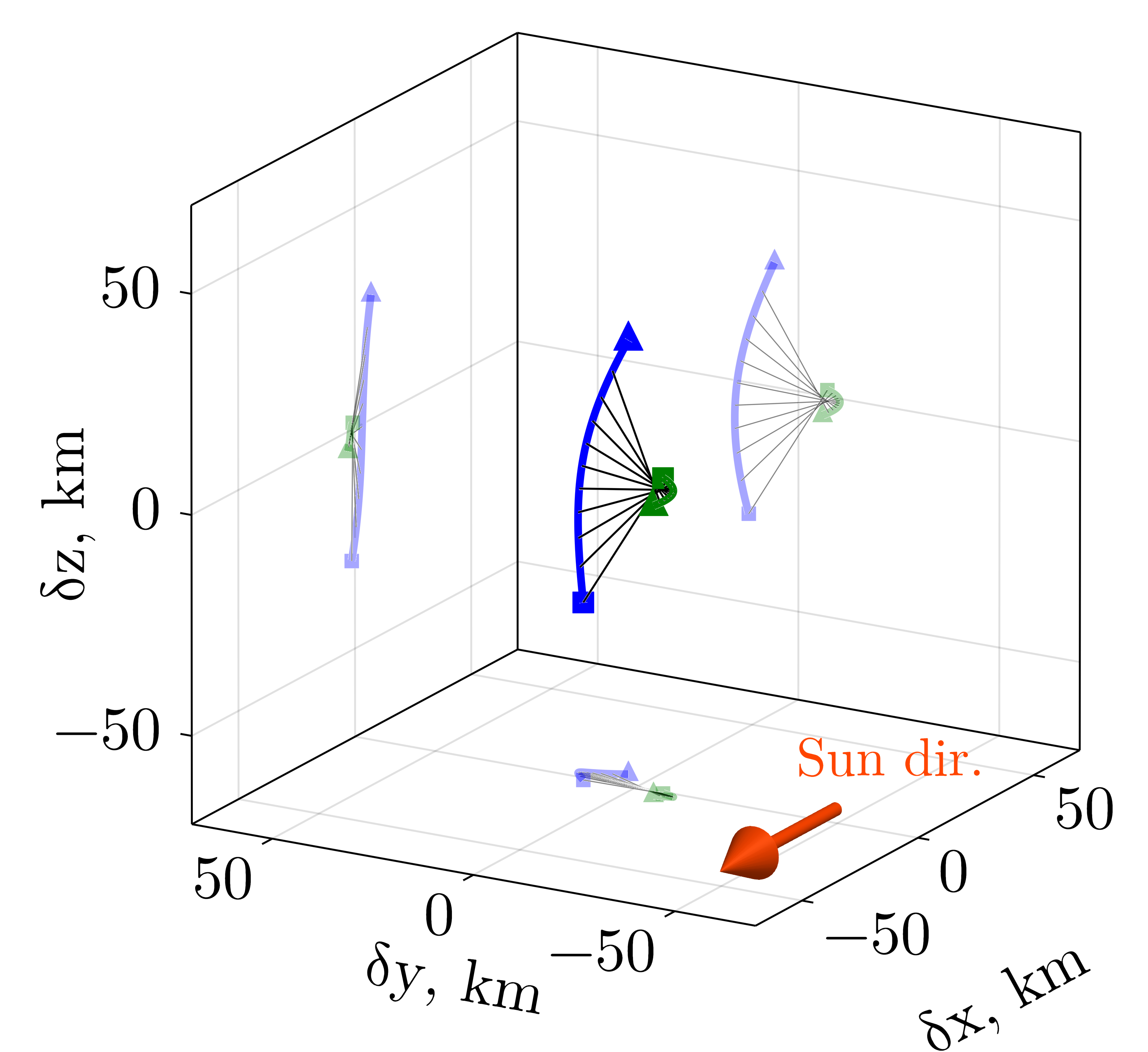}
        \caption{Trajectory around $2^{\rm nd}$ apolune}
         \label{fig:traj_sun_avoidance_mc10_idx13}
     \end{subfigure}
     \\
     \begin{subfigure}[b]{0.375\textwidth}
         \centering
         \includegraphics[width=\textwidth]{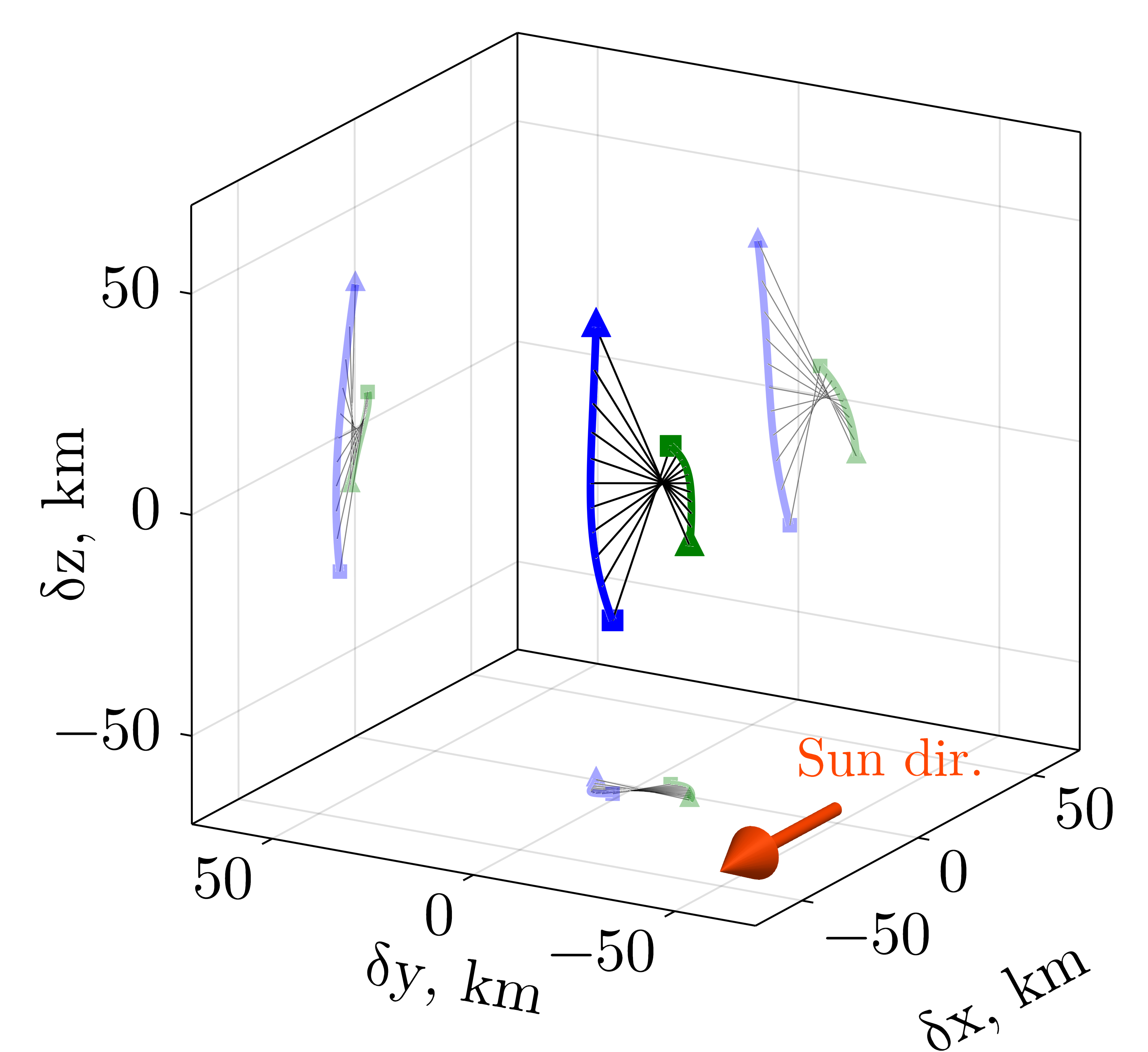}
        \caption{Trajectory around $3^{\rm rd}$ apolune}
         \label{fig:traj_sun_avoidance_mc10_idx15}
     \end{subfigure}
     \hfill %\\
     \begin{subfigure}[b]{0.375\textwidth}
         \centering
         \includegraphics[width=\textwidth]{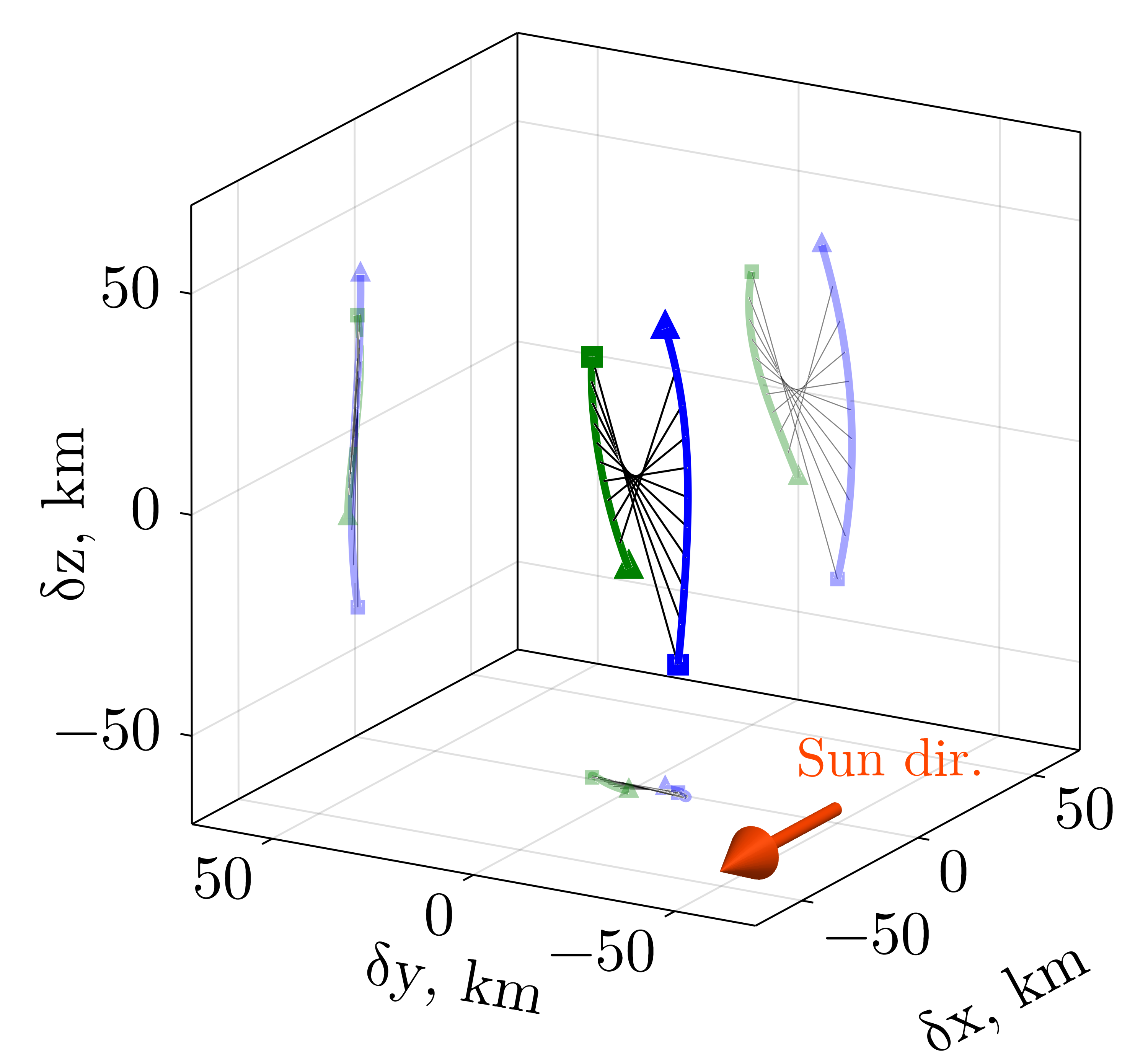}
        \caption{Trajectory around $4^{\rm th}$ apolune}
         \label{fig:traj_sun_avoidance_mc10_idx17}
     \end{subfigure}
    \caption{Sample reference-relative trajectory along apolune segments in Sun-Moon rotating frame}
    \label{fig:traj_sun_avoidance}
\end{figure}

Reference-relative trajectories in~$\mathcal{F}_{\rm SM}$ along four consecutive apolune segments from a Monte Carlo sample are shown in Figure~\ref{fig:traj_sun_avoidance}.
In Figure~\ref{fig:traj_sun_avoidance}, the square and triangle markers respectively indicate initial and final locations during the segment, and black lines indicate lines-of-sight between the two spacecraft.
Recalling the frame definition from Section~\ref{sec:background_frames}, the $-x$ direction points to the Sun, as indicated by the orange arrow; the blue and green paths correspond to the motion of each spacecraft, and the black, thin lines are the lines-of-sight between the two spacecraft.
Due to the 9:2 resonance of the reference NRHO, the orientation of $\mathcal{F}_{\rm RSW}$ with respect to $\mathcal{F}_{\rm SM}$ rotates approximately $360^{\circ} \times 2 / 9 = 80^{\circ}$ between consecutive apolune segments, completing a full periodic cycle after 9 revolutions.
The station-keeping scheme accommodates this apolune-to-apolune variation by exhibiting a variety of relative geometry in~$\mathcal{F}_{\rm SM}$, as shown by the four windows in Figure~\ref{fig:traj_sun_avoidance}, that still satisfy the bounded relative Sun phase angle constraints continuously during each apolune segment.
Notably, the reference-relative motion may traverse similar distances, as in Figure~\ref{fig:traj_sun_avoidance_mc10_idx11} and~\ref{fig:traj_sun_avoidance_mc10_idx13}, or one spacecraft may traverse a larger distance than the other, as in Figure~\ref{fig:traj_sun_avoidance_mc10_idx17}.

% trajectory

\subsection{\rev{Monte Carlo Experiments with Three-Spacecraft Formation}}
\rev{We consider station-keeping formations with a larger number of spacecraft.
We conduct Monte Carlo experiments with $M = 3$ spacecraft for experiments I and II from Table~\ref{tab:summary_mc_experiments}, where we enforce bounds on inter-spacecraft range.
Experiments III and IV are omitted for the $M=3$ case, as bounding the relative Sun phase angle essentially requires the formation to lie close to a plane orthogonal to the Sun direction: through preliminary experiments, this was found to be challenging to satisfy with the two maneuvers around apolune segments alone and with the level of uncertainties assumed in our analyses.
We use the constraint tightening parameter in Table~\ref{tab:OCP_parameters_separation_constraints}.
Since these have been tuned for the 2-spacecraft case, we expect the tightening to be potentially insufficient for the 3-spacecraft case, resulting in CT path constraint violation or infeasible solutions; by keeping the same parameter values, we assess their effect as $M$ increases.}

\begin{figure}[t]
     \centering
     \begin{subfigure}[b]{0.33\textwidth}
         \centering
         \includegraphics[width=\textwidth]{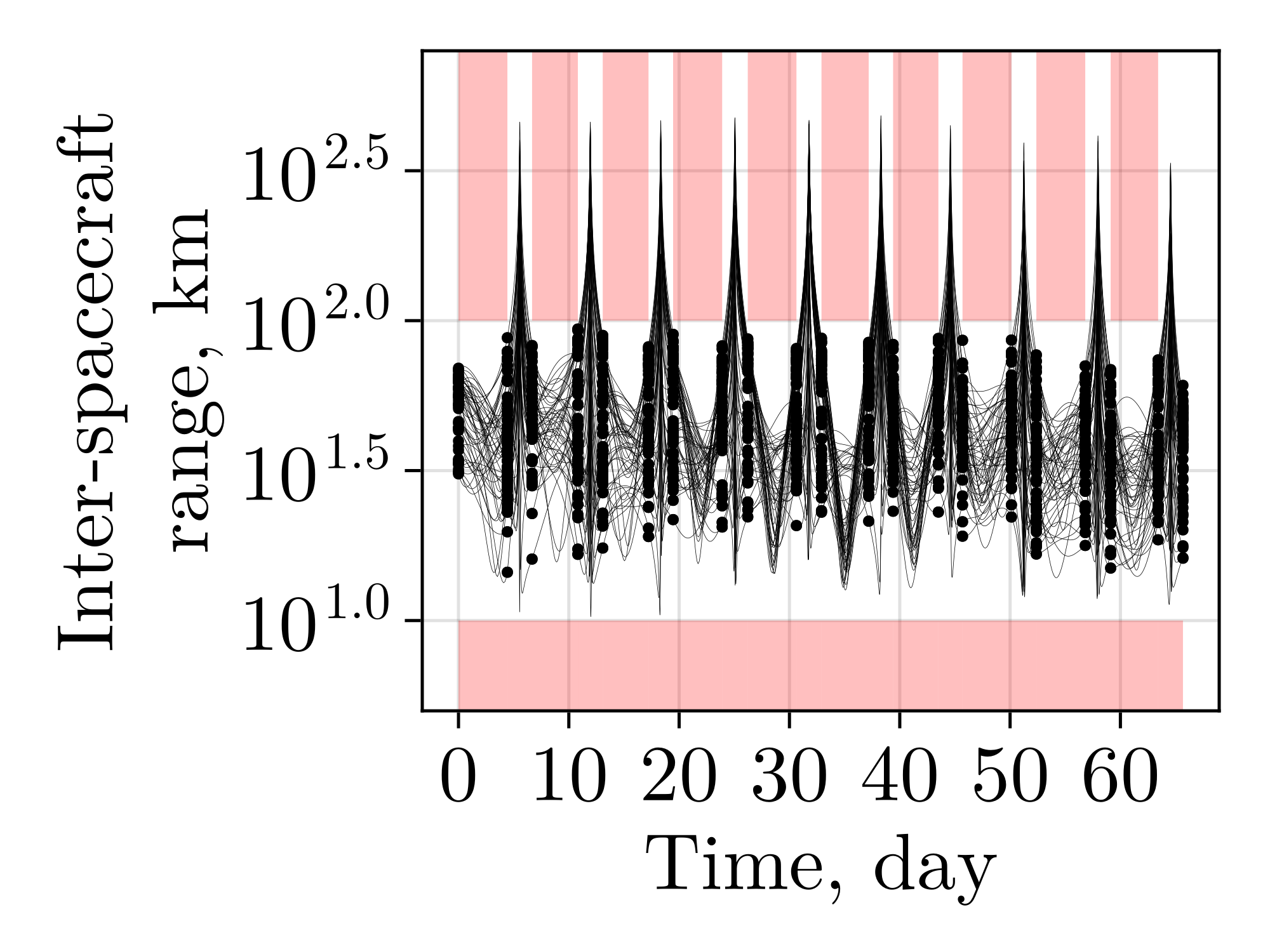}
        \caption{Spacecraft pair $(1,2)$}
        \label{fig:isoperimetric_sep_constrained_range_history_n3_comb1-2}
     \end{subfigure}
     \hfill %\\
     \begin{subfigure}[b]{0.33\textwidth}
         \centering
         \includegraphics[width=\textwidth]{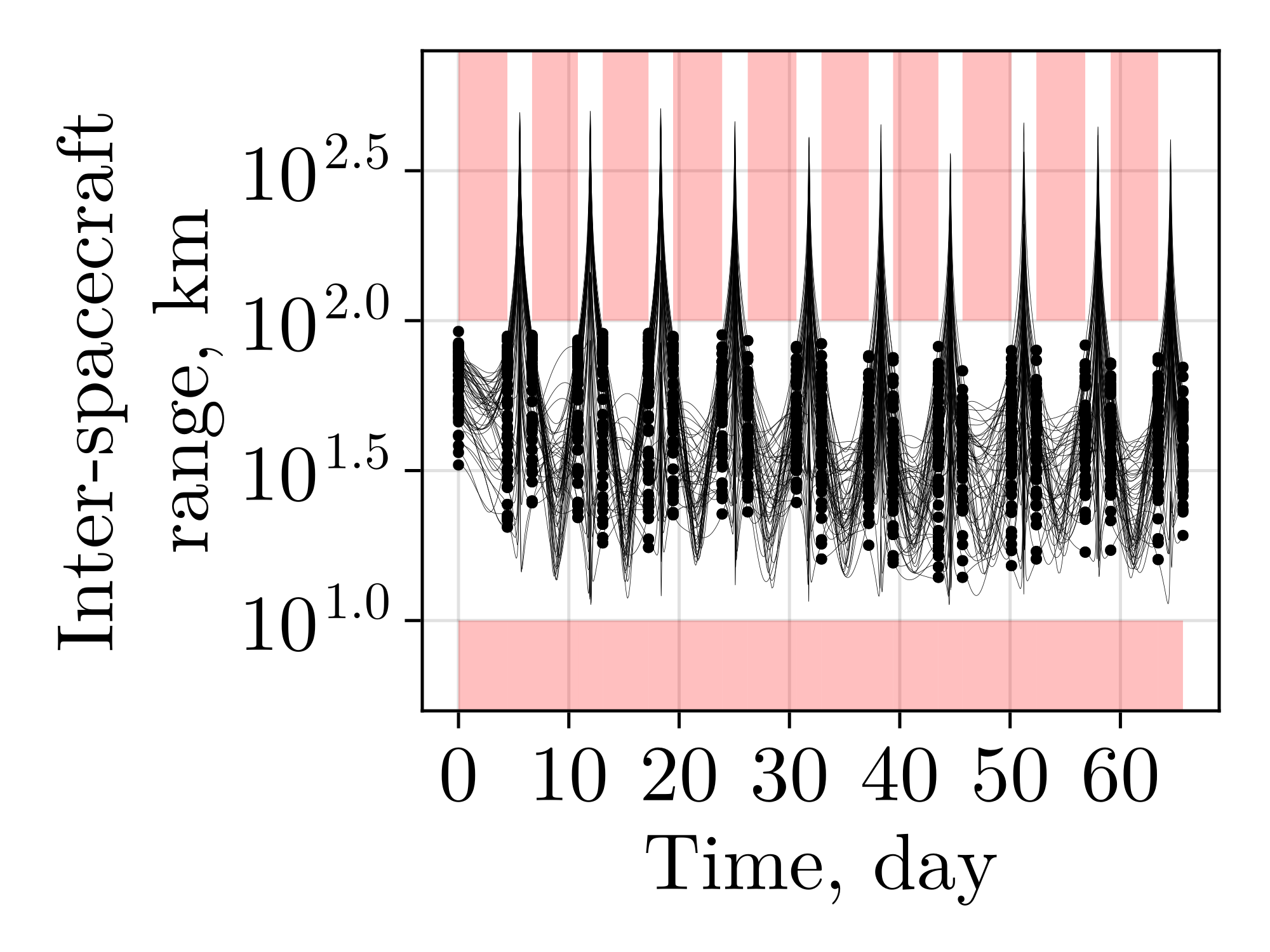}
        \caption{Spacecraft pair $(1,3)$}
        \label{fig:isoperimetric_sep_constrained_range_history_n3_comb1-3}
     \end{subfigure}
     \hfill %\\
     \begin{subfigure}[b]{0.33\textwidth}
         \centering
         \includegraphics[width=\textwidth]{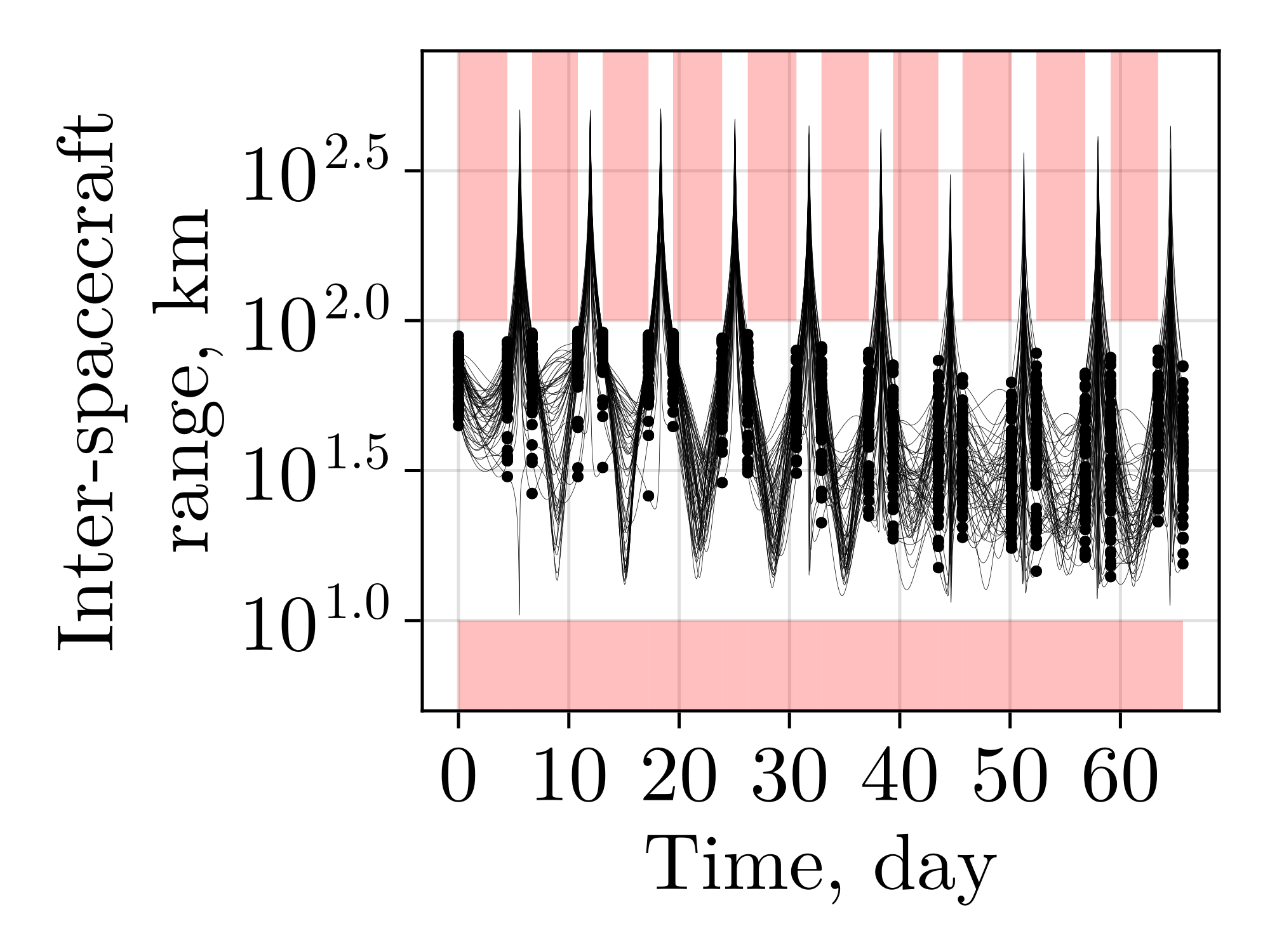}
        \caption{Spacecraft pair $(2,3)$}
        \label{fig:isoperimetric_sep_constrained_range_history_n3_comb2-3}
     \end{subfigure}
    \caption{Experiment I ($M=3$): inter-spacecraft range continuously enforced}
    \label{fig:n3_range_isoperimetric}
\end{figure}

\begin{figure}[t]
     \centering
     \begin{subfigure}[b]{0.33\textwidth}
         \centering
         \includegraphics[width=\textwidth]{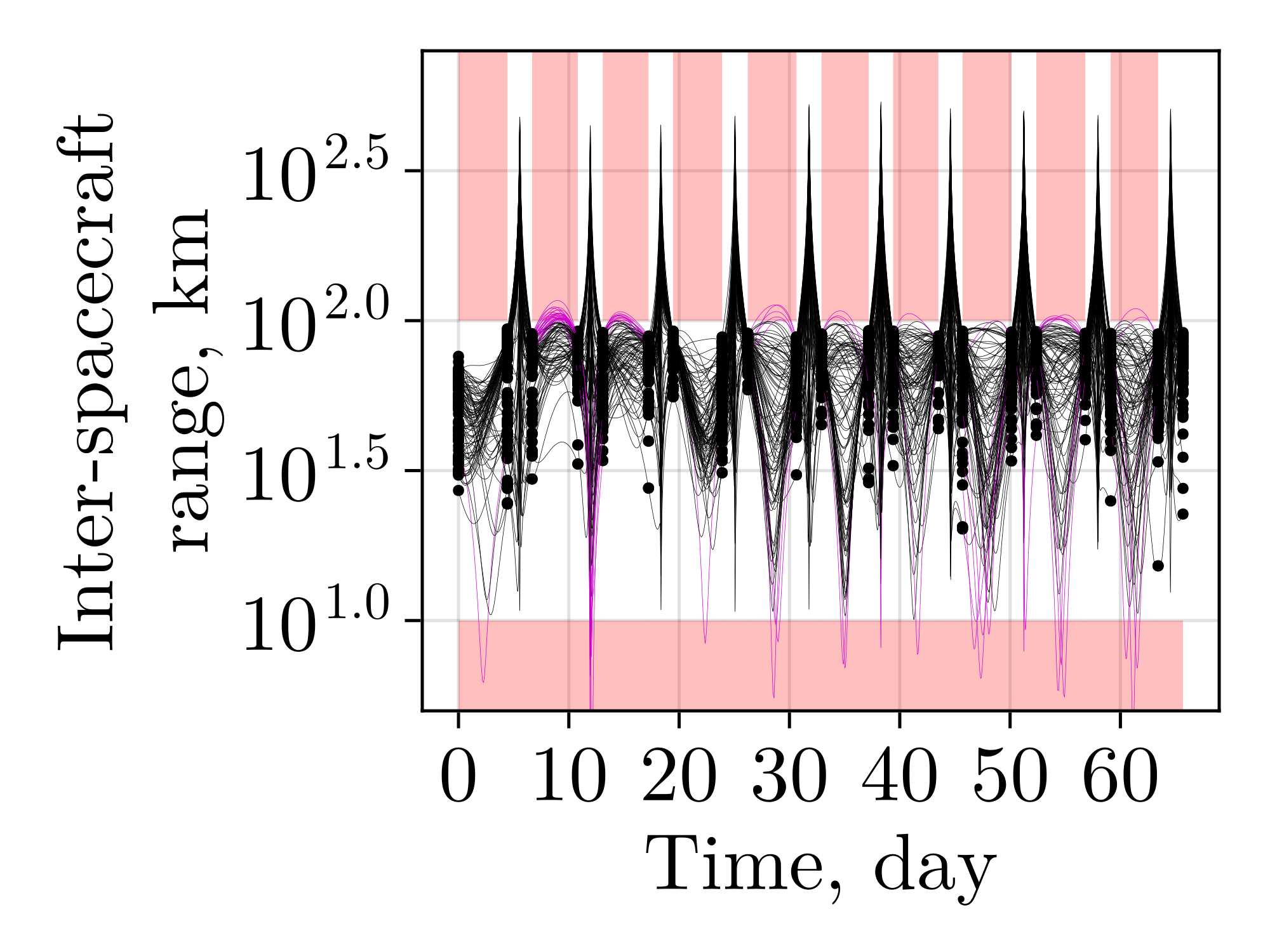}
        \caption{Spacecraft pair $(1,2)$}
        \label{fig:nodes_sep_constrained_range_history_n3_comb1-2}
     \end{subfigure}
     \hfill %\\
     \begin{subfigure}[b]{0.33\textwidth}
         \centering
         \includegraphics[width=\textwidth]{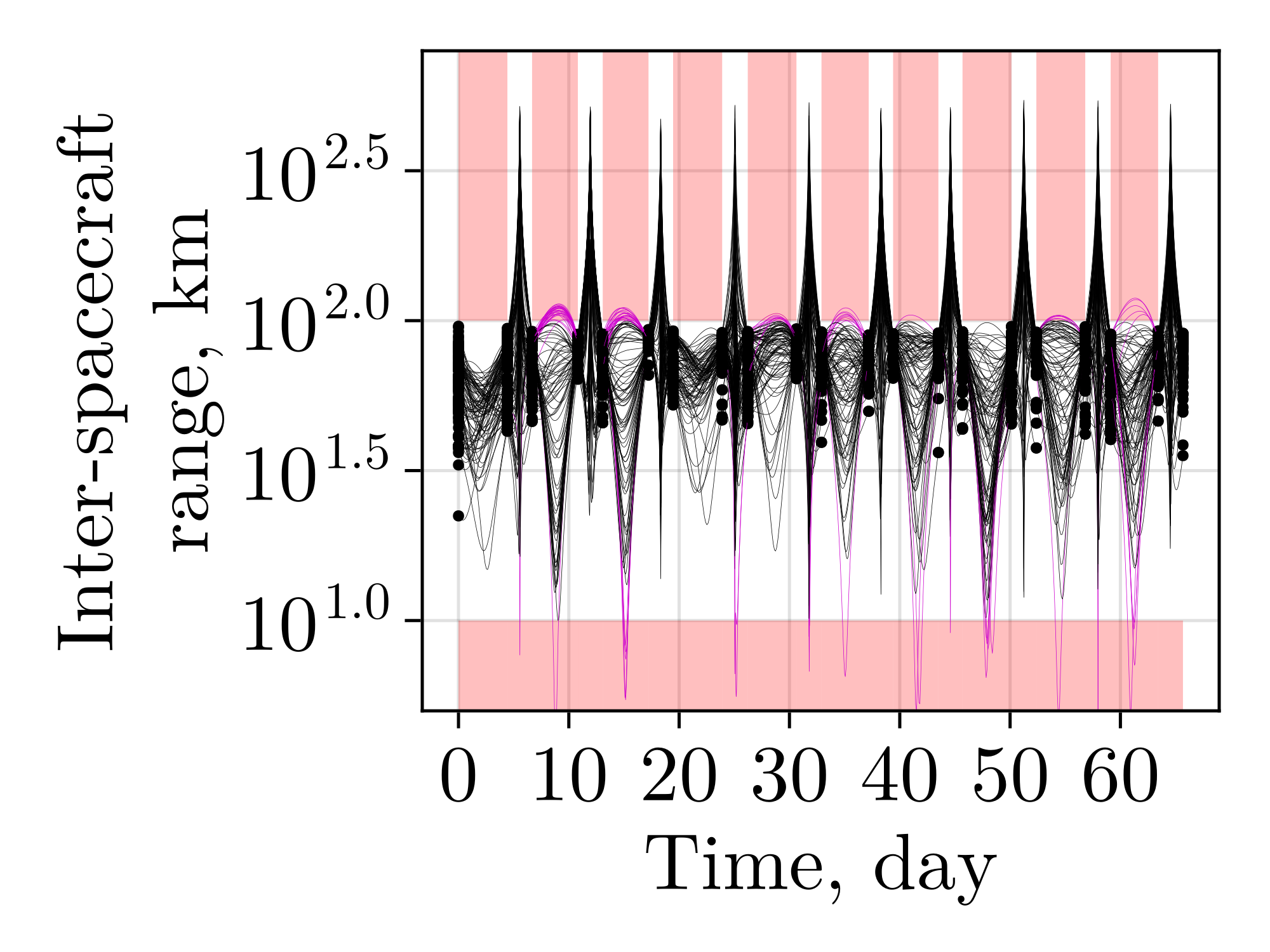}
        \caption{Spacecraft pair $(1,3)$}
        \label{fig:nodes_sep_constrained_range_history_n3_comb1-3}
     \end{subfigure}
     \hfill %\\
     \begin{subfigure}[b]{0.33\textwidth}
         \centering
         \includegraphics[width=\textwidth]{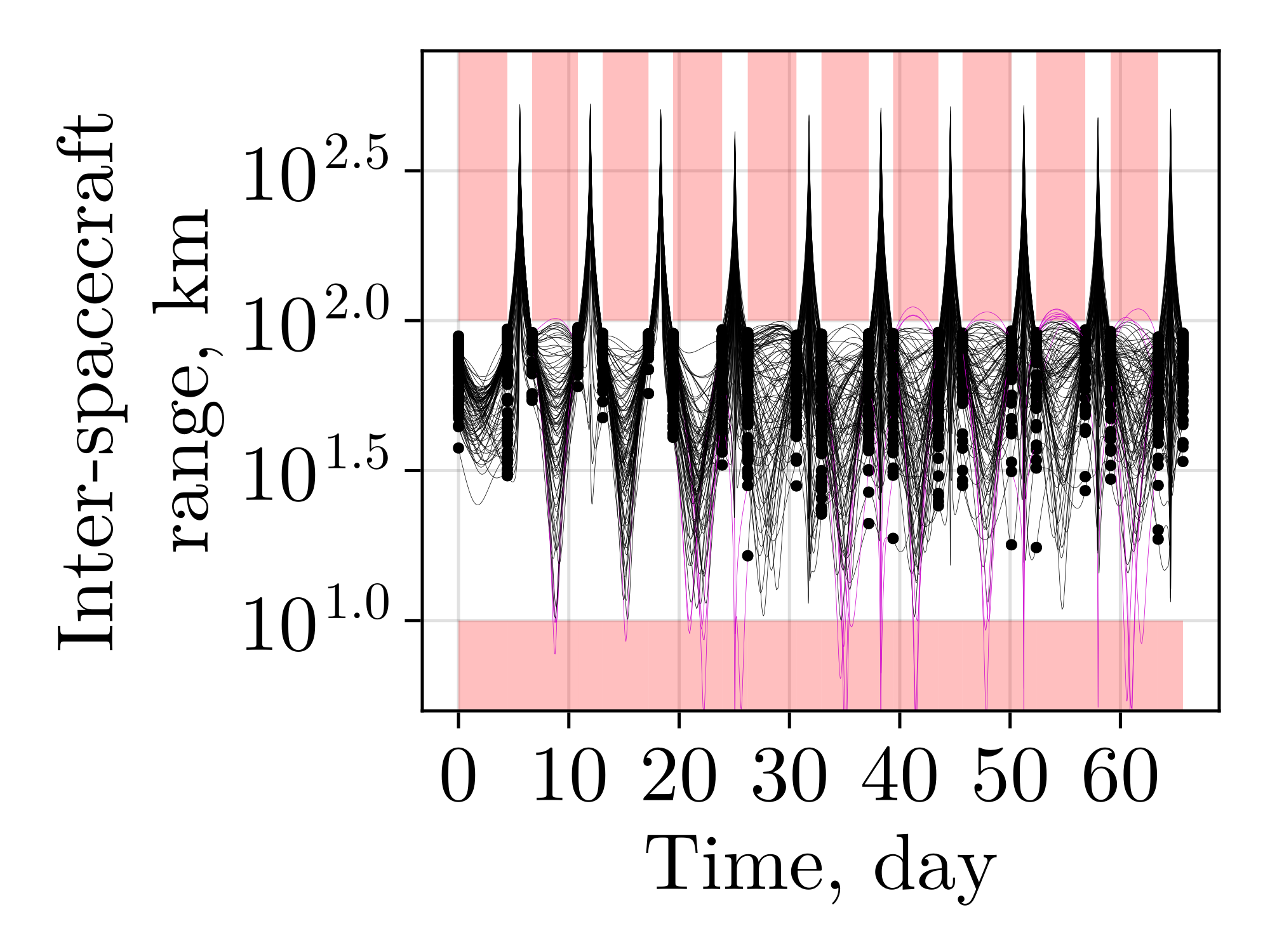}
        \caption{Spacecraft pair $(2,3)$}
        \label{fig:nodes_sep_constrained_range_history_n3_comb2-3}
     \end{subfigure}
     \caption{Experiment II ($M=3$): inter-spacecraft range enforced at nodes}
    \label{fig:n3_range_nodes}
\end{figure}

\rev{
The cumulative control cost statistics with the $M=3$ constellation are reported in~Table~\ref{tab:cost_statistics_3spacecraft}.
We now encounter a few cases where the MVOCP becomes infeasible.
As hypothesized, with the same tightening parameters as in the $M=2$ case and the same levels of uncertainty, spacecraft pairs are more likely to reach/violate the bounds on the range.
In the Table, the maximum CT path constraint violation $\mathrm{err}_{\Delta r}$ as well as statistics on the cost are computed from the MVOCP feasible cases.
We find that in 95/100 cases from experiment I, the original constraint tightening still yields successful formation station keeping, with no $\mathrm{err}_{\Delta r}$.
Figures~\ref{fig:n3_range_isoperimetric} and~\ref{fig:n3_range_nodes} show the time histories of $\|\rbold_1(t) - \rbold_2(t)\|_2$ and $\phi_{12}(t)$ between each spacecraft pair.
The per-spacecraft control cost is noticeably higher, at around $\times 1.5 \sim \times 2$ of the per-spacecraft cost from the 2-spacecraft formation in Table~\ref{tab:cost_statistics}.
}

\begin{figure}[]
     \centering
     \begin{subfigure}[b]{0.49\textwidth}
         \centering
         \includegraphics[width=\textwidth]{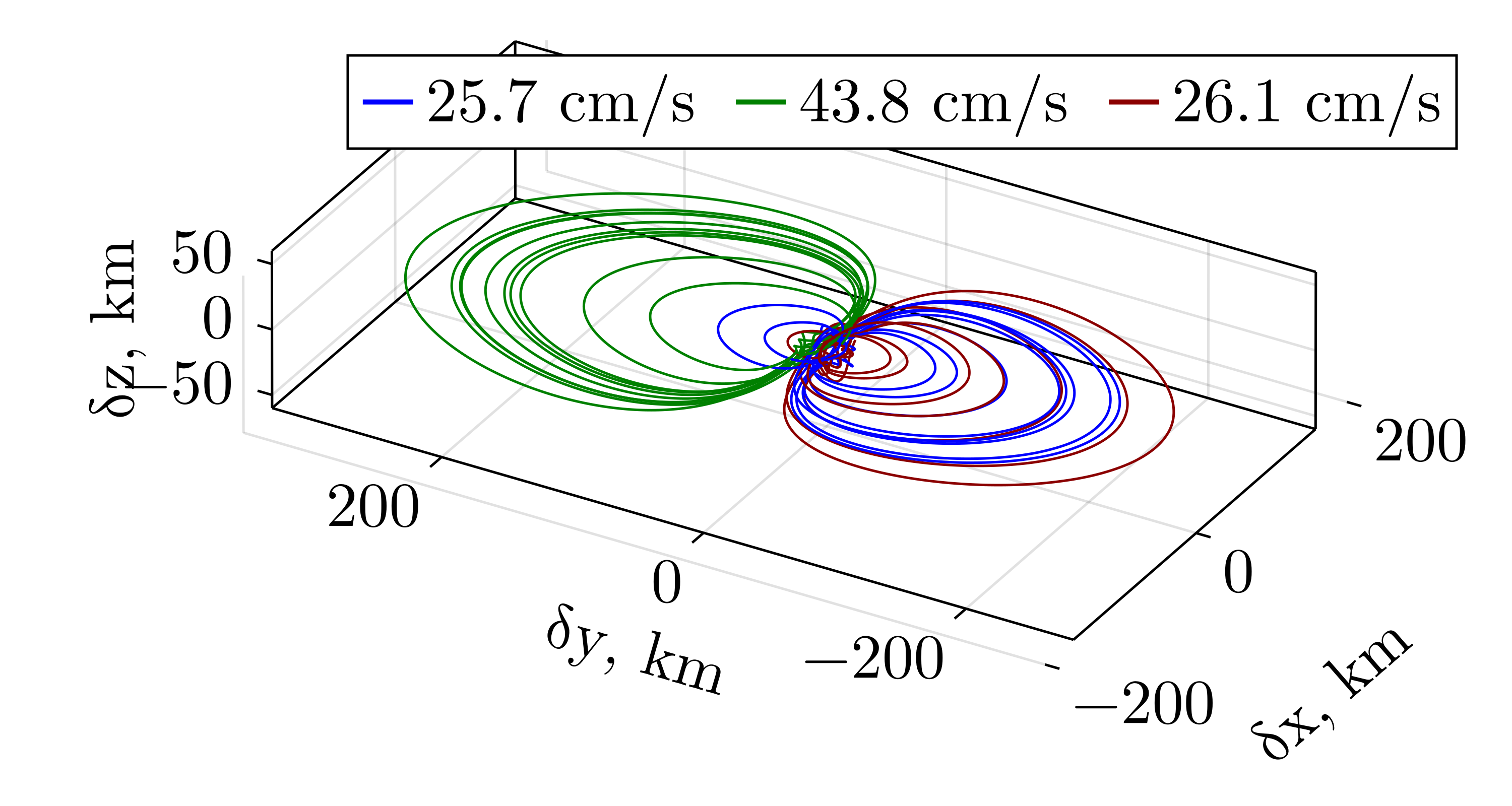}
        \caption{Perilune segments separated by RW-plane}
         \label{fig:relative_trajectory_RTN_mc46_n3}
     \end{subfigure}
     \hfill %\\
     \begin{subfigure}[b]{0.49\textwidth}
         \centering
         \includegraphics[width=\textwidth]{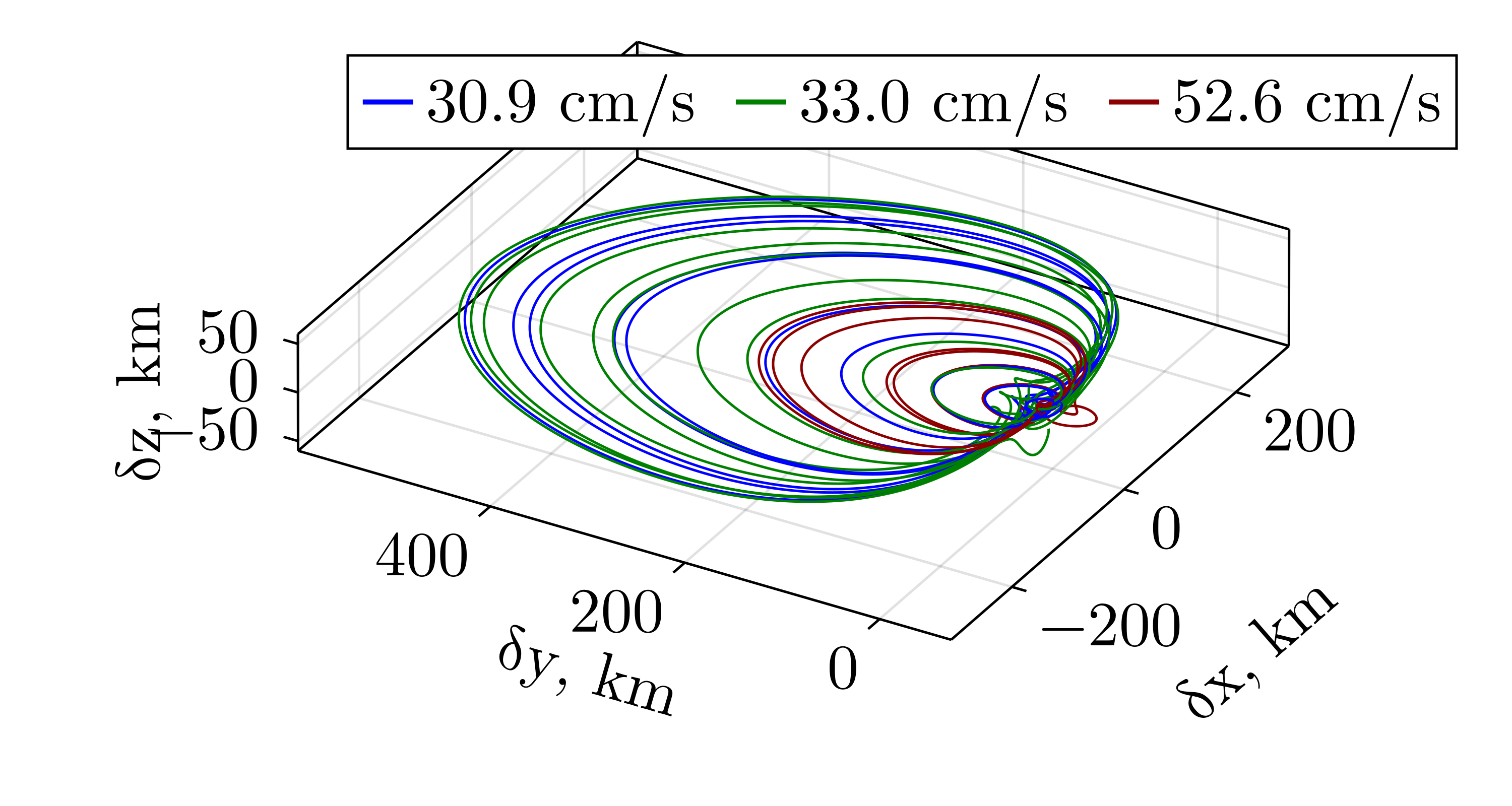}
        \caption{Perilune segments on one side of RW-plane}
         \label{fig:relative_trajectory_RTN_mc9_n3}
     \end{subfigure}
    \caption{Sample reference-relative trajectories in 3-spacecraft formation in RSW frame over 10 revolutions}
    \label{fig:relative_trajectory_RTN_n3}
\end{figure}

\rev{Figure~\ref{fig:relative_trajectory_RTN_n3} shows sample reference-relative trajectories over the 10 revolution durations.
As with the 2-spacecraft case, we see a variety of perilune segment orientations across the Monte Carlo runs.
As an example, Figure~\ref{fig:relative_trajectory_RTN_mc46_n3} shows a case where the paths during perilune passage spread on either side of the RW-plane, while Figure~\ref{fig:relative_trajectory_RTN_mc9_n3} shows a case where all spacecraft fly on the same side of the RW-plane.
As with the 2-spacecraft case, note there are no constraints applied to prescribe a certain geometry on the perilune passages.
}

% Table of costs with M3
\begin{table}[t]
\centering
\footnotesize
\caption{MPC performance summary over 10 revolutions ($\sim 65.5$ \SI{}{days}) for 3-spacecraft formation}
\begin{tabular}{@{}llllllllllll@{}}
\toprule
\multirow{2}{*}{Exp.} &
\multirow{2}{*}{\begin{tabular}[c]{@{}l@{}}MVOCP\\feasible\end{tabular}} &
\multirow{2}{*}{\begin{tabular}[c]{@{}l@{}}CT\\feasible\end{tabular}} &
\multirow{2}{*}{\begin{tabular}[c]{@{}l@{}}$\mathrm{err}_{\Delta r}$,\\\SI{}{km}\end{tabular}}  & 
\multicolumn{2}{c}{Spacecraft 1} & 
\multicolumn{2}{c}{Spacecraft 2} & 
\multicolumn{2}{c}{Spacecraft 3} & 
\multicolumn{2}{c}{Formation Total} 
\\ 
& & &
  & Mean,~\SI{}{cm/s}& $3$-$\sigma$,~\SI{}{cm/s}
  & Mean,~\SI{}{cm/s}& $3$-$\sigma$,~\SI{}{cm/s}
  & Mean,~\SI{}{cm/s}& $3$-$\sigma$,~\SI{}{cm/s}
  & Mean,~\SI{}{cm/s}& $3$-$\sigma$,~\SI{}{cm/s}
\\ \midrule
I &  95/100 &  95/100 & 0.00 & 27.53 & 50.68 & 36.37 & 62.02 & 39.65 & 75.50 & 103.54 & 150.79 \\
II &  100/100 &  (6/100) & 18.85 & 114.06 & 153.93 & 73.49 & 148.09 & 115.96 & 200.76 & 303.51 & 481.19 \\
\bottomrule
\end{tabular}
\label{tab:cost_statistics_3spacecraft}
\end{table}

% n=3, range history for isoperimetric

% n=3, range history for nodes

% trajectory

\subsubsection{Extensions to Formations Beyond Three Spacecraft}
\rev{There are two effects that negatively affect the station keeping problem for formations with increasing number of satellites.
First, even from a deterministic point of view, increasing the number of spacecraft results in a decreasing feasible solution space of the MVOCP; that is, there is a limit to the number of spacecraft that can sustain a bounded relative distance among one another and remain in the vicinity of the reference~\cite{Brown2026-sr}.
While the introduction of maneuvers helps alleviate this issue by introducing controlled trajectories as options for the spacecraft, the two maneuver per revolution cadence may not be sufficient to sustain such a trajectory.
Second, the effect of uncertainties, even at a level consistent with the 2-spacecraft formation, makes constraint violation more likely due to the higher occupancy of the volume around the reference NRHO, thus requiring stricter constraint tightening, and possibly more frequent control maneuvers.
}

% ===================================================================== %
\section{Conclusions}
\label{sec:conclusions}
In this work, we developed a model predictive control (MPC) framework for formation flight on libration point orbits (LPOs),~\rev{incorporating continuous-time (CT) path constraints via isoperimetric reformulation with constraint tightening.}
Through numerical experiments, we demonstrated that our proposed framework provides successful guidance for the formation over an extended duration in the presence of realistic uncertainties.
Using the multi-vehicle optimal control problem (MVOCP) with CT path constraints, we found that a 2-spacecraft formation satisfying both bounded separation and relative Sun phase angle constraints at apolune segments can be achieved at \rev{a yearly per-spacecraft $3$-$\sigma$ control cost of $280 \sim 310$~\SI{}{cm/s}}.
We illustrated the need for isoperimetric reformulation, as \rev{naively enforcing the path constraints at the discretized control nodes results in frequent inter-sample constraint violations.}
We also reported the variability on the apolune segments that satisfy the relative Sun phase angle constraints as the relative orientation of $\mathcal{F}_{\rm RSW}$ and $\mathcal{F}_{\rm SM}$ undergoes its roughly 2-month cycle.
\rev{Finally, we extended the experiment into a 3-spacecraft formation. With no additional tuning of constraint tightening parameters from the 2-spacecraft formation case, the MPC framework still provides CT feasible station-keeping for 95\% of the cases, with a station-keeping cost that is $\times 1.5 \sim \times 2$ of the 2-spacecraft formations.}

% ===================================================================== %
\section{Appendix}
\subsection{Jacobian Expressions}
\label{appendix:jacobian}
% We also introduce a linear approximation model for the propagation of state perturbation, to be used within convex subproblems of the SCP.
% Under first-order approximation, the additive perturbation on the state at some initial time $t_k$, denoted by $\delta \xbold_i(t_k)$, is linearly mapped to some future time $t \geq t_k$ via the state-transition matrix (STM) $\Phibold_i \in \mathbb{R}^{6 \times 6}$,
% \begin{equation}
%     \delta \xbold_i(t) = \Phibold_i(t,t_k) \delta \xbold_i(t_k),
% \end{equation}
% where $\Phibold(t,t_k)$ is obtained by integrating the matrix-valued initial value problem (IVP),
% \begin{equation}    \label{eq:STM_propagation}
%     \begin{aligned}
%         \dot{\boldsymbol{\Phi}}_i (t,t_k)
%         &= \dfrac{\partial \fbold(\xbold_i(t), t)}{\partial \xbold_i} \boldsymbol{\Phi}(t,t_k)
%         ,
%         \\
%         \boldsymbol{\Phi}_i(t_k,t_k) &= \Ibold_{6}
%         .
%     \end{aligned}
% \end{equation}
Computing $\boldsymbol{\Psi}_{j+1,j|k}$ requires integrating the matrix-valued IVP
\begin{equation}
    \begin{aligned}
        \dot{\boldsymbol{\Psi}}_i (t,t_k)
        &=
        \left. \dfrac{\partial \fbold_{\rm aug}(\Zbold(t),t)}{\partial \Zbold} \right|_{\bar{\Zbold}(t)}
        \boldsymbol{\Psi}(t,t_k)
        ,
        \\
        \boldsymbol{\Psi}_i(t_k,t_k) &= \Ibold_{6M + L}
        ,
    \end{aligned}
\end{equation}
along with $\Zbold$, which necessitates the Jacobian $\partial \fbold_{\rm aug}(\Zbold(t),t)/\partial \Zbold \in \mathbb{R}^{(6M+L) \times (6M+L)}$,
\begin{equation}
    \dfrac{\partial \fbold_{\rm aug}(\Zbold(t),t)}{\partial \Zbold}
    =
    \begin{bmatrix}
        \dfrac{\partial \fbold_{\rm concat}(\Xbold(t),t)}{\partial \Xbold} & \boldsymbol{0}_{6M \times L}
        \\
        \dfrac{\partial \fbold_{\rm slack}(\Xbold(t),t)}{\partial \Xbold} & \boldsymbol{0}_{L \times L}
    \end{bmatrix}
    .
\end{equation}
The upper-left block is the Jacobian of the concatenated dynamics,
\begin{equation}
    \dfrac{\partial \fbold_{\rm concat}(\Xbold(t),t)}{\partial \Xbold} 
    = \operatorname{blkdiag}{\left(
        \dfrac{\partial \fbold(\xbold_0(t),t)}{\partial \xbold_0},
        \ldots,
        \dfrac{\partial \fbold(\xbold_{M-1}(t),t)}{\partial \xbold_{M-1}}
    \right)} 
    .
\end{equation}
The $\ell^{\rm th}$ row of the lower-left block ${\partial \fbold_{\rm slack}(t)}/{\partial \Xbold} \in \mathbb{R}^{1 \times 6M}$ contains the sensitivities of the $\ell^{\rm th}$ path constraint $g_{\ell}$ with respect to $\Xbold$,
\begin{equation}
    \left(\dfrac{\partial \fbold_{\rm slack}(t)}{\partial \Xbold}\right)_{[\ell,:]}
    =
    2 W_{\ell} \max{\left[ 0, g_{\ell}(\Xbold(t),t) \right]} \dfrac{\partial g_{\ell}(\Xbold(t),t)}{\partial \Xbold}
    .
\end{equation}
When re-initializing the linearized dynamics~\eqref{eq:MVOCP_ct_convex_subproblem_dynamics} with an updated reference solution $\bar{\Zbold},\bar{\Ubold}$, both the augmented state $\bar{\Zbold}$ and STMs $\boldsymbol{\Psi}$ are integrated together (applying controls $\bar{\Ubold}$ along the way), resulting in a set of $(6M + L) + (6M + L)^2$ ordinary differential equations (ODE).
The high dimension of the ODE, together with the nonlinearity and time-dependence of the HFEM, makes this process by far the most computationally expensive part of the SCP algorithm.
Nevertheless, the integration between each pair of adjacent control times $t_{j|k}$ and $t_{j+1|k}$ is decoupled and can be embarrassingly parallelized.

\bibliography{references}

\end{document}